\documentclass{aa} 
\usepackage[english]{babel}
\usepackage{amsmath}
\usepackage{bm}
\usepackage{txfonts}
\usepackage{color} % Benutzung von Farben
\usepackage{graphicx} % Bilder, Abbildungen, Graphiken  [draft], [dvips], [pdftex]
\usepackage{subfig}
\usepackage[round]{natbib}
\bibpunct{(}{)}{;}{a}{}{,} %to follow the A&A style

\begin{document}
%\graphicspath{{Images/}{../}}

%\definecolor{lightyellow}{rgb}{1,1,.7}
%\definecolor{darkblue}{rgb}{0,0,.5}
%\definecolor{lightblue}{rgb}{.8,.85,1}
%\definecolor{red}{rgb}{1,0,0}
%\definecolor{green}{rgb}{0,1,0}
%\definecolor{blue}{rgb}{0,0,1}
%\definecolor{pink}{rgb}{1,0,1}
%\definecolor{testcolor}{rgb}{0.8,0.8,1}
%\definecolor{black}{rgb}{0,0,0}

\authorrunning{O. Kuzmychov \& S.V. Berdyugina}
\titlerunning{Paschen-Back effect in CrH and its astrophysical application}

\title{Paschen-Back effect in the CrH molecule\\
       and its application for magnetic field measurements\\
       on stars, brown dwarfs, and hot exoplanets}

%\subtitle{}

\author{O. Kuzmychov\inst{1} \and S.V. Berdyugina\inst{1,2}}

\institute{
   Kiepenheuer-Institut f\"ur Sonnenphysik, Sch\"oneckstr. 6, 79104
   Freiburg, Germany\\
   \email{oleksii.kuzmychov@kis.uni-freiburg.de, sveta@kis.uni-freiburg.de}
\and
   NASA Astrobiology Institute, University of Hawaii, 2680 Woodlawn Dr.,
   Honolulu, HI, USA
}

\date{Received xxx x, 2012; accepted xxx x, 2012}

\offprints{O. Kuzmychov, \email{oleksii.kuzmychov@kis.uni-freiburg.de}}  

%\today

\abstract
 {} 
{
We investigated the Paschen-Back effect in the (0,0) band of the A$^6\Sigma^+$--X$^6\Sigma^+$ system of the CrH molecule, and we examined its potential for estimating magnetic fields on stars and substellar objects, such as brown dwarfs and hot exoplanets.
} 
{
%We diagonalize the Hamiltonian matrix represented in the wave functions of Hund's coupling case (a) and obtain numerically the eigenvalues and eigenfunctions in Hund's coupling case (b). Taking into account the matrix elements of the magnetic Hamiltonian represented in the wave functions in Hund's coupling case (b), we diagonalize the Hamiltonian matrix again to obtain the eigenvalues and eigenfunctions in a Paschen-Back regime at different magnetic field strengths. Using the Unno solution of the polarized radiative transfer equation in stellar atmospheres we obtain the Stokes profiles for the CrH bands.\\
We carried out quantum mechanical calculations to obtain the energy level structure of the electronic-vibrational-rotational states considered both in the absence and in the presence of a magnetic field. Level mixing due to magnetic field perturbation (the Paschen-Back effect) was consistently taken into account. Then, we calculated frequencies and strengths of transitions between magnetic sublevels. Employing these results and solving numerically a set of the radiative transfer equations for polarized radiation, we calculated Stokes parameters for both the individual lines and the \mbox{(0,0)} band depending on the strength and orientation of the magnetic field.
}
{
We demonstrate that magnetic splitting of the individual CrH lines shows a significant asymmetry due to the Paschen-Back effect already at $1$~G field. This leads to a considerable signal in both circular and linear polarization, up to $30$\% at the magnetic field strength of $\geq 3$~kG in early L dwarfs. The polarization does not cancel out completely even at very low spectral resolution and is seen as broad-band polarization of a few percent. Since the line asymmetry depends only on the magnetic field strength and not on the filling factor, CrH lines provide a very sensitive tool for direct measurement of the stellar magnetic fields on faint cool objects, such as brown dwarfs and hot Jupiters, observed with low spectral resolution. 
}
{}

% max 6 issues

\maketitle

%%%%%%%%%%%%%%%%%%%%%%%%%%%%%%%%%%%%%%%%%%%%%%%%%%%%%%%%%%%%%%%%%%%%
\section{Introduction}
%%%%%%%%%%%%%%%%%%%%%%%%%%%%%%%%%%%%%%%%%%%%%%%%%%%%%%%%%%%%%%%%%%%%
 
During the last decade the spectropolarimetry of diatomic molecules became an important tool for studying stellar atmospheres and magnetism \citep[see review by][]{berdyugina2011}. At temperatures below $4000$~K, a number of diatomic molecules readily exist in the atmospheres of cool stars \citep[see review on molecules of astrophysical interest by][]{bernath2009}. Because molecules have a more complex energy level structure then atoms, they provide a powerful tool for measuring stellar magnetic fields via the Zeeman and Paschen-Back effects \citep{berdyugina2002, berdyugina2003, berdyugina2005}.

Brown dwarfs are substellar objects with surface temperatures below about $2200$~K and masses in the range of $13-80$ Jupiter masses. Since \citet{nakajima1995} confirmed the existence of these objects, their spectra posed a challenge to understanding the physics of cool substellar objects. The spectra of L-type brown dwarfs show metal hydride bands, mainly FeH and CrH, as their dominant molecular feature \citep[][]{kirkpatrick1999}. For example, the \mbox{(1,0)} and \mbox{(0,0)} bands of the FeH, which are located at $8692$~\AA\, and $9896$~\AA\, respectively, are very prominent in the M and early-L dwarfs. The CrH \mbox{(0,0)} band from the A$^6\Sigma$ -- X$^6\Sigma$ electronic system appears at $8610$~\AA\ and is seen in all L-type dwarfs, reaching its maximum strength at mid-L dwarfs. A quantum mechanical study of the FeH molecule aimed at the investigation of its capability for measuring the stellar magnetic fields was first carried out by \citet{afram2008}. The authors simulated the polarization signals in the individual FeH lines from the F$^4\Delta$--X$^4\Delta$ system and compared them with the observational data. A similar work, but dealing only with the intensity signals of the FeH lines, was done by \citet{shulyak2010}. 

In this paper we examine the interaction of the CrH molecule with a magnetic field to develop a new diagnostic for measuring magnetic fields in cool substellar objects. The range of objects this diagnostic can be applied to include sunspots \citep[e.g.,][]{engvold1980,sriramachandran2011}, starspots \citep[e.g.,][]{berdyugina2011}, M- and L-type  dwarfs \citep[e.g.,][]{pavlenko1999}, and hot Jupiters of the same temperature range as cool dwarfs are. Even though CrH has not yet been detected in hot Jupiters, it is plausible that this molecule can be an important source of opacity in their atmospheres. Therefore, its high magnetic sensitivity (as we show in this paper) can be useful for measuring magnetic fields in hot Jupiters too.

The angular momenta coupling in the CrH A$^6\Sigma$ and X$^6\Sigma$ electronic states indicate that the appropriate limiting situation is Hund's case~\textit{b}. An important feature of case~\textit{b} is that the $P$ and $R$ rotational branches lie farther apart in wavelength as compared to case~\textit{a}, and they also have opposite polarities in the Paschen-Back regime \citep[hereafter PBR,][]{berdyugina2005}. In other words, the net-polarization signal that comes from the individual lines due to the Paschen-Back effect accumulates within a rotational branch and results in the \emph{broad-band polarization}, which can be seen at lower spectral resolution \citep[cf. the work on the CH molecule in magnetic white dwarfs by][]{berd2007}. This makes the CrH molecule attractive for polarimetric observations of faint, magnetized substellar atmospheres which cannot yet be observed with high spectral resolution.

The outline of the paper is as follows.

First, we calculate the rotational level structure of the \mbox{(0,0)} vibrational band of the A$^6\Sigma$ and X$^6\Sigma$ electronic states using the effective Hamiltonian for a $^6\Sigma$ state given by \citet{ram1993}. Using the rotational energy values obtained and taking into account the matrix elements of the magnetic Hamiltonian represented in the case~\textit{b} wave functions, as in \citet{berdyugina2005}, we calculate the magnetic level structure in both electronic states for different magnetic field strengths (Sect.~\ref{sec: energy_levels}). Strengths of the transitions between the magnetic sublevels are calculated in Sect.~ \ref{sec: strengths}.

In Sect.~\ref{sec: stokes}, we calculate the synthetic Stokes profiles for both the individual transitions and the entire \mbox{(0,0)} band by employing the results from Sects.~\ref{sec: energy_levels} and \ref{sec: strengths} and solving numerically a set of the radiative transfer equations. We employ three model atmospheres from the Phoenix-Drift grid by \citet{witte2009} with the effective temperatures of $2500$~K, $2000$~K, and $1500$~K. These three models represent roughly the physical conditions in the atmospheres of late-M, early- and mid-L dwarfs, as well as of a possible hot Jupiter.

We summarize our results in Sect. \ref{sec: conclusions}. We conclude that the CrH A$^6\Sigma^+$--X$^6\Sigma^+$ system is a sensitive diagnostic tool for studying the magnetic fields in cool substellar objects. Measurable signals can be observed even at fields of a few G and with low spectral resolution, which is advantageous for faint objects, such as brown dwarfs and hot Jupiters. This occurs thanks to the Paschen-Back effect which causes an asymmetry of the magnetic components of a spectral line. This is in contrast to the Zeeman regime (hereafter ZR), where splitting of a spectral line and strengths of the individual magnetic components are always symmetric.

%\textcolor{green}{At the end of the introduction the outline of the paper may be described.}

%\section{Theory / Calculations}% or mathematical derivations

%%%%%%%%%%%%%%%%%%%%%%%%%%%%%%%%%%%%%%%%%%%%%%%%%%%%%%%%%%%%%%%%%%%%
\section{Energy level structure}\label{sec: energy_levels}
%%%%%%%%%%%%%%%%%%%%%%%%%%%%%%%%%%%%%%%%%%%%%%%%%%%%%%%%%%%%%%%%%%%%

%-------------------------------------------------
\subsection{Rotational level structure}\label{subsec: non-mag}
%-------------------------------------------------

We calculate the rotational level structure of the electronic states X$^6\Sigma^+$ and A$^6\Sigma^+$ of the CrH molecule following \citet{ram1993}, who employed the basis wave functions in Hund's case~\emph{a}. Earlier, a similar work was done by \citet{kleman1959}, who worked from the basis wave functions in case~\emph{b}.

The total Hamiltonian of the CrH molecule in the absence of an external magnetic field, $\hat H_\mathrm{b}$, can be partitioned as
\begin{equation}
\hat{H}_\mathrm{b}=\hat{H}_{0}+\hat{V},
\label{eq: ham1}
\end{equation} 
where $\hat{H}_0$ represents all the electronic and vibrational terms, and $\hat{V}$ is a perturbation operator. The subscript ``b'' indicates that, after taking account of all interactions contained in $\hat H_\mathrm{b}$, the electronic states considered occur in case~\textit{b}. The perturbation ope\-ra\-tor $\hat{V}$ is given by
\begin{equation}
\hat V=\hat H_\mathrm{rot}+\hat H_\mathrm{cd}+\hat H_\mathrm{so}+\hat H_\mathrm{sr}+\hat H_\mathrm{ss},
\label{eq: hprime}
\end{equation} 
where $\hat H_\mathrm{rot}$ is the rotational Hamiltonian of the nuclei; $\hat H_\mathrm{cd}$ takes into account the centrifugal distortion of the molecule; and $\hat H_\mathrm{so}$, $\hat H_\mathrm{sr}$, and $\hat H_\mathrm{ss}$ arise respectively from the spin-orbital, spin-rotational, and spin-spin interactions. We note that $\hat{H}_\mathrm{so}$ is zero for case~\emph{b}.

Representing the total Hamiltonian $\hat H_\mathrm{b}$ in the wave functions in case~\textit{a}, $|\mathrm{a}\rangle$, we have
\begin{equation} 
\mathcal H_\mathrm{b}=\langle \mathrm{a}|\hat{H}_0|\mathrm{a}\rangle+\langle \mathrm{a}|\hat{V}|\mathrm{a}\rangle+\mathcal O\left(\langle \mathrm{a}|\hat{V}^2|\mathrm{a}\rangle\right)+\ldots\, .
\label{eq: heff}
\end{equation}
As a result of the perturbation $\hat V$, the Hamiltonian matrix $\mathcal H_\mathrm{b}$ becomes non-diagonal. However, the first term is diagonal with respect to the electronic and vibrational state and gives the unperturbed (and still degenerate) energy level $E_0$. The se\-cond and the third terms are the corrections to $E_0$ of the first and of the second order, respectively. The dots refer to the higher-order corrections to $E_0$. We considered here up to the third-order and forth-order corrections arising from the spin-rotational and spin-spin interactions, respectively.

We diagonalize numerically the effective Hamiltonian \citep[see, e.g.,][]{brown2003} for the rotational levels of a $^6\Sigma$ state that contains the $\hat V$-dependent terms from the right-hand part of Eq.~(\ref{eq: heff}) and that is given by \citet[Table I]{ram1993}. As a result, we obtain the eigenvalues $E_{\Sigma J}$, and the eigenvectors of $\hat H_\mathrm{b}$ in the absence of an external magnetic field. The representation $\langle \mathrm{b}|\hat{H}_\mathrm{b}|\mathrm{b}\rangle$ in the wave functions in case~\emph{b} is then diagonal with respect to the rotational state $|\Sigma J\rangle$.

After the energy values $E_{\Sigma J}$ in both the upper A$^6\Sigma$ and the lower X$^6\Sigma$ electronic states have been obtained, one can calculate the energies of all possible electronic transitions in the \mbox{(0,0)} band allowed by the quantum mechanical selection rules. The line positions calculated are in a good agreement with those analyzed by \citet{ram1993} and computed by \citet{burrows2002}.

%The energy values $E_{\Sigma J}$ in the absence of an external magnetic field were calculated earlier by \citet{kleman1959}, who worked from the wave functions in case~\textit{b}, and \citet{ram1993}.

%-------------------------------------------------
\subsection{Magnetic level structure}
%-------------------------------------------------

Now we consider the CrH molecule in the presence of an external magnetic field and examine its impact on the rotational structure calculated in Sect. \ref{subsec: non-mag}. 

When neglecting the interaction between the magnetic levels, the energy shift for a level $M$ at the magnetic field strength $H$, with respect to its energy value in the absence of an external magnetic field, obtains as follows:
\begin{equation}
\Delta E=g\mu_0 MH,
\label{eq: zeeman_regime}
\end{equation}
where $g$ and $\mu_0$ are the Land\'{e} factor and the Bohr magneton, respectively. The case, when the expression (\ref{eq: zeeman_regime}) holds, is called \textit{Zeeman regime} \citep[see, e.g.,][]{berdyugina2002}. However, as we show in Sect. \ref{subsec: stokes_profiles}, the ZR is practically not applicable for the CrH $^6\Sigma$ electronic states (especially for the lower state), so the interaction between magnetic levels cannot be neglected.

To take into account the interaction between the magnetic levels, i.e., to calculate the Paschen-Back effect, we consider the total Hamiltonian of the molecule in the form
\begin{equation}
\hat H_\mathrm{mag}=\hat H_\mathrm{b}+\hat H_H,
\label{eq: ham2}
\end{equation}
where the interaction of the molecule with an external magnetic field $\hat H_H$ is considered to be smaller than the intrinsic molecular interactions $\hat H_\mathrm{b}$. More precisely, we consider here the case when $\hat{H}_H$ is significantly smaller than the sum $\hat{H}_\mathrm{rot}+\hat{H}_\mathrm{cd}$, but it can be comparable or larger than $\hat{H}_\mathrm{sr}$ and $\hat{H}_\mathrm{ss}$. This implies that we do not take into account the perturbation $\hat{H}_H$ on the rotational structure, and we consider its impact on the fine level structure only. We note that the values of $\hat{H}_\mathrm{rot}$, $\hat{H}_\mathrm{cd}$, $\hat{H}_\mathrm{sr}$, and $\hat{H}_\mathrm{ss}$ depend on the angular momentum quantum numbers $J$ and $N$, so the degree of the level mixing for a given magnetic field strength varies within the fine structure (here, it decreases for higher rotational levels). In other words, for two adjacent rotational levels the mixing occurs at a certain magnetic field strength, which can be deduced from measurements when the entire molecular band is modeled.

To obtain the eigenvalues of $\hat H_\mathrm{mag}$, we proceed in the same way as we did for the rotational level structure. Since $\hat H_\mathrm{b}$ is diagonal with respect to a rotational state $|\Sigma J\rangle$, we express $\hat H_\mathrm{mag}$ in the eigenfunctions of $\hat H_\mathrm{b}$,%, namely $|\mathrm{b}\rangle$, and we obtain the effective Hamiltonian as follows:
\begin{equation}  
\mathcal H_\mathrm{mag}=\langle \mathrm{b}|\hat H_\mathrm{b}|\mathrm{b}\rangle+\langle \mathrm{b}|\hat H_H|\mathrm{b}\rangle,
\label{eq: hmageff}
\end{equation}
where the second term is non-diagonal with respect to the rotational state $|\Sigma J\rangle$. We limit our investigation of the molecular level structure to the first-order correction to the energy value $E_{\Sigma J}$, which is given by the second term in the right-hand part of Eq.~(\ref{eq: hmageff}). Thus, the Hamiltonian matrix $\mathcal H_\mathrm{mag}$ will be specified as
%\footnote{The interaction of the magnetic levels belonging to the different rotational levels $N$ will not be taken into account.}
%(see Fig. \ref{fig: eflevels} and \ref{fig: maglevels})
\begin{equation}
 \begin{array}{c|cccccc}
 &\rotatebox{90}{$\left|- \frac{5}{2}\right\rangle$}&\rotatebox{90}{$\left|- \frac{3}{2}\right\rangle$}&\rotatebox{90}{$\left|- \frac{1}{2}\right\rangle$}&\rotatebox{90}{$\left|+ \frac{1}{2}\right\rangle$}&\rotatebox{90}{$\left|+ \frac{3}{2}\right\rangle$}&\rotatebox{90}{$\left|+ \frac{5}{2}\right\rangle$}\\ \hline
 \left\langle -\frac{5}{2}\right|&\mathrm{i}&\mathrm{ii}&&&&\\
 \left\langle -\frac{3}{2}\right|&\mathrm{ii}&\mathrm{i}&\mathrm{ii}&&&\\
 \left\langle -\frac{1}{2}\right|&&\mathrm{ii}&\mathrm{i}&\mathrm{ii}&&\\
 \left\langle +\frac{1}{2}\right|&&&\mathrm{ii}&\mathrm{i}&\mathrm{ii}&\\
 \left\langle +\frac{3}{2}\right|&&&&\mathrm{ii}&\mathrm{i}&\mathrm{ii}\\
 \left\langle +\frac{5}{2}\right|&&&&&\mathrm{ii}&\mathrm{i},
 \end{array}
\label{eq:hammag}
\end{equation}
where the first column and the first row refer to the eigenfunctions in case~\textit{b}. The perturbation matrix has a trigonal form because, according to the first-order approximation, only the perturbation of the adjacent levels within the multiplet structure (fine structure) needs to be considered. Consequently, only the matrix ele\-ments forming the main and secondary diagonals are distinct from zero. They are \citep[][Table A.4]{berdyugina2005}
\begin{align*}
\mathrm{i:}\; &E_{\Sigma J}+\mu_0\frac{MH}{J(J+1)}\{J(J+1)+S(S+1)-N(N+1)\},\label{eq: matrh1}\\ 
\mathrm{ii:}\; &E_{\Sigma J}+\mu_0\frac{H}{J+1}\times \\ 
   &\sqrt{\frac{(J+1)^2-M^2}{(2J+1)(2J+3)}}\left\{ \sqrt{(J+S+1)(J+S+2)-N(N+1)}\right.\times \\ 
   &\left. \sqrt{N(N+1)-(J-S)(J-S+1)}\right\}, 
%\label{eq: matrh}
\end{align*}
where the quantities $\mu_0$, $M$, $H$, and $S$ are, respectively, the Bohr magneton, the magnetic quantum number, the magnetic field strength, and the total electron spin. The rotational quantum numbers $N$ and $J$ obey the relation $N=J+\Sigma$, where $\Sigma$ is the projection of $S$ on the inter-nuclear axis. Each rotational level $N$ consists of six fine-structure levels, denoted by $J$ ($J=N+5/2,\, N+3/2,\, \ldots,\, N-5/2$), owing to six possible projections of the total spin on the inter-nuclear axis. They are degenerate in the absence of an external magnetic field, while in the presence of a magnetic field each level $J$ splits into $2J+1$ magnetic levels $M$. 

By diagonalizing the Hamiltonian matrix (\ref{eq:hammag}) we obtain the eigenvalues $E_{\Sigma JM}$ and the eigenvectors $C_{\Sigma JM}$ of $\hat H_\mathrm{mag}$. In Sect.~\ref{subsec: pbr_strengths} we will make use of $C_{\Sigma JM}$ to calculate the strengths of the magnetic transitions.

\begin{figure}
  \centering
  \subfloat{\label{fig: N_u=4}\resizebox{\hsize}{!}{\includegraphics{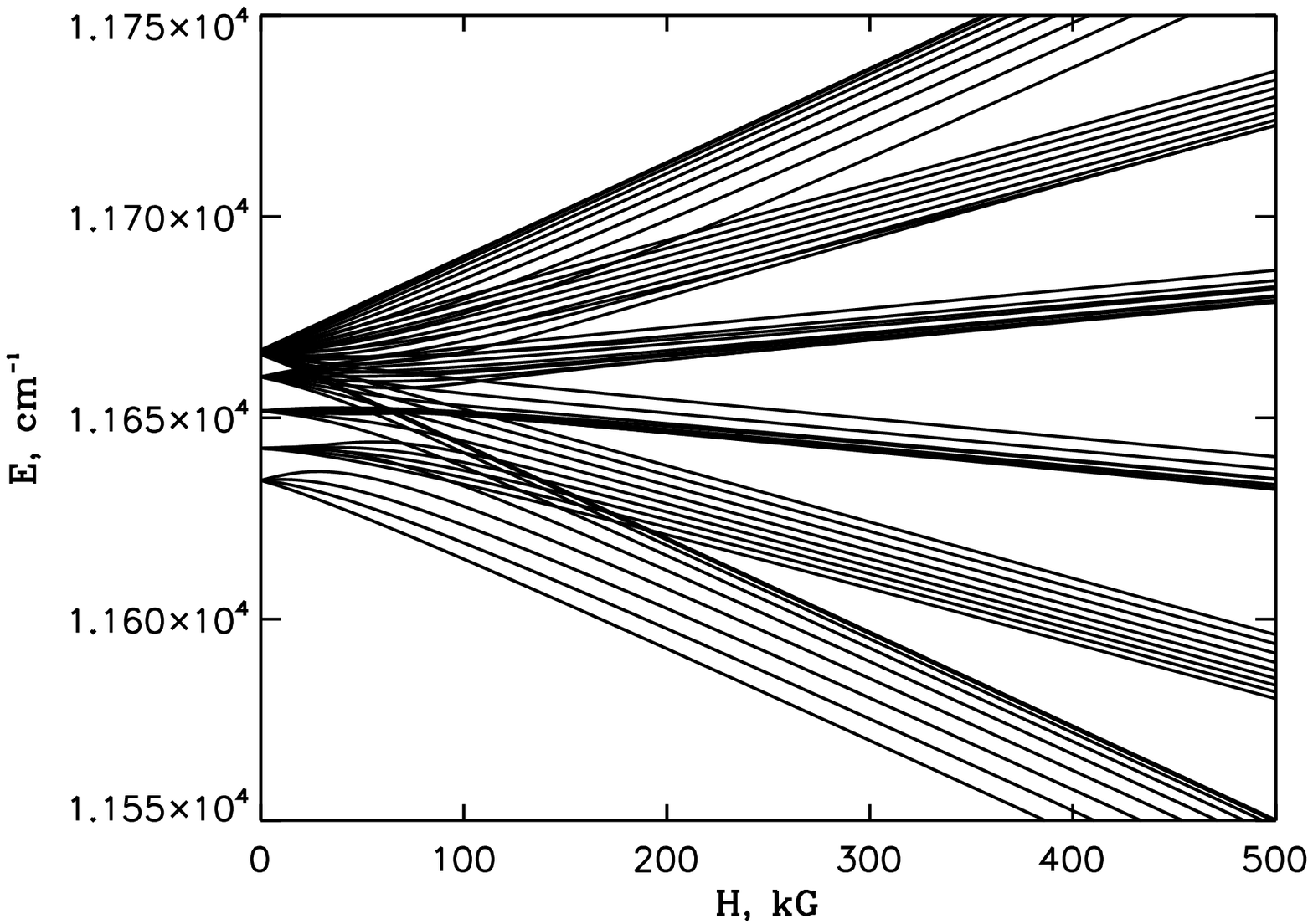}}}\\                
  \subfloat{\label{fig: N_l=3}\resizebox{\hsize}{!}{\includegraphics{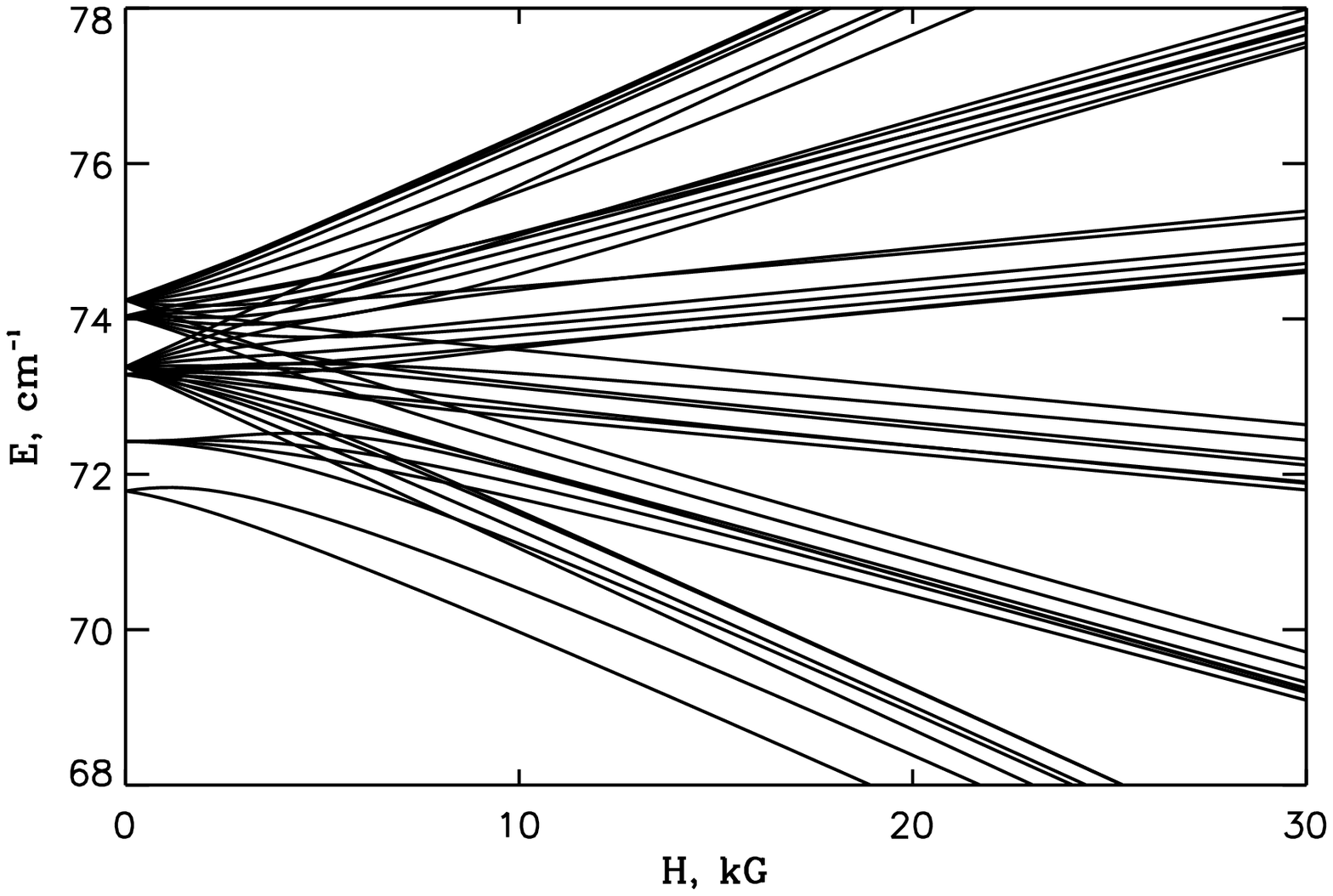}}}
  %\subfloat[A mouse]{\label{fig:mouse}\includegraphics[width=0.3\textwidth]{mouse}}
  \caption{Fine structure of the rotational levels $N^\prime=4$ (top) and $N^{\prime\prime}=3$ (bottom) in the presence of an external magnetic field. Each of the six fine structure levels $J$ splits into $2J+1$ magnetic levels $M$,  which are perturbed by nearby levels with the same $M$ number.}
  \label{fig: magfield}
\end{figure}

Figure~\ref{fig: magfield} shows splitting of the fine structure le\-vels of the upper and lower rotational levels $N^\prime=4$ and $N^{\prime\prime}=3$, respectively~\footnote{Throughout this paper, we use single and double primes to indicate upper and lower levels, respectively.}, 
depending on the magnetic field strength. At a weak magnetic field ($\lesssim\!100$ G for $N^\prime=4$ and $\lesssim\!10$ G for  $N^{\prime\prime}=3$), the splitting is linear with the field strength and the interaction between the magnetic le\-vels with the same quantum number $M$ is negligible (ZR). As the field strength increases, the magnetic levels spread out and the levels with the same quantum number $M$ repel each other. As a result, the magnetic levels form a blended structure in the PBR. Subsequently, at stronger magnetic fields ($\gtrsim 300$ kG for $N^\prime=4$ and $\gtrsim 15$ kG for $N^{\prime\prime}=3$), the $M$ levels are rearranged into a new multiplet structure corresponding to the six spin projections on the magnetic field direction (complete PBR).

%Thus, starting from $\approx 250$~\mbox{kG}, the complete Paschen-Back regime occurs for the rotational level $N^\prime=4$ (see again Fig.~\ref{fig: magfield}, top). The magnetic structure of the rotational level $N^{\prime\prime}=3$ entries the Paschen-Back regime at the field strengths $\approx~15$ \mbox{kG} (see Fig.~\ref{fig: magfield}, bottom).

Once the energy values $E_{\Sigma JM}$ for both the upper A$^6\Sigma$ and the lower X$^6\Sigma$ electronic states have been obtained, magnetic transition energies can be calculated as their difference. Transition wavelengths are then obtained by correcting for the air refraction index according to \citet{edlen1966}. We note that in addition to the main branch transitions ($\Delta J = \Delta N$), we also consider here the satellite ($\Delta J \ne \Delta N$) and forbidden transitions as discussed by \citet{berdyugina2005}.

%\begin{figure}%[tp]
%      \resizebox{\hsize}{!}{\includegraphics{N_u=4.pdf}}%PBH1cut.png
%      \caption{Dependence of the magnetic level structure (A$^6\Sigma$, $v^\prime=0$, $N^\prime=4$) on an external magnetic field. Zeeman regime is an appropriate limiting situation up to few \mbox{kGs}. Starting from $\approx 250$~\mbox{kGs}, the complete Paschen-Back regime occurs.}
%      \label{fig: PBH1cut}
%\end{figure}
%\begin{figure}%[bp]
%		  \resizebox{\hsize}{!}{\includegraphics{N_l=3.pdf}}%PBH2cut.png
%      \caption{Dependence of the magnetic level structure (X$^6\Sigma$, $v^{\prime\prime}=0$, $N^{\prime\prime}=4$) on an external magnetic field. Zeeman regime is an appropriate limiting situation up to few hundred \mbox{Gs}. Starting from $\approx 15$ kGs, the complete Paschen-Back regime occurs.}
%      \label{fig: PBH2cut}
%\end{figure}

%%%%%%%%%%%%%%%%%%%%%%%%%%%%%%%%%%%%%%%%%%%%%%%%%%%%%%%%%%%%%%%%%%%%
\section{Transition strengths}\label{sec: strengths}
%%%%%%%%%%%%%%%%%%%%%%%%%%%%%%%%%%%%%%%%%%%%%%%%%%%%%%%%%%%%%%%%%%%%

Intensity of a spectral line is proportional to the Einstein coefficient $A$, which in turn is proportional to the sum of squares of the electric dipole operator matrix elements $R_{J^\prime J^{\prime\prime}}$,
% $\hat P$:
\begin{equation}
A \propto \sum |R_{J^\prime J^{\prime\prime}}|^2,
\label{eq: einst_coeff}
\end{equation}
where the sum is taken over all levels contributing to the transition. Hence, our goal now is to compute the matrix elements $R_{J^\prime J^{\prime\prime}}$ in the PBR.

Following the discussion in Sect.~\ref{sec: energy_levels}, we first express the matrix elements $R_{J^\prime J^{\prime\prime}}$ in the case~\textit{a} basis functions, then transform them into case~\textit{b} to obtain the strengths of transitions in the ZR. Finally, we compute the strengths of transitions in the PBR using the eigenvectors of the magnetic Hamiltonian matrix~(\ref{eq:hammag}).

%-------------------------------------------------
\subsection{ZR transition strengths}
%-------------------------------------------------

We employ the \emph{Born-Oppenheimer approximation} to express the case~\textit{a} wave functions $|\Lambda S\Sigma;\, v;\, \Omega J M\rangle$  as a product of the electronic $|\Lambda S\Sigma\rangle$, vibrational $|v\rangle$, and rotational parts $|\Omega J M\rangle$ \citep[e.g.,][]{schadee1978}:
\begin{equation}
|\Lambda S\Sigma;\, v;\, \Omega J M\rangle=|\Lambda S\Sigma\rangle\; |v\rangle\; |\Omega J M\rangle. 
\label{eq: wfa}
\end{equation}
Correspondingly, the right-hand part of Eq.~(\ref{eq: einst_coeff}) can be expressed as a product of three numbers,
\begin{equation}
\sum_{M^\prime M^{\prime\prime}} |R_{J^\prime J^{\prime\prime}}|^2=f_{e}\; q_{v^\prime v^{\prime\prime}} \sum_{M^\prime M^{\prime\prime}} q^2_{\Omega^\prime \Omega^{\prime\prime}J^\prime J^{\prime\prime}M^\prime M^{\prime\prime}},
\label{eq: sqdipmatrel}
\end{equation}
where $f_{e}$ is the electronic oscillator strength, $q_{v^\prime v^{\prime\prime}}$ is the Franck-Condon factor (both are constants for a given vibrational band), and $\sum q^2_{\Omega^\prime \Omega^{\prime\prime}J^\prime J^{\prime\prime}M^\prime M^{\prime\prime}}$ is the 
% $q^2_{\Omega^\prime \Omega^{\prime\prime}J^\prime J^{\prime\prime}M^\prime M^{\prime\prime}}$ are the strengths (or 
H\"onl-London factor. The matrix elements $q^2_{\Omega^\prime \Omega^{\prime\prime}J^\prime J^{\prime\prime}M^\prime M^{\prime\prime}}$ and $q_{\Omega^\prime \Omega^{\prime\prime}J^\prime J^{\prime\prime}M^\prime M^{\prime\prime}}$ are called, respectively, \emph{strength} and \emph{amplitude} of a Zeeman component. It is the value of this amplitude that distinguishes the ZR (in both cases, \textit{a} and \textit{b}) from the PBR.

In the ZR, the matrix elements $q_{\Omega^\prime \Omega^{\prime\prime}J^\prime J^{\prime\prime}M^\prime M^{\prime\prime}}$ can be further expressed as a pro\-duct of two factors to separate the dependence on the quantum number $M$:
\begin{equation}
q_{\Omega^\prime \Omega^{\prime\prime}J^\prime J^{\prime\prime}M^\prime M^{\prime\prime}}=q_{\Omega^\prime \Omega^{\prime\prime}J^\prime J^{\prime\prime}}\;q_{J^\prime J^{\prime\prime}M^\prime M^{\prime\prime}}.
\label{eq: ampl_zeem}
\end{equation} 
Expressions for the amplitudes $q_{\Omega^\prime \Omega^{\prime\prime}J^\prime J^{\prime\prime}}$ and $q_{J^\prime J^{\prime\prime}M^\prime M^{\prime\prime}}$ are given by \citet[][Table~I]{schadee1978}.
%\footnote{Since only $\Sigma$ states of the CrH are considered and for them the quantum numbers $\Omega$ and $\Sigma$ are identical, we will write $\Sigma$ instead of $\Omega$ from now on.} 

%Since $q_{\Omega^\prime \Omega^{\prime\prime}J^\prime J^{\prime\prime}M^\prime M^{\prime\prime}}$ are known, our concern is to obtain the amplitudes of the electronic transitions in the presence of an external magnetic field, $q^\mathrm{mag}_{\Sigma^\prime \Sigma^{\prime\prime}J^\prime J^{\prime\prime}M^\prime M^{\prime\prime}}$.
%\footnote{If we need to emphasize that the matrix elements $q_{\Omega^\prime \Omega^{\prime\prime}J^\prime J^{\prime\prime}M^\prime M^{\prime\prime}}$ are expressed in the wave functions in case~\textit{a}, we will write $q^\mathrm{a}_{\Omega^\prime \Omega^{\prime\prime}J^\prime J^{\prime\prime}M^\prime M^{\prime\prime}}$.} 

%This is a well known quantum mechanical problem, where a transformation of the matrix elements of a physical quantity from one representation to another should be undertaken.

%In light of the discussion in Sect.~\ref{sec: energy_levels}, 
Now we can obtain the matrix elements $q^\mathrm{b}_{\Sigma^\prime \Sigma^{\prime\prime}J^\prime J^{\prime\prime}M^\prime M^{\prime\prime}}$ represented in the case~\textit{b} wave functions.
%, and then transform them into the PBR to get $q^\mathrm{mag}_{\Sigma^\prime \Sigma^{\prime\prime}J^\prime J^{\prime\prime}M^\prime M^{\prime\prime}}$, as shown below. %We make use here of the eigenvectors from both the effective Hamiltonian (\ref{eq: heff}) and (\ref{eq: hmageff}), as shown below.
First, 
%for the transition amplitude between two electronic states, 
for the $M$-independent part of the transition amplitude
%the upper $|\Sigma^\prime J^\prime\rangle$ and the lower $|\Sigma^{\prime\prime} J^{\prime\prime}\rangle$, in case~\textit{b} 
we obtain
\begin{equation}
q^\mathrm{b}_{\Sigma^\prime\Sigma^{\prime\prime}J^\prime J^{\prime\prime}}=\sum_{\Sigma^\prime_i}\sum_{\Sigma^{\prime\prime}_j}C^{\mathrm T}_{\Sigma^\prime_i J^\prime}\;q^\mathrm{a}_{\Omega^\prime\Omega^{\prime\prime}J^\prime J^{\prime\prime}}\;C_{\Sigma^{\prime\prime}_j J^{\prime\prime}},
\label{eq: ampl_ab}
\end{equation}
where $C_{\Sigma_{i,j} J}$ are the eigenvectors obtained by diagonalizing the Hamilton matrix (\ref{eq: heff}), and $C^{\mathrm T}_{\Sigma_{i,j} J}$ are transposed vectors. Indices $i$ and $j$ relate to different spin projections for the level with the total angular momentum quantum number $J$; $\Sigma_i$ and $\Sigma_j$ take values $-5/2$, $-3/2$, \ldots, $5/2$. The second factor in Eq.~(\ref{eq: ampl_zeem}), $q_{J^\prime J^{\prime\prime}M^\prime M^{\prime\prime}}$, is not affected by the transformation between the cases~\textit{a} and \textit{b}, and remains the same in both coupling cases.
Hence, the total ZR amplitude in the case~\textit{b} wave functions is
 \begin{equation}
 q^\mathrm{ZR}_{\Sigma^\prime \Sigma^{\prime\prime}J^\prime J^{\prime\prime}M^\prime M^{\prime\prime}} = q^\mathrm{b}_{\Sigma^\prime\Sigma^{\prime\prime}J^\prime J^{\prime\prime}}\;q_{J^\prime J^{\prime\prime}M^\prime M^{\prime\prime}}.
 \label{eq: zr_ampl}
\end{equation}

\begin{figure*}%[t!]
\centering
%\hfill %
\subfloat[Zeeman pattern composed of three groups of lines that belong to the transitions with $\Delta M=0,\,\pm 1$. The horizontal axis is dimensionless; it depicts the energy deviation, scaled by the factor $\mu_0H$, of the magnetic transitions with respect to the transition in the absence of a magnetic field. The height of the lines indicates the strength of the particular magnetic transition. The group of lines in the middle that belongs to the transitions with $\Delta M=-1$ is plotted downwards for the sake of clarity. \label{fig: Bild1}]{\includegraphics[width=0.49\textwidth]{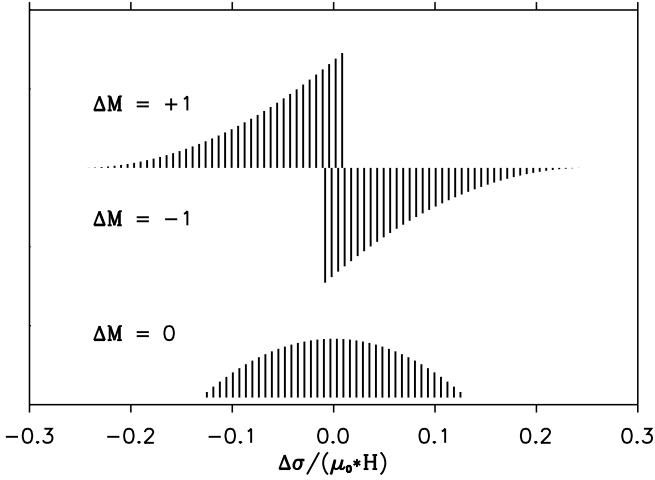}}
\hfill %\hspace{0.3cm}%\hfill % alternativ auch \hspace{1cm} fŸr genaue Angaben
\subfloat[Stokes profiles $I/I_c$, $V/I_c$, and $Q/I_c$. The horizontal axis depicts the deviation in angstroms from the line center, which is set to $0$. The vertical axis shows a strength of the particular Stokes signal and is different for each of the three panels. Again, the profiles here were obtained from the energy deviations scaled by the factor $\mu_0H$ (see text in Fig. \ref{fig: Bild1}). \label{fig: Bild2}]{\includegraphics[width=0.49\linewidth]{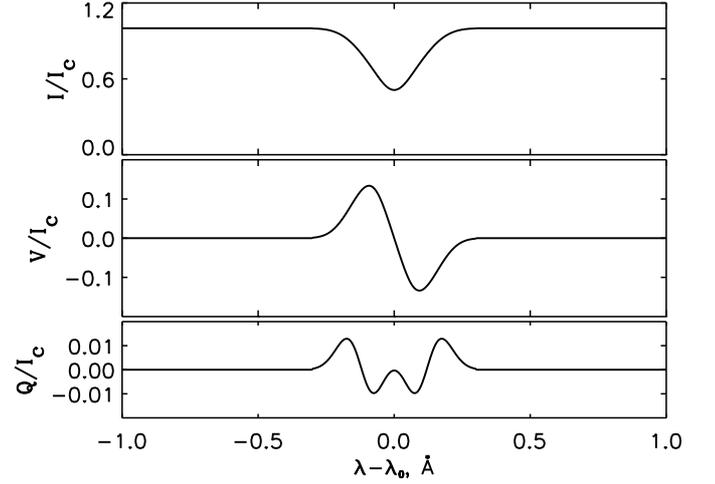}}
%\hfill %
\caption{Zeeman pattern and the corresponding Stokes profiles for the transition in Eq.~(\ref{eq: transition}) in the Zeeman regime.}% The electronic transition (\ref{eq: transition}) in the Zeeman regime.and the corresponding Stokes profiles in the Zeeman regime.}
\label{fig: zeeman}
\end{figure*}

By computing the energies and the strengths of all magnetic transitions between two rotational states in the presence of an external magnetic field we obtain a so-called \emph{Zeeman pattern}. In the ZR, it consists of three distinguished groups of lines which correspond to the transitions with $\Delta M=0,\, \pm 1$ and are called $\pi$- and $\sigma^\pm$-components.
% and plotting the strengths versus the energy shifts, relative to the energy value of the transition in the absence of a magnetic field, 
An example is shown in Fig.~\ref{fig: Bild1}, where a Zeeman pattern is calculated for the transition
\begin{equation}
|\Sigma^\prime=3/2,\, J^\prime=41/2\rangle \rightarrow |\Sigma^{\prime\prime}=3/2,\, J^{\prime\prime}=39/2 \rangle 
\label{eq: transition}
\end{equation}
in the $R$ spectroscopic branch assuming the ZR.
%By way of example, we present on Figs. \ref{fig: zeeman} (Zeeman regime) and \ref{fig: pb} (Paschen-Back regime) the Zeeman patterns predicted for the electronic transition

As Eq.~(\ref{eq: zeeman_regime}) for the ZR predicts, the energy shifts are linear with the magnetic field strength. Therefore, the Zeeman pattern in Fig. \ref{fig: Bild1} is symmetric. If the ZR holds for this line, its shape does not change with the magnetic field strength. However, this is not the case for the CrH lines under consideration, and we need to proceed with the following step to find an expression for the transition strengths in the PBR.

%-------------------------------------------------
\subsection{PBR transition strengths}\label{subsec: pbr_strengths}
%-------------------------------------------------

The transformation of the amplitudes from the ZR in case~\textit{b} to the PBR is carried out with the help of the eigenvectors $C_{\Sigma_{i,j} JM}$ obtained by diagonalizing the Hamiltonian matrix~(\ref{eq:hammag}),
%The amplitude of the transition between two magnetic levels, the upper $|\Sigma^\prime J^\prime M^\prime\rangle$ and the lower $|\Sigma^{\prime\prime} J^{\prime\prime} M^{\prime\prime}\rangle$, is:
\begin{equation}
q^\mathrm{PBR}_{\Sigma^\prime \Sigma^{\prime\prime}J^\prime J^{\prime\prime}M^\prime M^{\prime\prime}}=\sum_{\Sigma^\prime_i}\sum_{\Sigma^{\prime\prime}_j}C^{\mathrm T}_{\Sigma^\prime_i J^\prime M^\prime}\;q^\mathrm{ZR}_{\Sigma^\prime\Sigma^{\prime\prime}J^\prime J^{\prime\prime} M^\prime M^{\prime\prime}}\;C_{\Sigma^{\prime\prime}_j J^{\prime\prime}M^{\prime\prime}},
\label{eq: ampl_bmag}
\end{equation}
where 
%the eigenvectors $C_{\Sigma_{k,l} JM}$ have been obtained by diagonalizing the Hamiltonian matrix (\ref{eq:hammag}). 
the indices $i$ and $j$ again relate to the different levels $J$ that form a fine structure of a rotational level $N$; $\Sigma_i$ and $\Sigma_j$ take values $-5/2$, $-3/2$, \ldots, $5/2$. In contrast to the ZR amplitude (\ref{eq: zr_ampl}), the PBR amplitude (\ref{eq: ampl_bmag}) is no longer a product of two factors, but a linear combination of the ZR transition amplitudes.

In Figs. \ref{fig: pb}a, \ref{fig: pb}c, \ref{fig: pb}e, \ref{fig: pb}g, and \ref{fig: pb}i we show the Zeeman patterns for the transition in Eq.~(\ref{eq: transition}) in the PBR for the magnetic field strengths $0.001$, $0.01$, $0.1$, $1$, and $10$~kG, respectively. The Zeeman pattern in Fig.~\ref{fig: pb}a appears quite symmetric; in other words, the upper and lower $N$ levels of the molecule seem to be in the ZR at $0.001$~kG. However, the corresponding Stokes $Q/I_c$ profile (Fig. \ref{fig: pb}b) is clearly asymmetric (see our discussion in Sect. \ref{subsec: stokes_profiles}). As the magnetic field strength increases, the pattern evolves rapidly, because the energy shifts no longer vary linearly, as Eq.~(\ref{eq: zeeman_regime}) predicts. At stronger magnetic fields the shape of the pattern changes dramatically (Figs. \ref{fig: pb}~e,g,i): the Zeeman shifts change their sign (the pattern appears twisted), and the Zeeman component strengths alter significantly. In this example of the $R$ spectroscopic branch ($\Delta J=1$), $\sigma^+$ components become stronger than $\sigma^-$ and $\pi$ components in a strong field regime (cf., Fig. \ref{fig: pb}g). This is in contrast to the $P$ branch ($\Delta J=-1$), where $\sigma^-$ transitions become dominant (not shown). Thus, considering a great number of rotational transitions we expect a \emph{net polarization} in the \mbox{(0,0)} band of the CrH molecule.% (\emph{broad-band polarization}).

%For the weak magnetic field ($0.01$~\mbox{kGs}) the Zeeman patterns look symmetric (Fig. \ref{fig: 1R}). It indicates that both the upper and the lower states occur in the Zeeman regime. At $0.1$~\mbox{kGs} the patterns are no longer symmetric due to the fact that the widening of the magnetic le\-vels is no longer linear with the field strength (Fig. \ref{fig: 2R}). While increasing the field strength, the interacting magnetic levels repel each other and the patterns for some polarization modes become ''twisted`` (Fig. \ref{fig: 3R}). The strengths of some Zeeman components begin to alter in comparison to the strengths in the weak field regime. For the strong magnetic field the look of the patterns changes dramatically. While some Zeeman components become stronger, others are being suppressed. 
%Thus, for the $R$ branch, the $\sigma^+$-components become stronger than $\sigma^-$ or $\pi$ (Fig. \ref{fig: 4R}), while for the $P$ branch we have the opposite case: the stronger components belong to the $\sigma^-$ polarization mode. Considering a great number of the electronic transitions, we expect that the $R$ branch will have predominantly $\sigma^+$ polarity and $P$ $\sigma^-$ polarity.

%%%%%%%%%%%%%%%%%%%%%%%%%%%%%%%%%%%%%%%%%%%%%%%%%%%%%%%%%%%%%%%%%%%%
\section{Radiative transfer calculations}\label{sec: stokes}
%%%%%%%%%%%%%%%%%%%%%%%%%%%%%%%%%%%%%%%%%%%%%%%%%%%%%%%%%%%%%%%%%%%%

The theoretical line parameters calculated as described in the previous sections have been used to synthesize the Stokes profiles for both individual lines and the entire (0,0) band with the code STOPRO \citep[described by][]{solanki1987, frutiger1999, berdyugina2003}. The code assumes local thermodynamic equilibrium (LTE) and solves numerically a set of polarized transfer equations in a model atmosphere.  Furthermore, it is assumed that a magnetized atmosphere acts on a spectral line (both atomic and molecular) through the  Zeeman or Paschen-Back effect. The polarization arising from scattering in the atmosphere is neglected. Number densities of about 300 atomic and molecular species are calculated under the assumption of the chemical equilibrium as described in \citet{berdyugina2003}.

%-------------------------------------------------
\subsection{Atmosphere models and condensate opacities}\label{subsec:atmos}
%-------------------------------------------------

We employ the original Drift-Phoenix atmospheric models calculated by \citet{witte2009} for $T_\mathrm{eff}=2500$~K, $2000$~K, and $1500$~K. The solar abundances of the chemical elements and the surface gravity of $\log g=5.0$ is assumed. These three atmospheric models correspond roughly to the physical parameters of late-M dwarfs, early- and mid-L dwarfs, and hot Jupiters.  

%Because the Drift-Phoenix atmospheric models we employed (or more precisely, opacities in the models) were calculated for $1.14$ $\mu$m and we are interested in the wavelength range of the \mbox{(0,0)} vibrational band of the CrH, we recalculated the opacities for the wavelength needed. Thus, we included the continuum gas opacity arising from: i)~Rayleigh scattering by the atoms H and He, and the molecules H$_2$, H$_2$O, CO and CH$_4$; ii)~bound-free and free-free transitions in H, He, Mg, Si, Fe, H$^-$, H$_2^-$, H$_2^+$ , He$^-$. We did not take into account the pressure induced opacity \citep[see e.g.][]{linsky1969}.

In atmospheres of cool brown dwarfs and hot exoplanets dust can considerably contribute to the total opacity. The Drift-Phoenix models, for example in contrast to the Phoenix model grid by \citet{allard1995}, employ the physics of dust formation and predict abundances of such condensate species as TiO$_2$, Al$_2$O$_3$, Fe, SiO$_2$, MgO, MgSiO$_3$, and Mg$_2$SiO$_4$. Since their significance as opacity sources increases for lower temperature and higher pressure, we want to estimate the effect of dust scattering and absorption on the CrH spectrum in the three selected models. To do this, we make use of an analytical solution of the Mie theory \citep{mie1908}.

%Here we employ an analytical solution of the Mie theory \citep{mie1908}. 
The opacity $\varkappa$ caused by the interaction of the radiation with dust particles is related to the interaction cross section $\sigma$ in the relation
\begin{equation}
\varkappa=N\sigma,
\label{eq: cond_opacity}
\end{equation}
where $\sigma =\sigma_\mathrm{sca}+\sigma_\mathrm{abs}$ is the sum of the cross sections due to scattering and absorption of the radiation, and $N$ is the number of the dust particles within a unit volume.

To calculate the interaction cross section $\sigma$, we employ an analytical expression obtained for scattering and absorption of electromagnetic waves by small (compared with the radiation wavelength) spherical particles \citep[see, e.g.,][]{landau8},
\begin{equation}
\sigma=\frac{128\pi^5}{3}\left|\frac{\varepsilon-1}{\varepsilon+2}\right|^2\frac{a^6}{\lambda^4}+24\pi^2\frac{a^3}{\lambda}\varepsilon^{\prime\prime} \left( \frac{1}{|\varepsilon|^2}+\frac{4\pi^2}{90}\frac{a^2}{\lambda^2}\right),
\label{eq: crosssec}
\end{equation}
where $a$ is the particle size (radius), $\lambda$ is the wavelength of the incident electromagnetic wave; $\varepsilon$ and $\varepsilon^{\prime\prime}$ are the relative dielectric function and its imaginary part of the particle material.

The Drift-Phoenix model grid provides the following depth-dependent dust parameters: average grain size, number of particles in the unit volume, and volume fraction for each dust species. Since we have the average particle size rather than the individual particle sizes of each species, we calculate the corresponding average (volume-fraction-weighted) relative dielectric function according to the effective medium theory  \citep[see, e.g.,][]{bohren1998}. Thus, Eq.~(\ref{eq: crosssec}) was evaluated using the average grain size and the average relative dielectric function. 

We have found that for the chosen model atmospheres the opacity due to light scattering on the dust particles is significantly larger than light absorption by these particles at the CrH band wavelength. Furthermore, the total condensate opacity in Eq.~(\ref{eq: cond_opacity}) becomes considerable (at a level of a few percent) with the gas opacity only for the $T_{\rm eff} = 1500$~K model and, respectively, for cooler atmospheres. We note that this condensate opacity contribution is already included in the atmospheric models used. 
%Nevertheless, we have consistently calculated CrH spectra using the total opacity and the corresponding optical depth scale, including both the gas and condensate contributions.

%-- end \bf for the dust opacity description 

\begin{figure*}%[]
 \centering
%use [scale=0.3] for the referee format, and [scale=0.37] for the normal one.
 \subfloat{\includegraphics[scale=0.37]{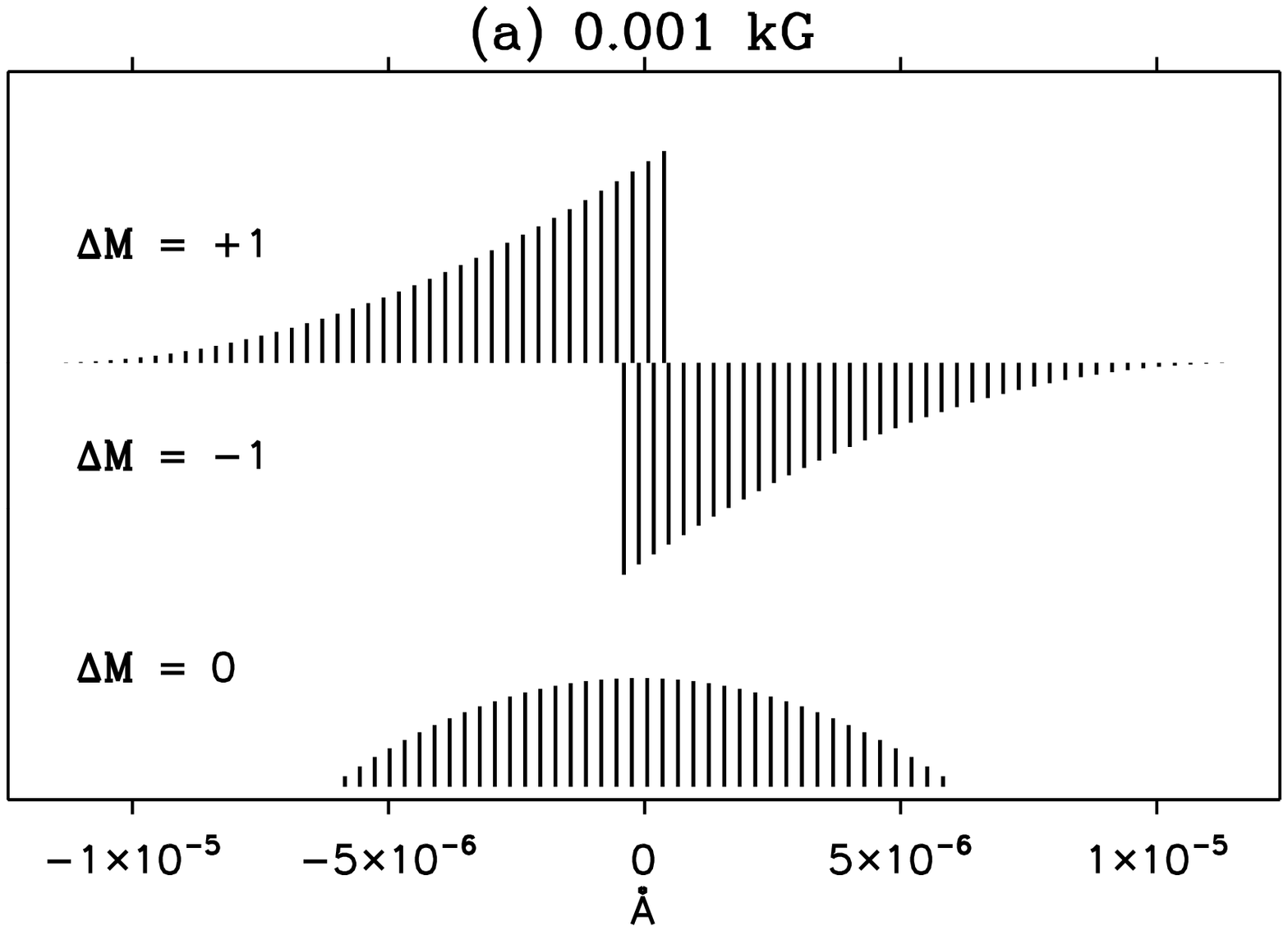}}%[scale=0.37] normal scale
 \hspace{2cm}
 \subfloat{\includegraphics[scale=0.37]{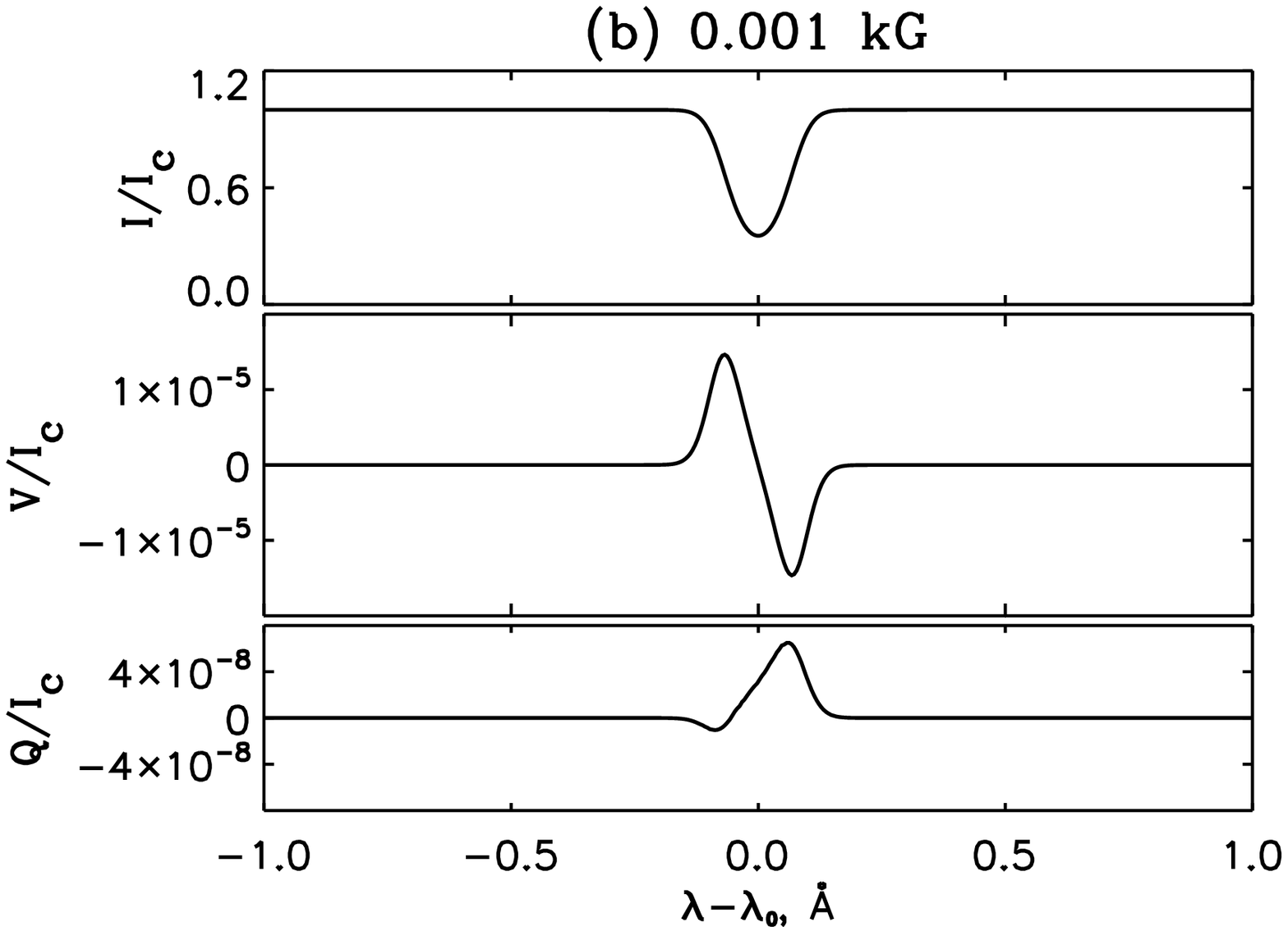}}\\ %[$0.001$ kG \label{fig: sp7}]
 \vspace{-0.25cm}
 \subfloat{\includegraphics[scale=0.37]{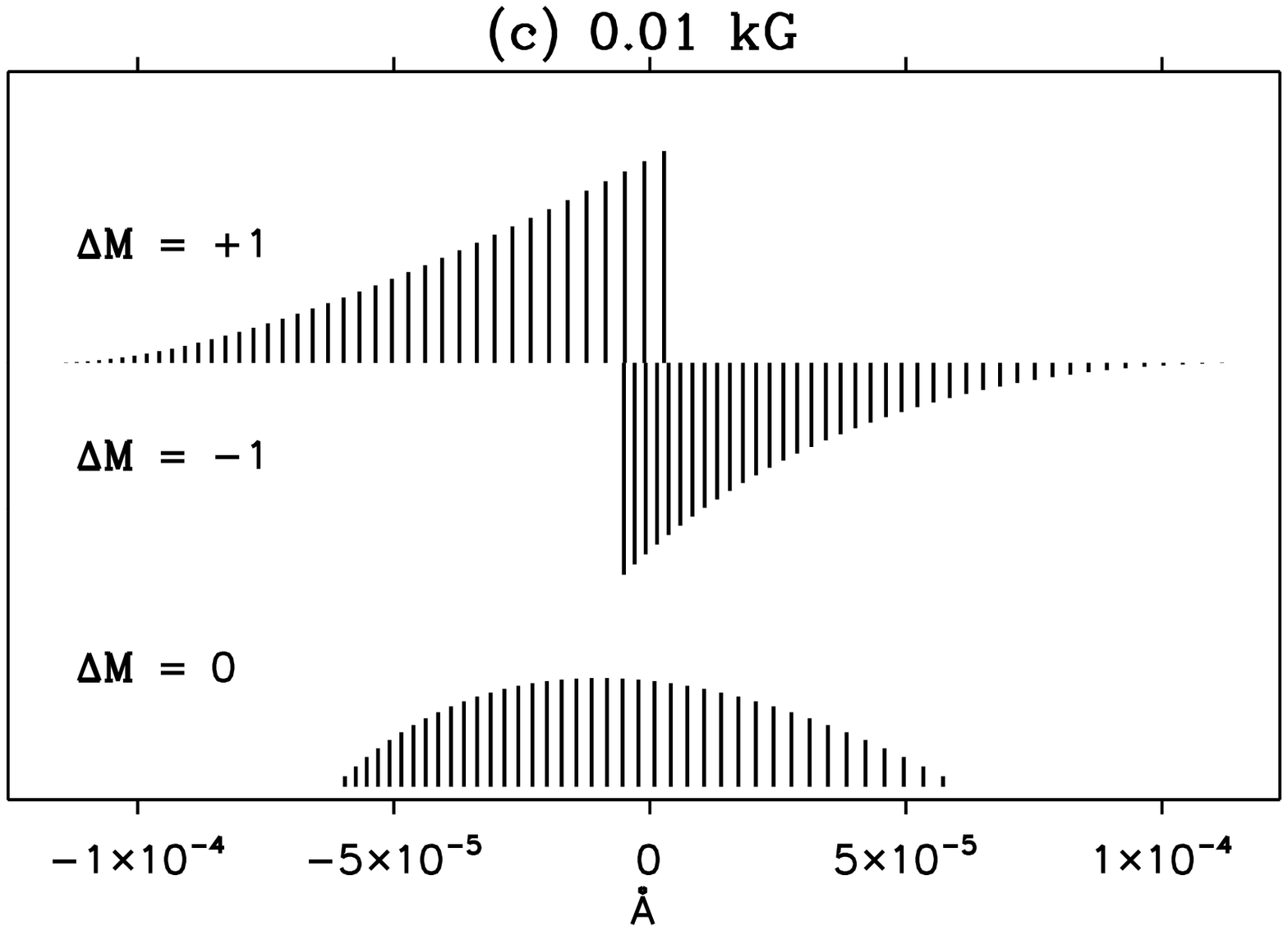}} %[$0.01$ kG\label{fig: zp7a}]
 \hspace{2cm}
 \subfloat{\includegraphics[scale=0.37]{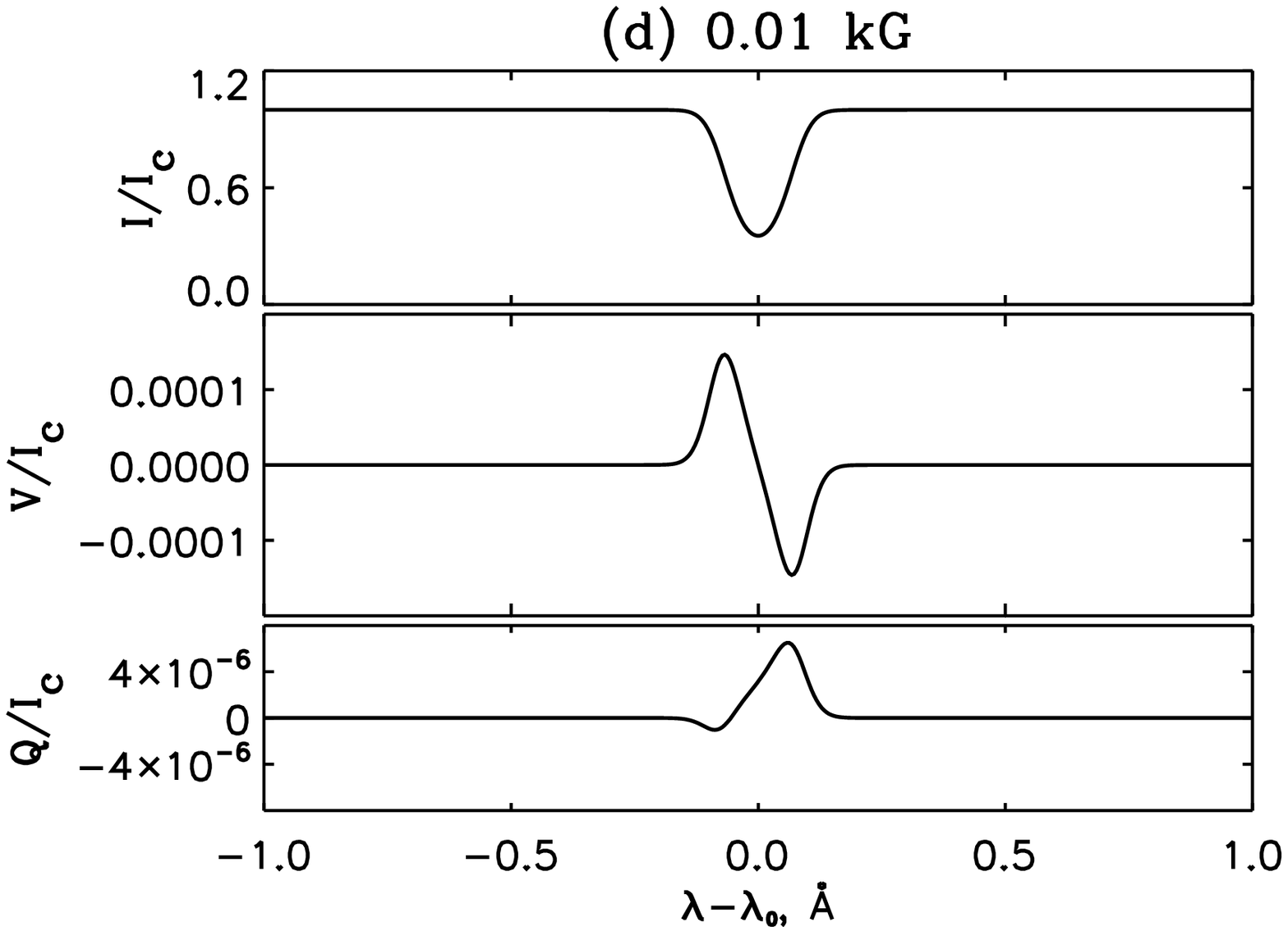}}\\ %[$0.01$ kG\label{fig: sp7a}]
 \vspace{-0.25cm}
 \subfloat{\includegraphics[scale=0.37]{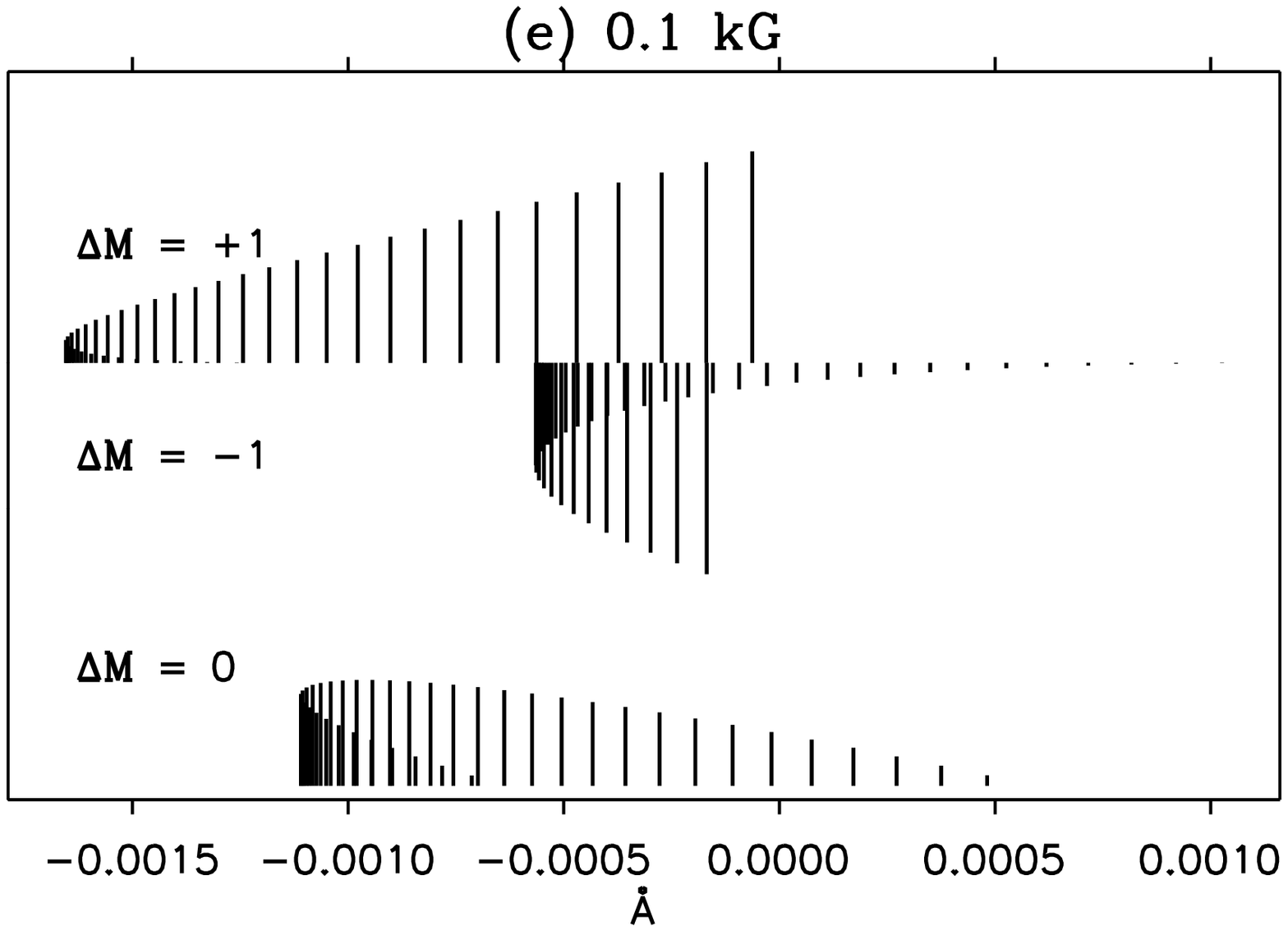}} %[$0.05$ kG\label{fig: zp8}]
 \hspace{2cm}
 \subfloat{\includegraphics[scale=0.37]{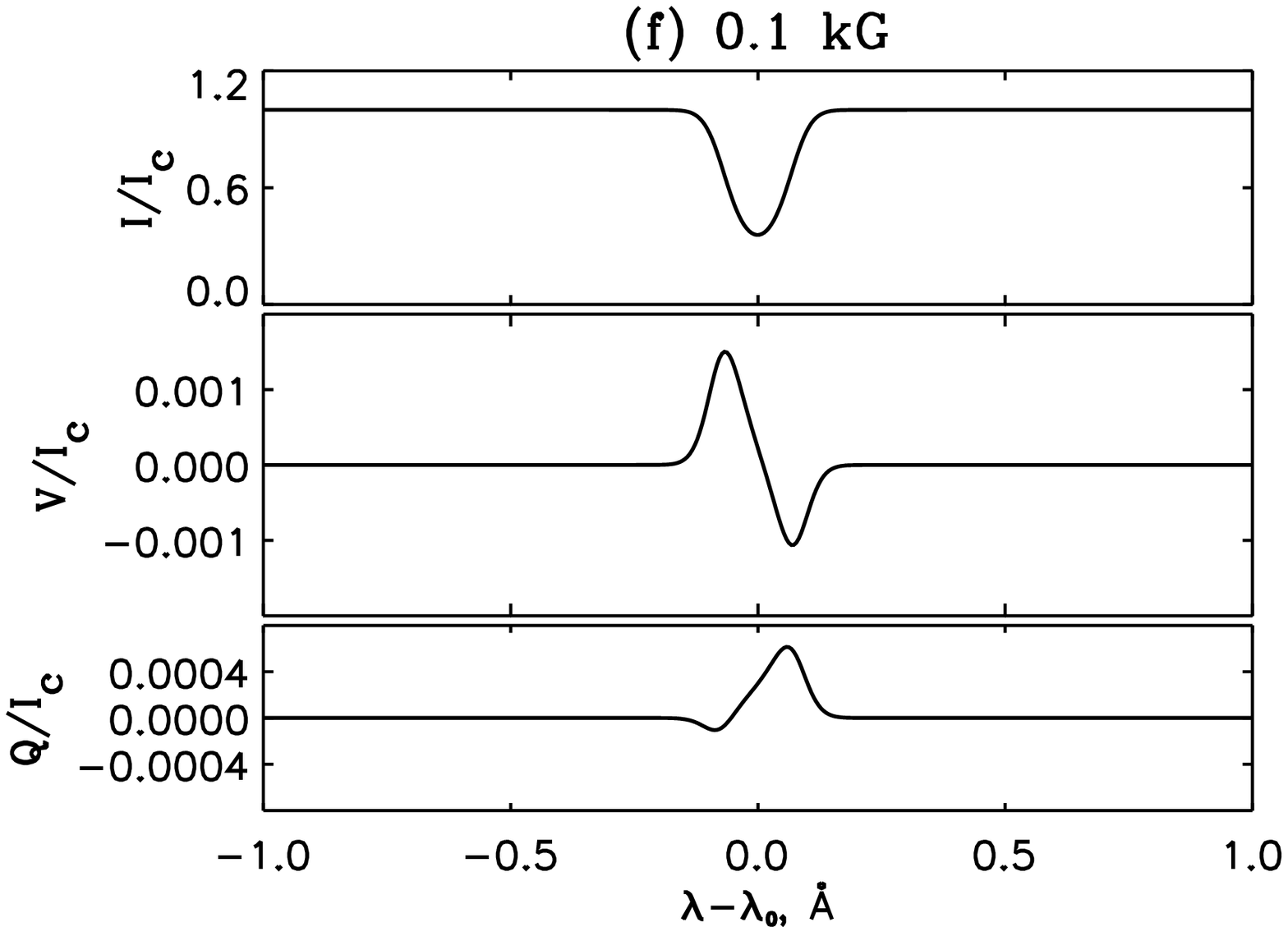}}\\ %[$0.05$ kG \label{fig: sp8}]
  \vspace{-0.25cm}
 \subfloat{\includegraphics[scale=0.37]{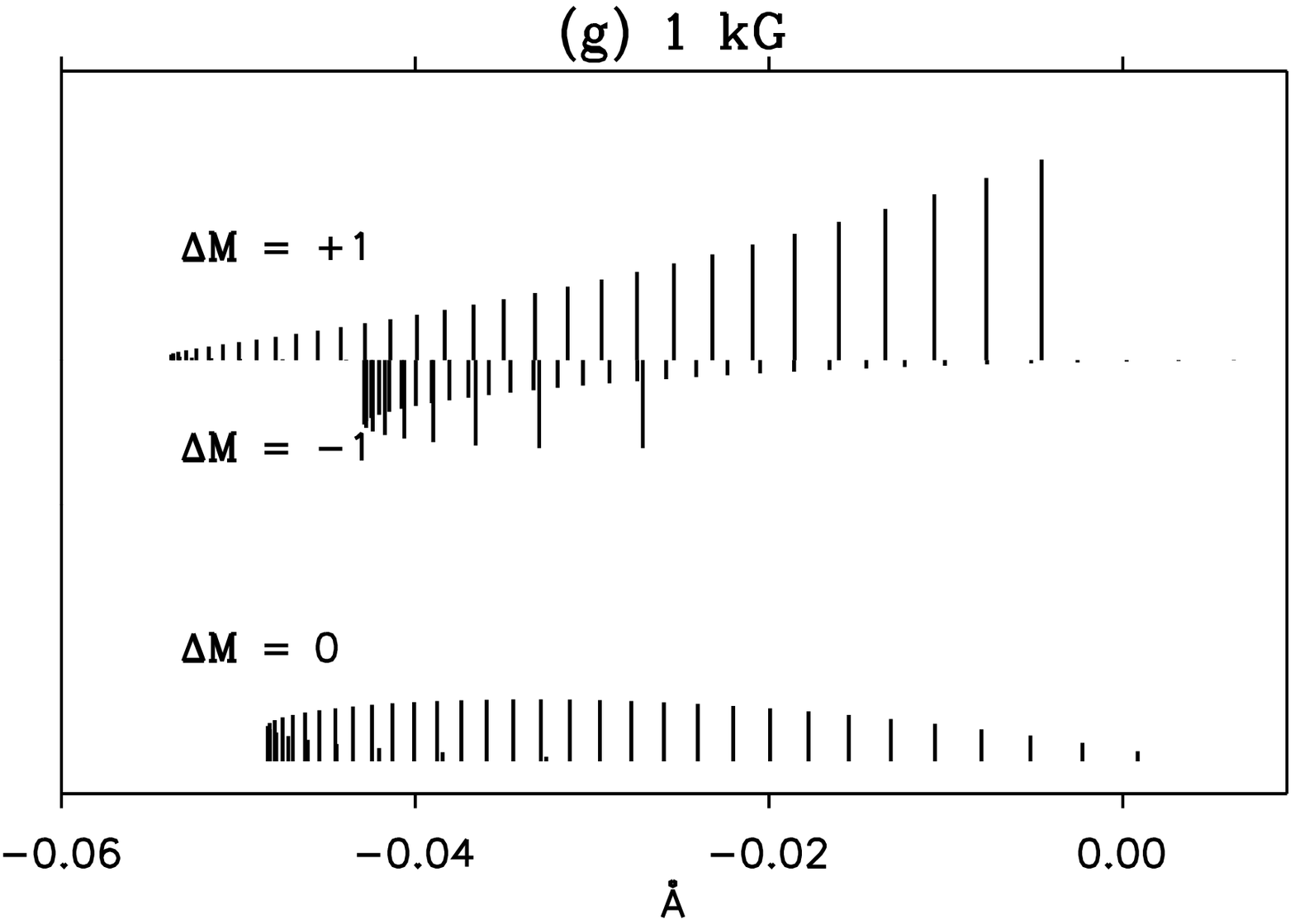}} %[$0.1$ kG\label{fig: zp8a}]
 \hspace{2cm}
  \subfloat{\includegraphics[scale=0.37]{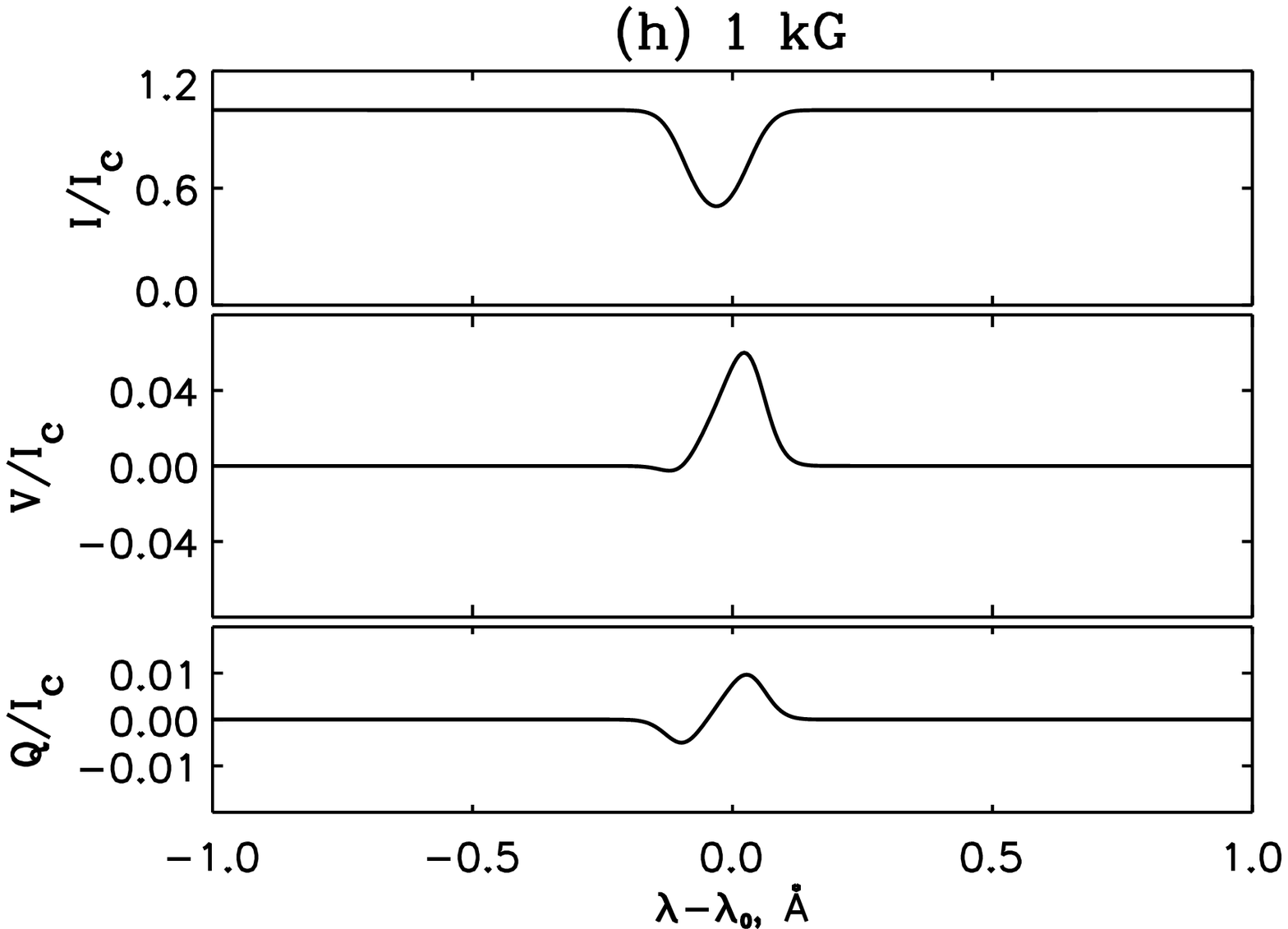}}\\ %[$0.1$ kG\label{fig: sp8a}]
   \vspace{-0.25cm}
 \subfloat{\includegraphics[scale=0.37]{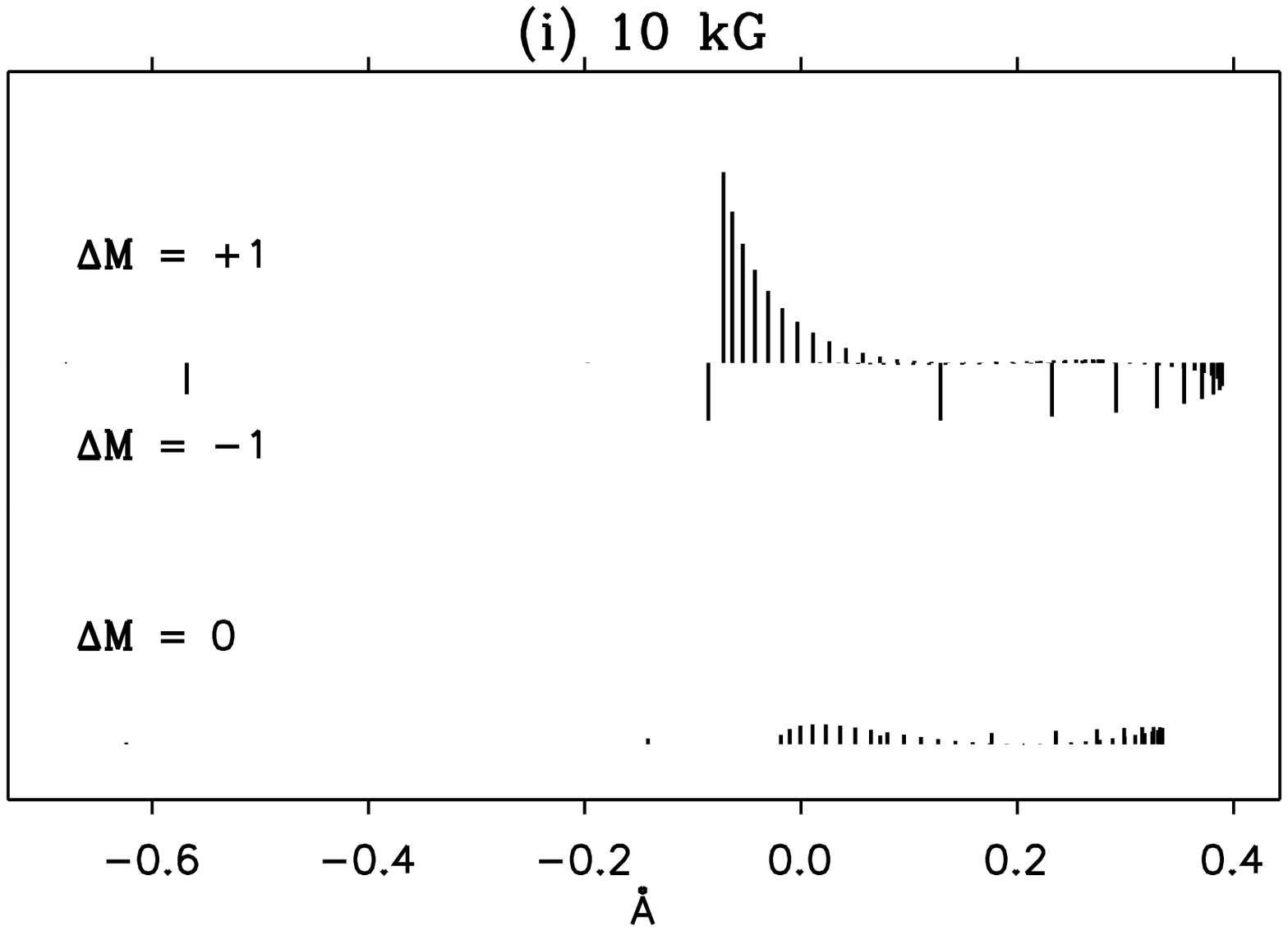}} %[$1$ kG\label{fig: zp9a}]
 \hspace{2cm}
 \subfloat{\includegraphics[scale=0.37]{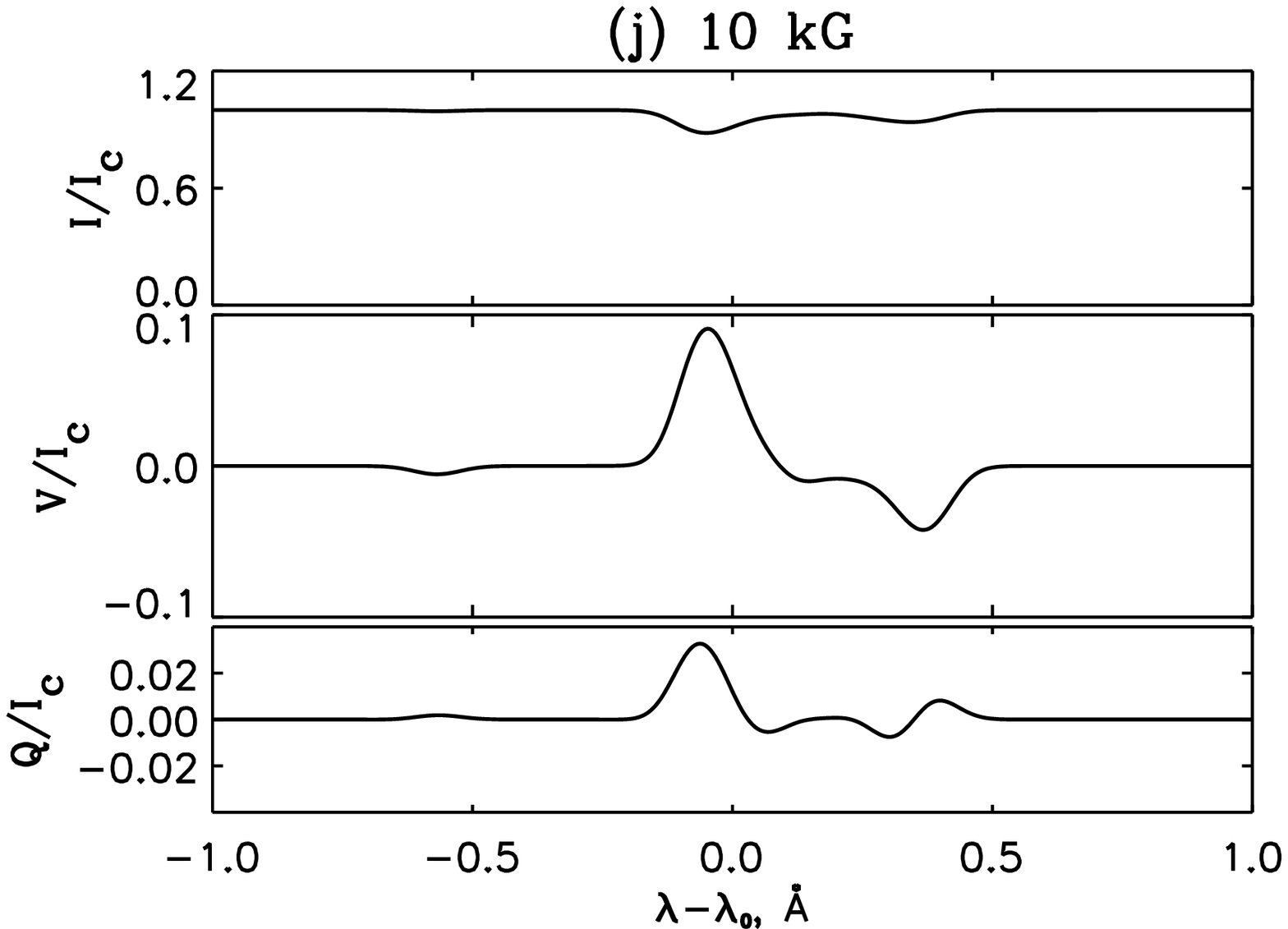}}%[$1$ kG\label{fig: sp9a}]

 \caption{Zeeman patterns and the corresponding Stokes profiles for the transition in Eq.~(\ref{eq: transition}) calculated for different magnetic field strengths.}
 \label{fig: pb}
\end{figure*}

\begin{figure*}%[]
 \centering
 %use [scale=0.3] for the referee format, and [scale=0.37] for the normal one.
 \subfloat{\includegraphics[scale=0.37]{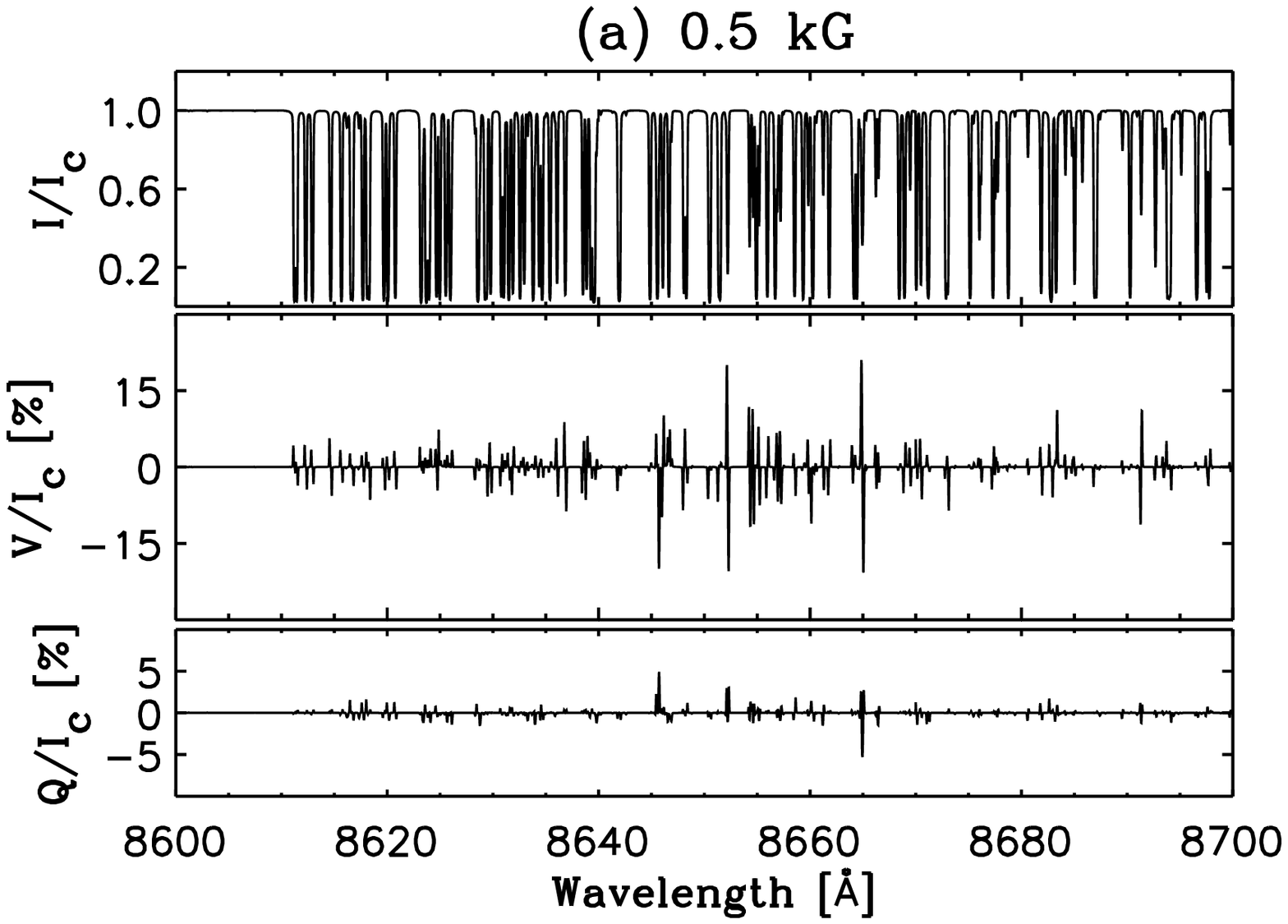}}%crh00_1kG_8600-8700_form.eps 
 \hspace{2cm}
 \subfloat{\includegraphics[scale=0.37]{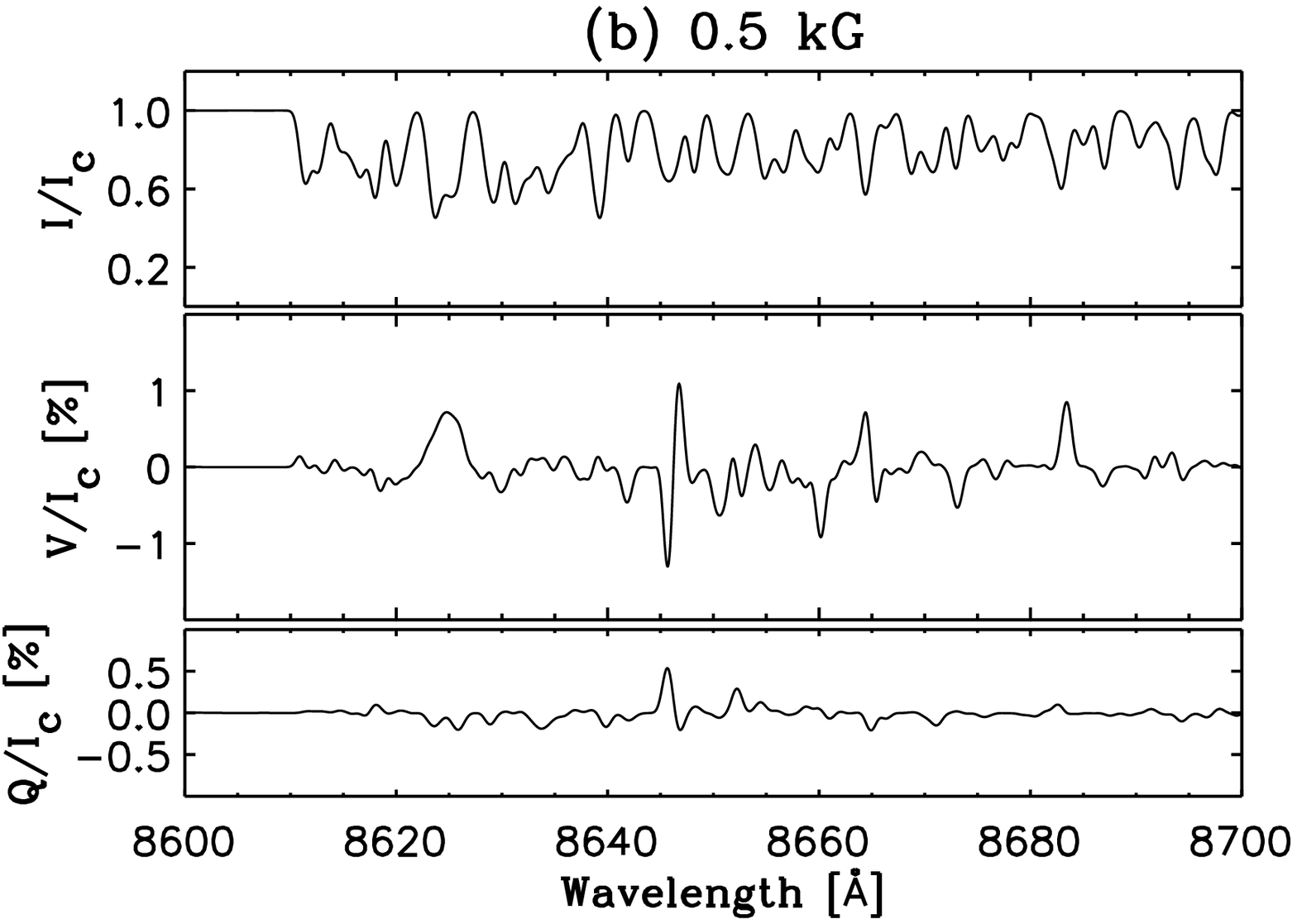}}\\ %[$0.001$ kG \label{fig: sp7}]
 \vspace{-0.25cm}
 \subfloat{\includegraphics[scale=0.37]{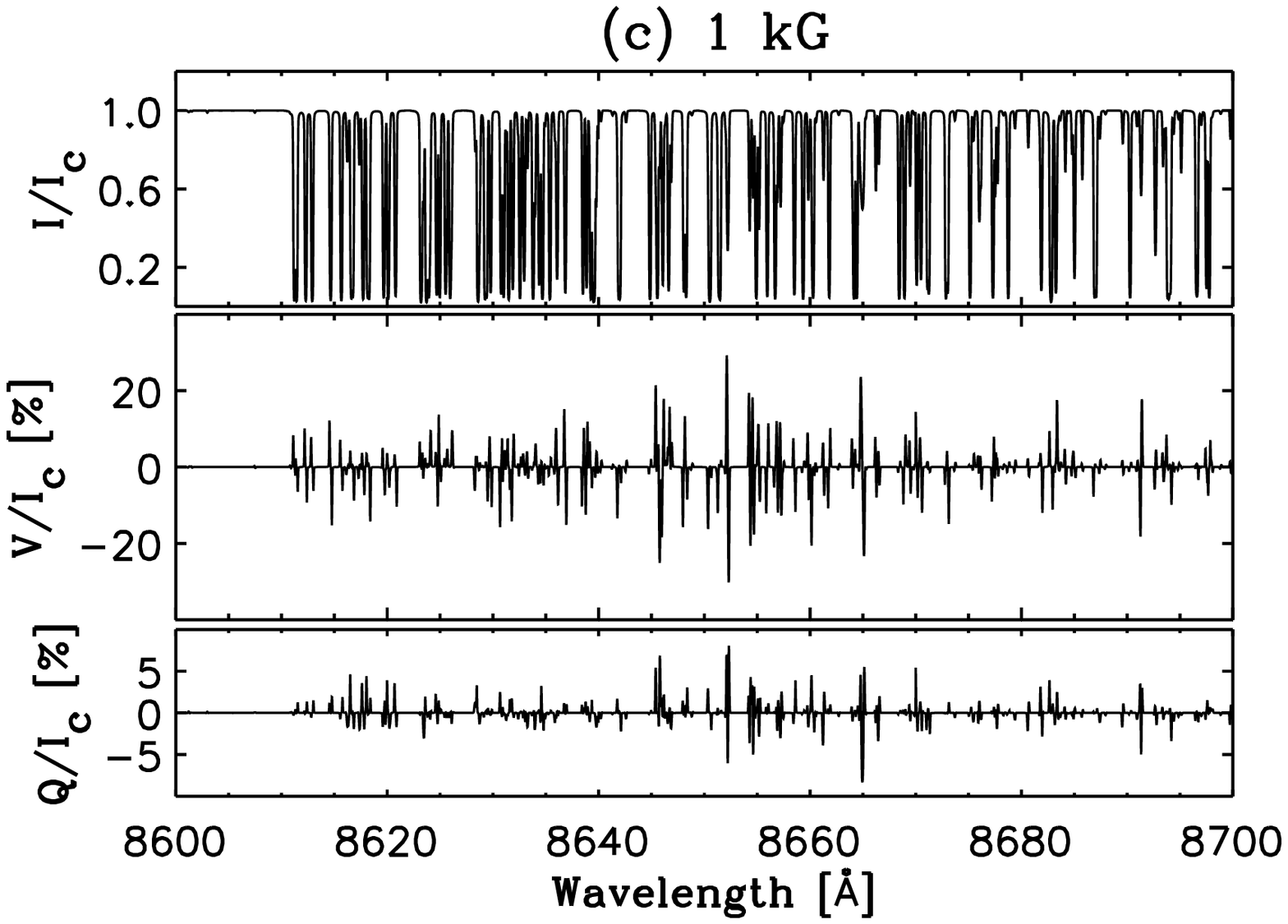}} %[$0.01$ kG\label{fig: zp7a}]
 \hspace{2cm}
 \subfloat{\includegraphics[scale=0.37]{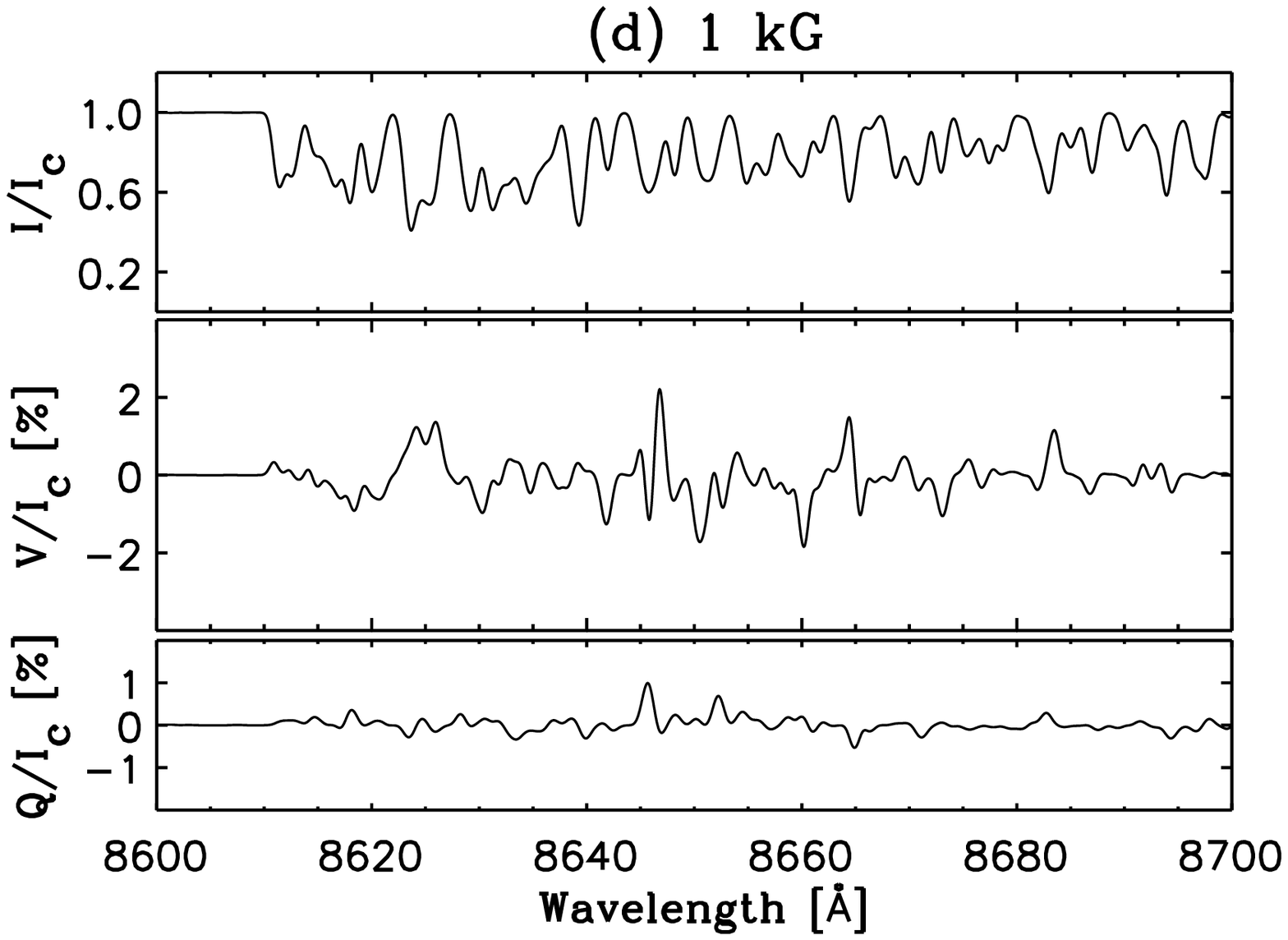}}\\ %[$0.01$ kG\label{fig: sp7a}]
 \vspace{-0.25cm}
 \subfloat{\includegraphics[scale=0.37]{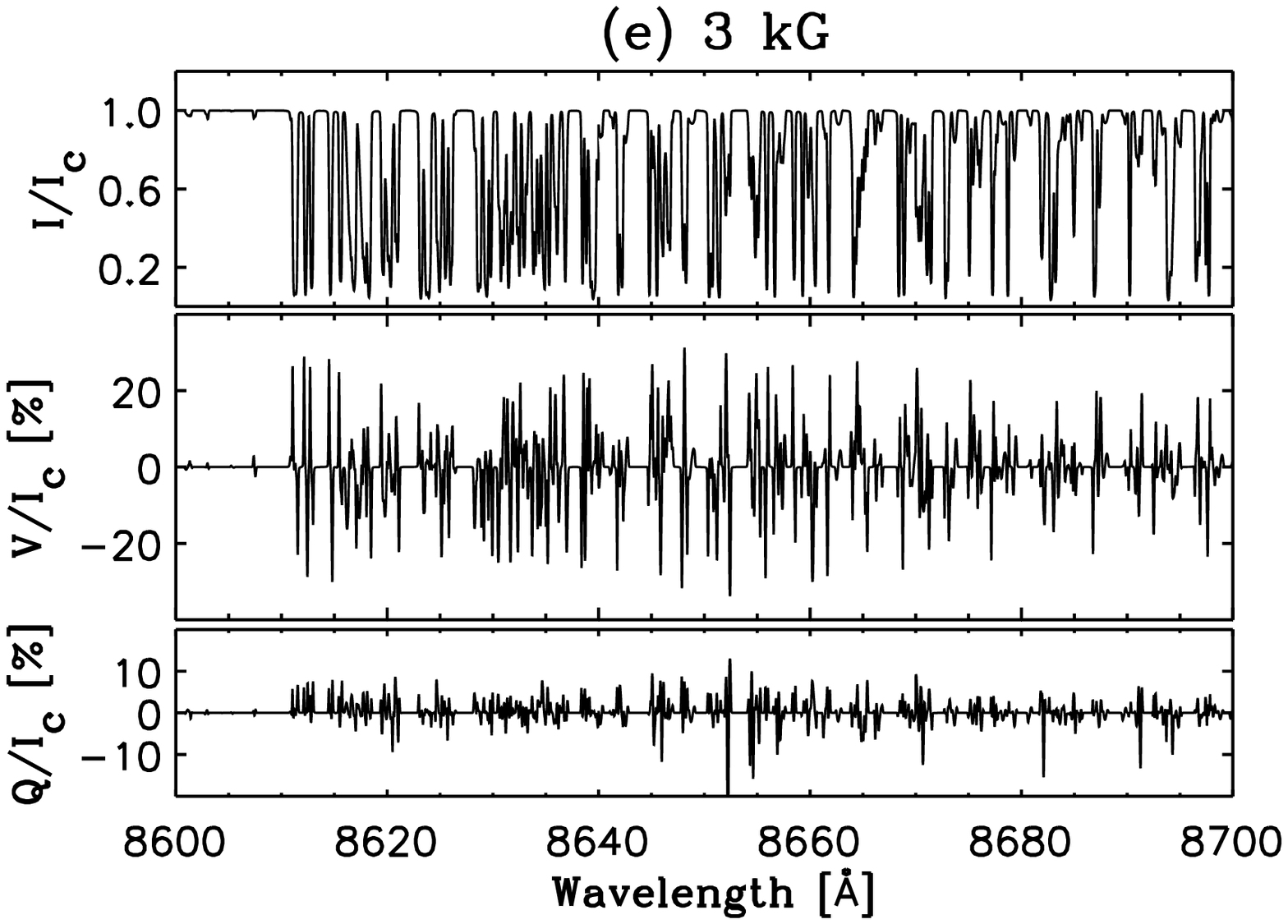}} %[$0.05$ kG\label{fig: zp8}]
 \hspace{2cm}
 \subfloat{\includegraphics[scale=0.37]{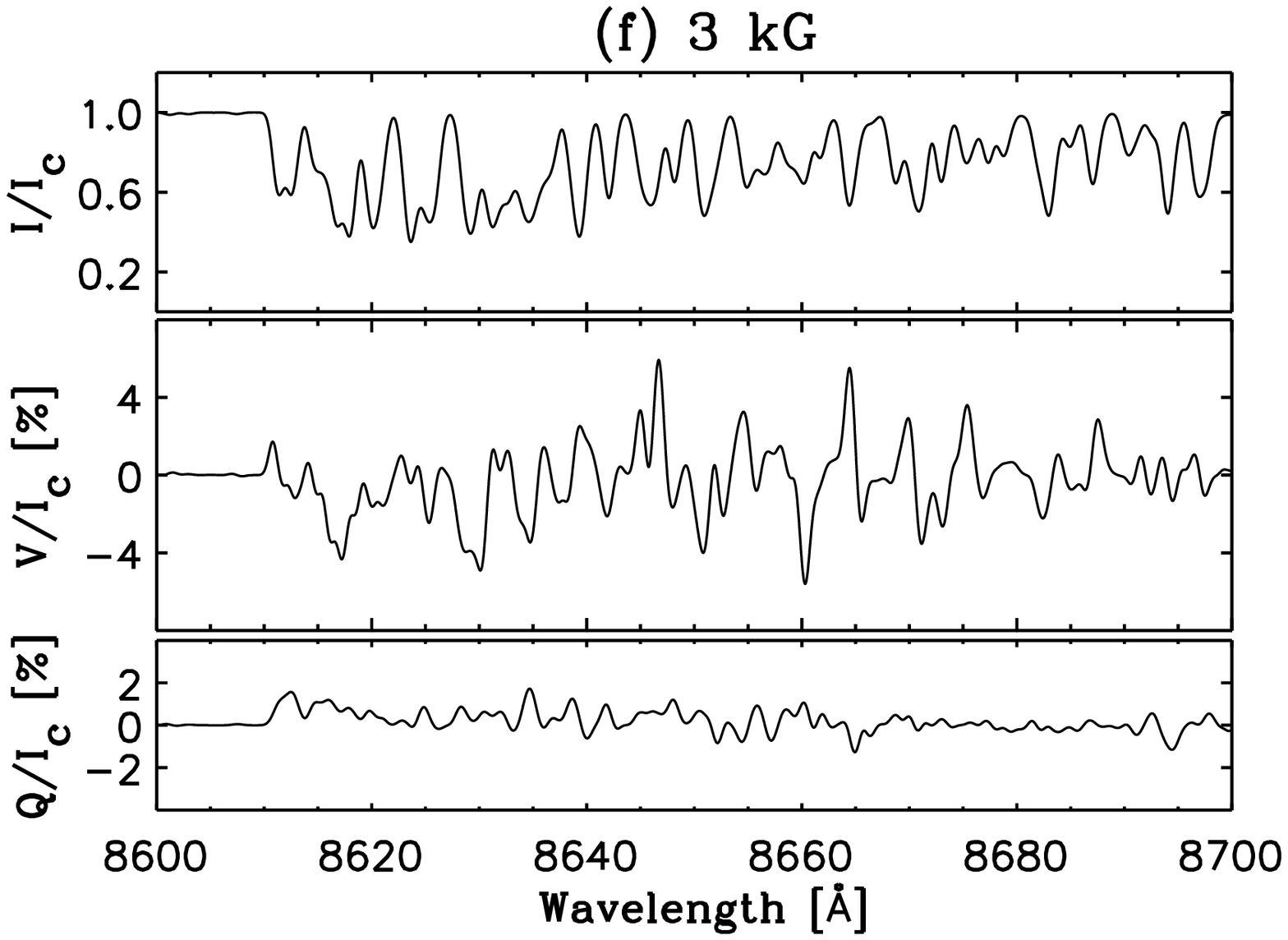}}\\ %[$0.05$ kG \label{fig: sp8}]
  \vspace{-0.25cm}
 \subfloat{\includegraphics[scale=0.37]{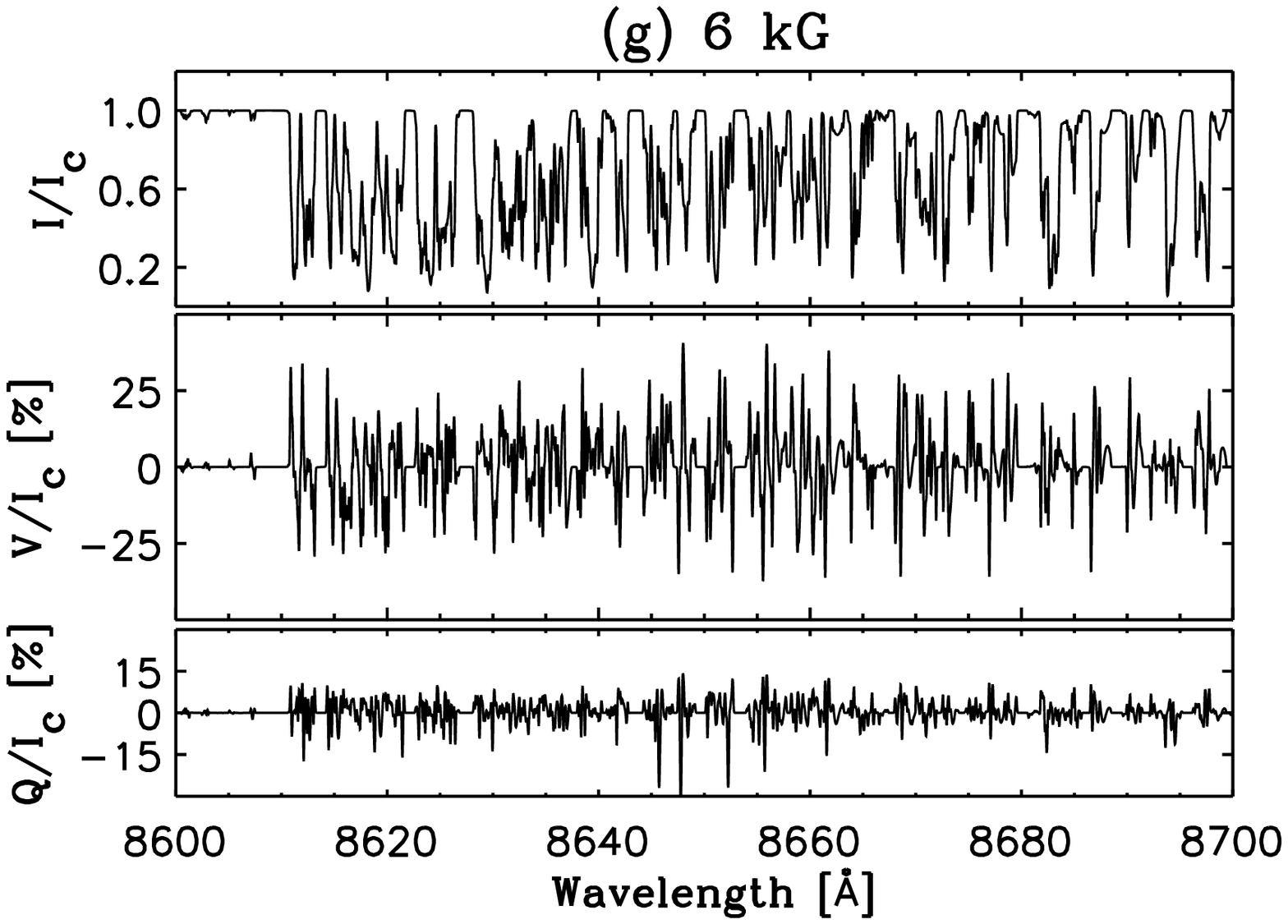}} %[$0.1$ kG\label{fig: zp8a}]
 \hspace{2cm}
 \subfloat{\includegraphics[scale=0.37]{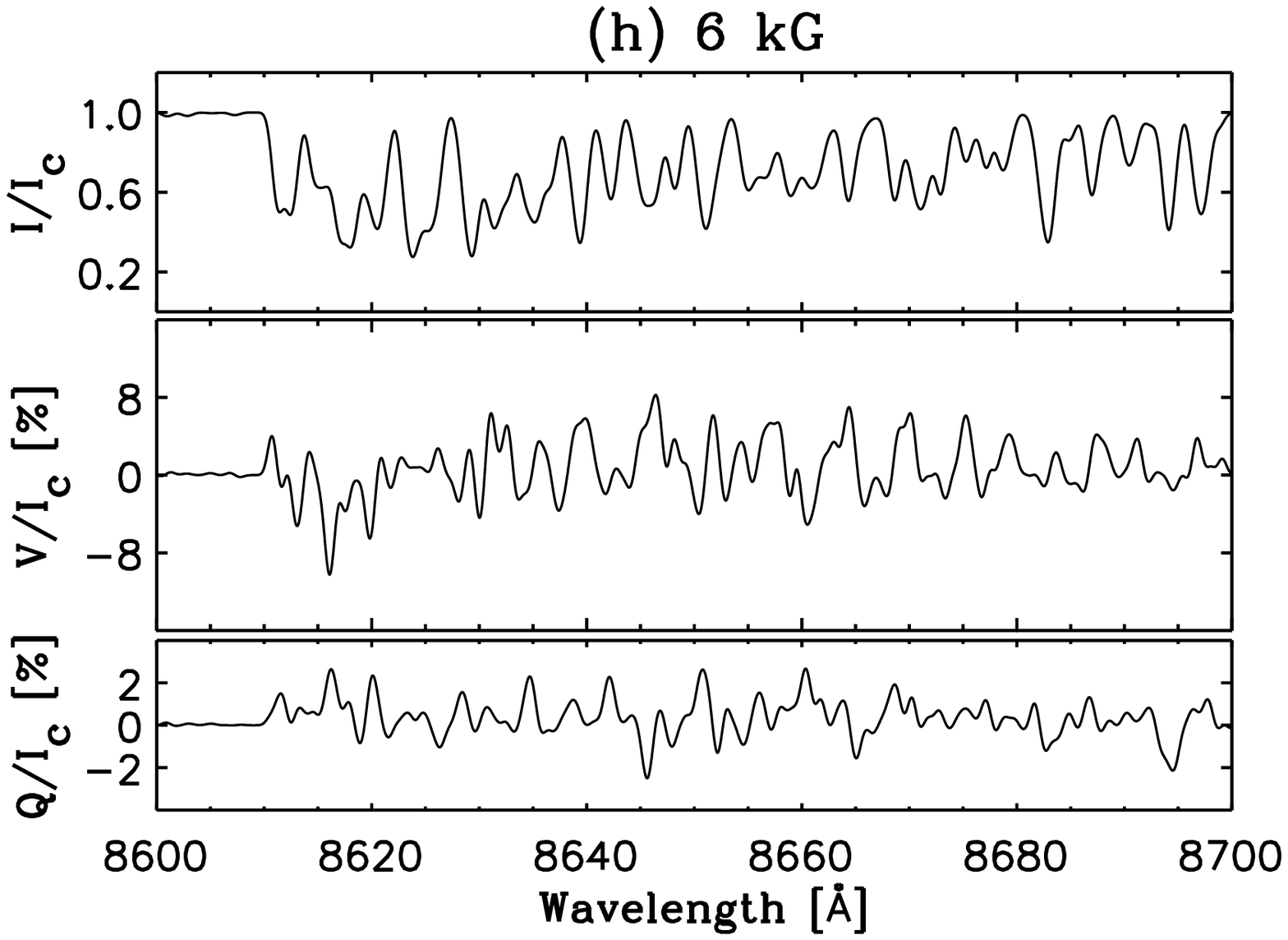}}\\ %[$0.1$ kG\label{fig: sp8a}]
   \vspace{-0.25cm}
 \subfloat{\includegraphics[scale=0.37]{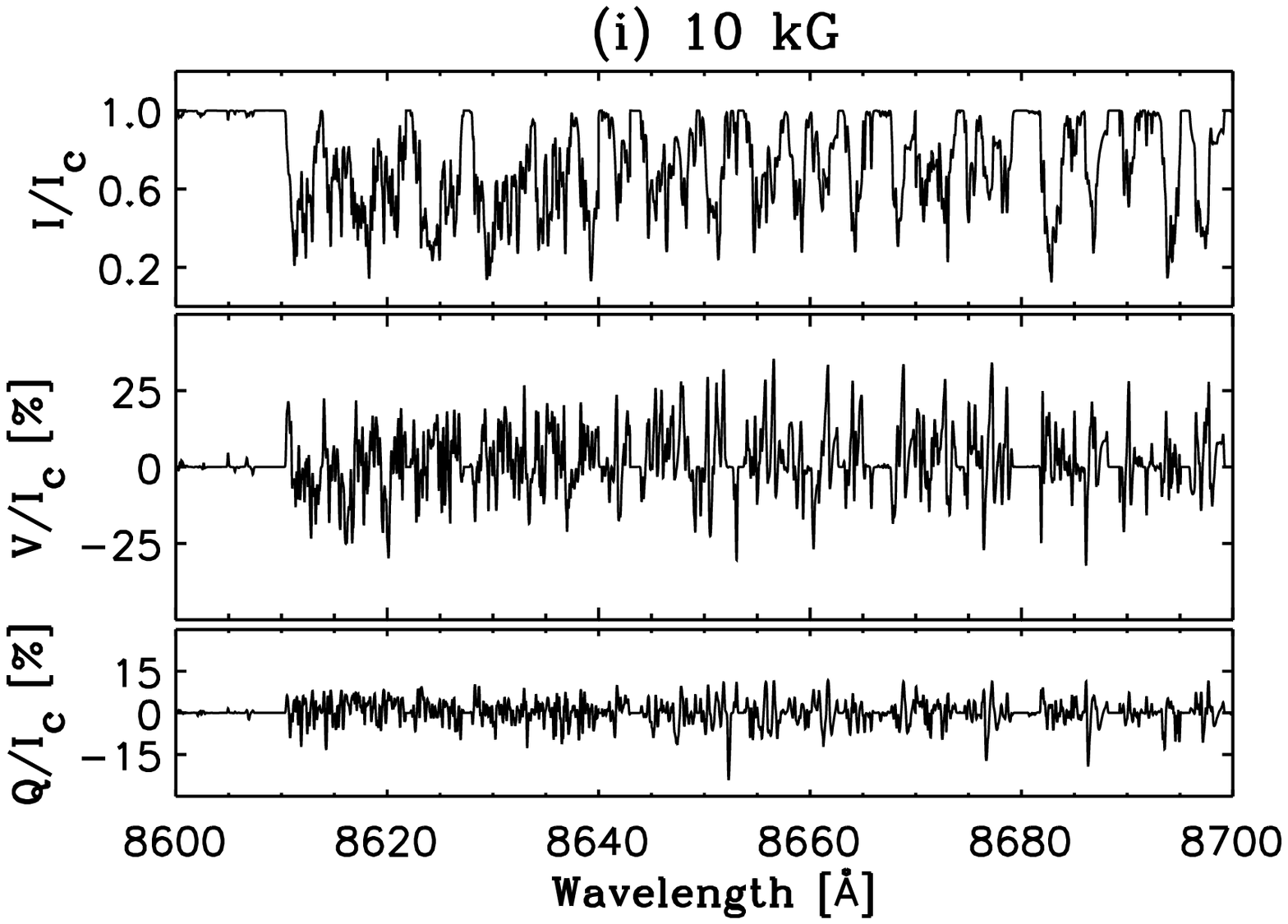}} %[$1$ kG\label{fig: zp9a}]
 \hspace{2cm}
 \subfloat{\includegraphics[scale=0.37]{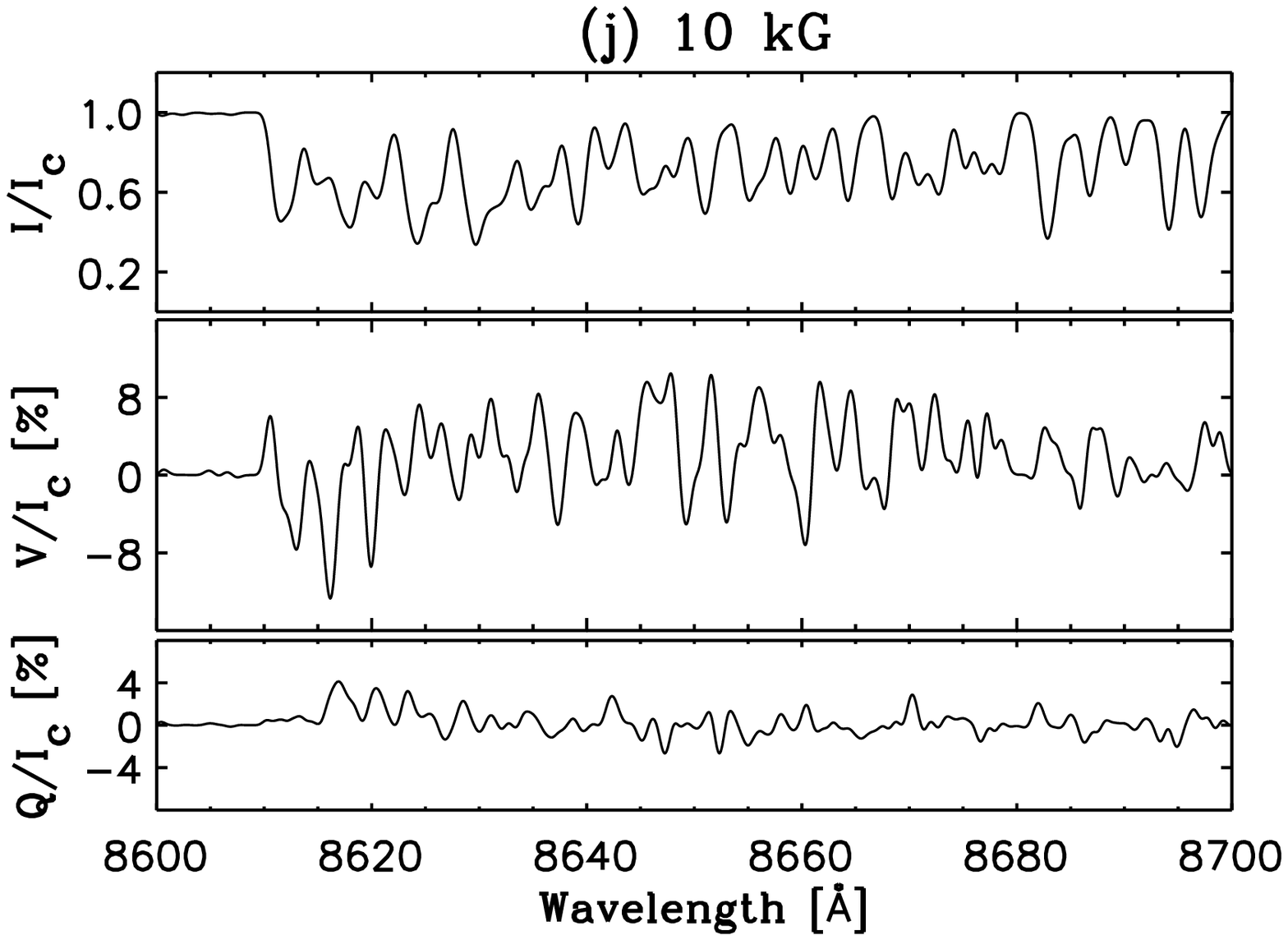}}%[$1$ kG\label{fig: sp9a}]

 \caption{Stokes profiles for the \mbox{(0,0)} band of the A$^6\Sigma^+$ - X$^6\Sigma^+$ system of the CrH calculated for $T_\mathrm{eff}=2500$~K at different magnetic field strengths. The spectra in the right column are calculated assuming a spectral resolution of $0.8$ \AA\, and a stellar rotation $v\sin i=20$~kms$^{-1}$.}
 \label{fig: crh00_2500K}
\end{figure*}

\begin{figure*}%[]
 \centering
 %use [scale=0.3] for the referee format, and [scale=0.37] for the normal one.
 \subfloat{\includegraphics[scale=0.37]{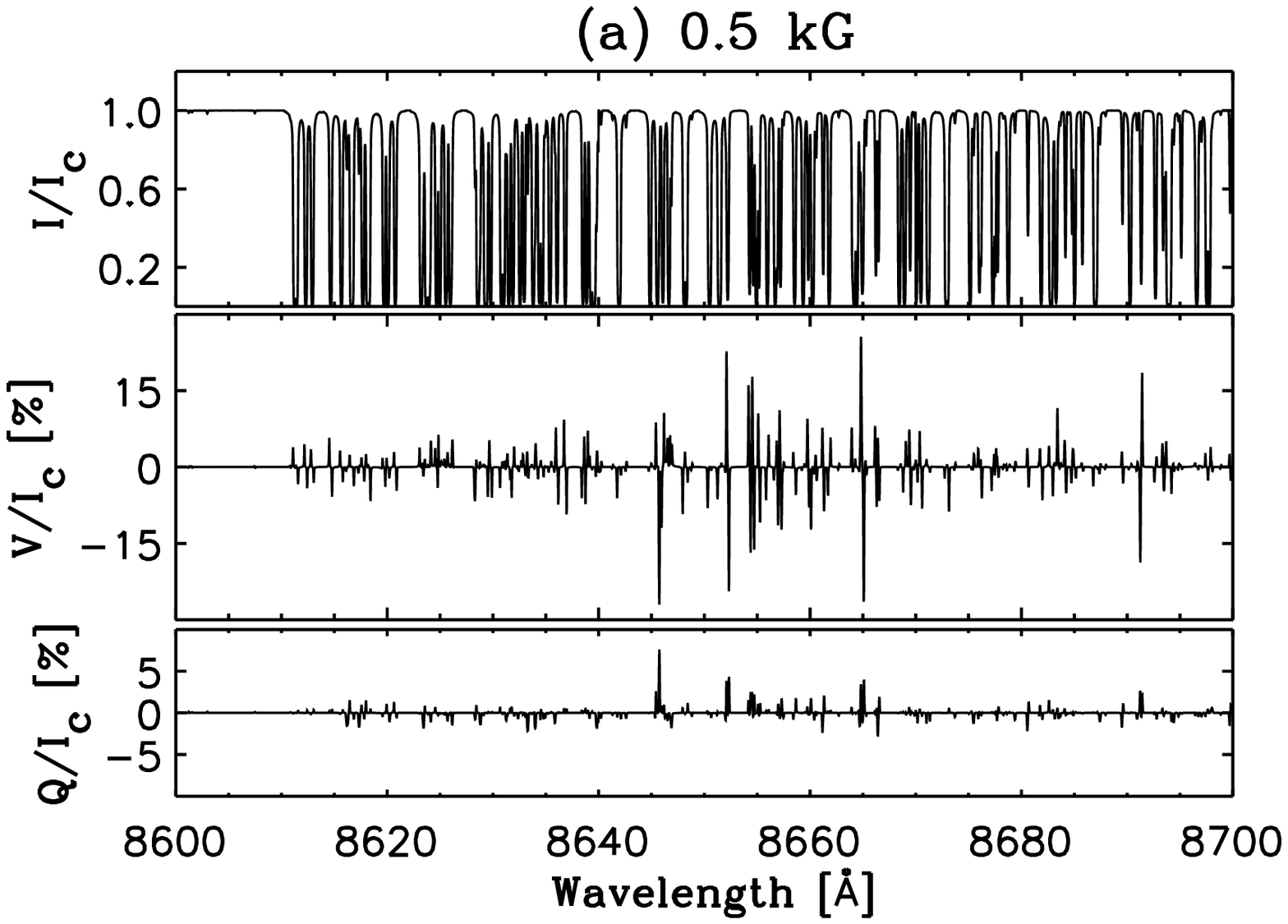}}%crh00_1kG_8600-8700_form.eps
 \hspace{2cm}
 \subfloat{\includegraphics[scale=0.37]{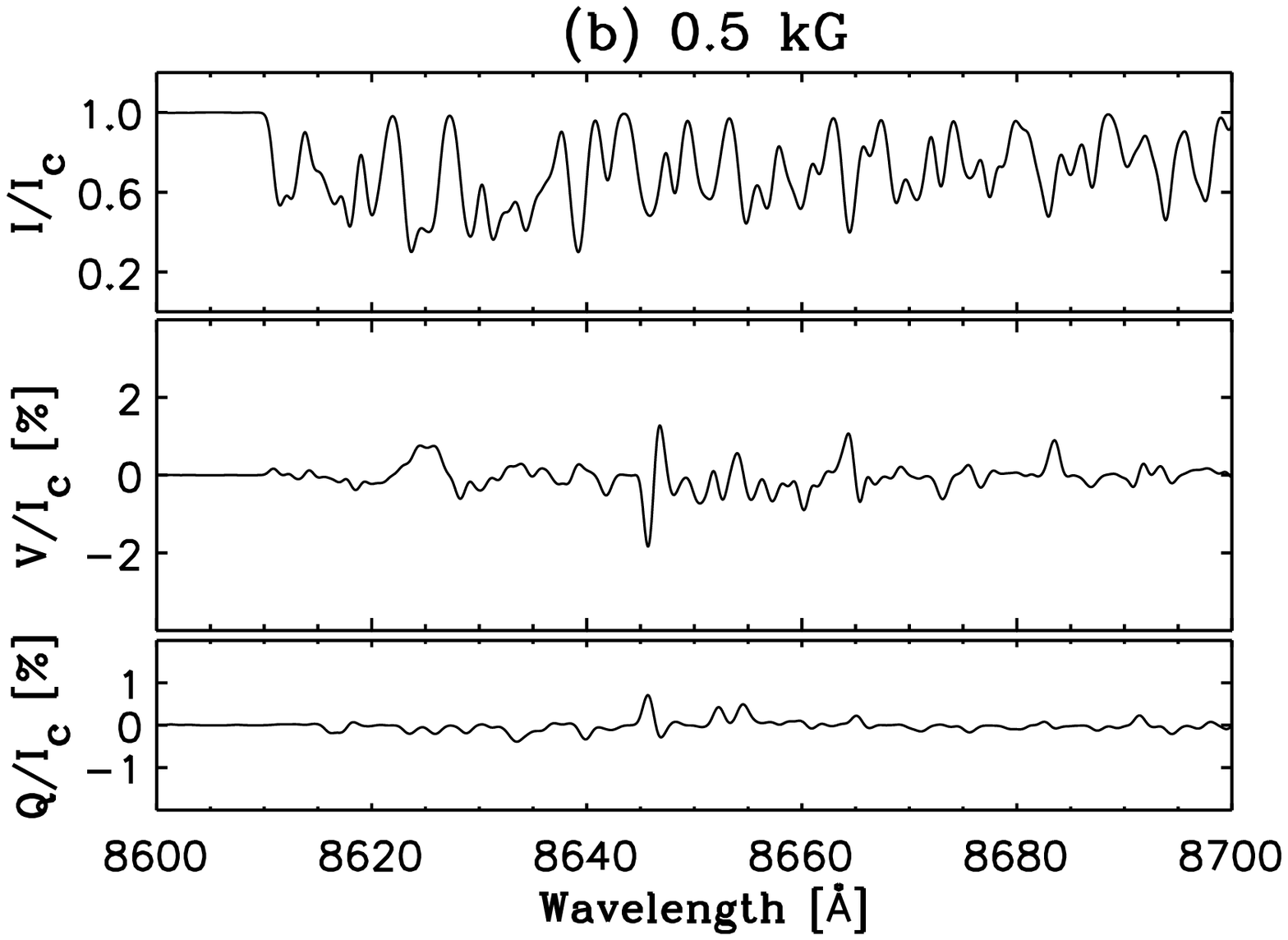}}\\ %[$0.001$ kG \label{fig: sp7}]
 \vspace{-0.25cm}
 \subfloat{\includegraphics[scale=0.37]{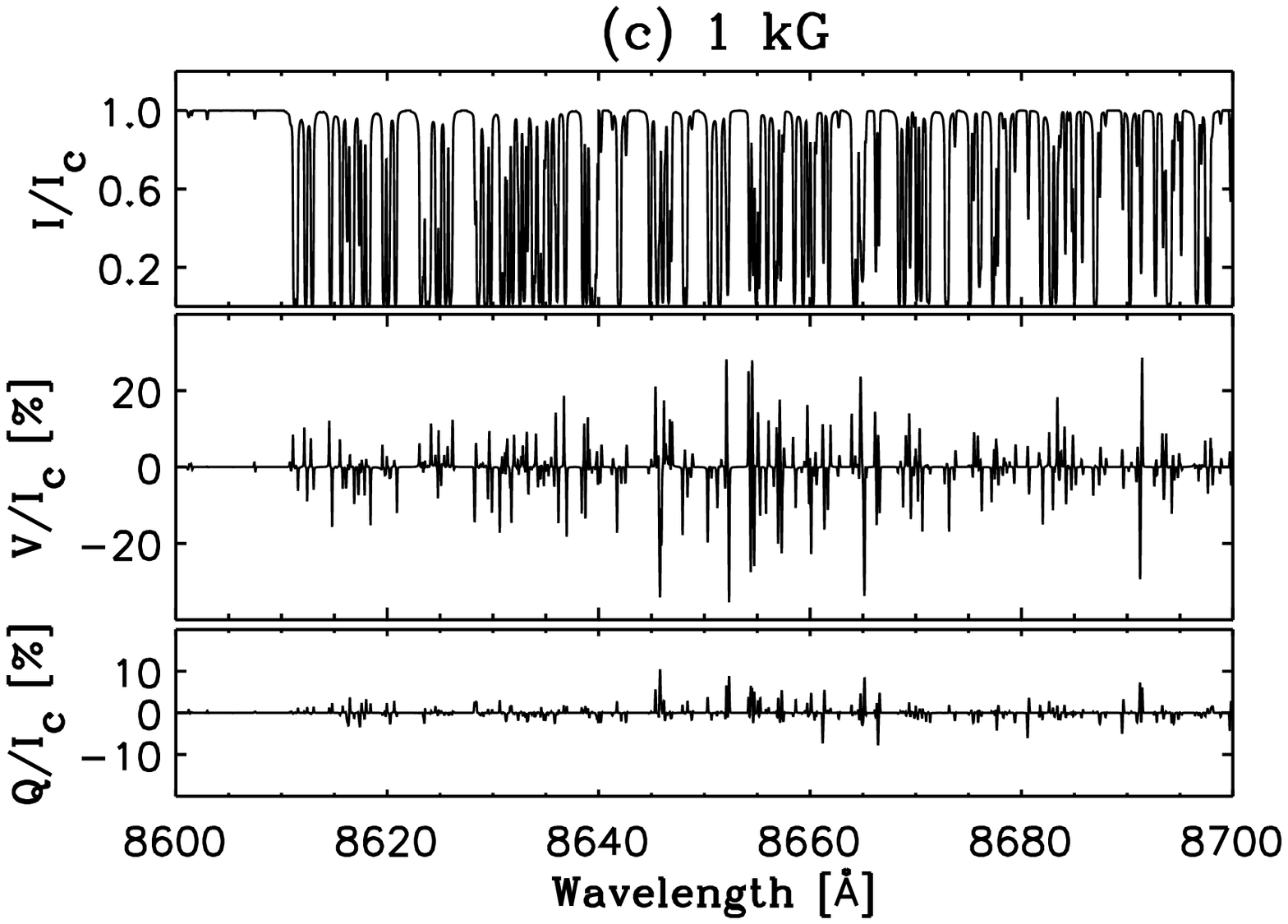}} %[$0.01$ kG\label{fig: zp7a}]
 \hspace{2cm}
 \subfloat{\includegraphics[scale=0.37]{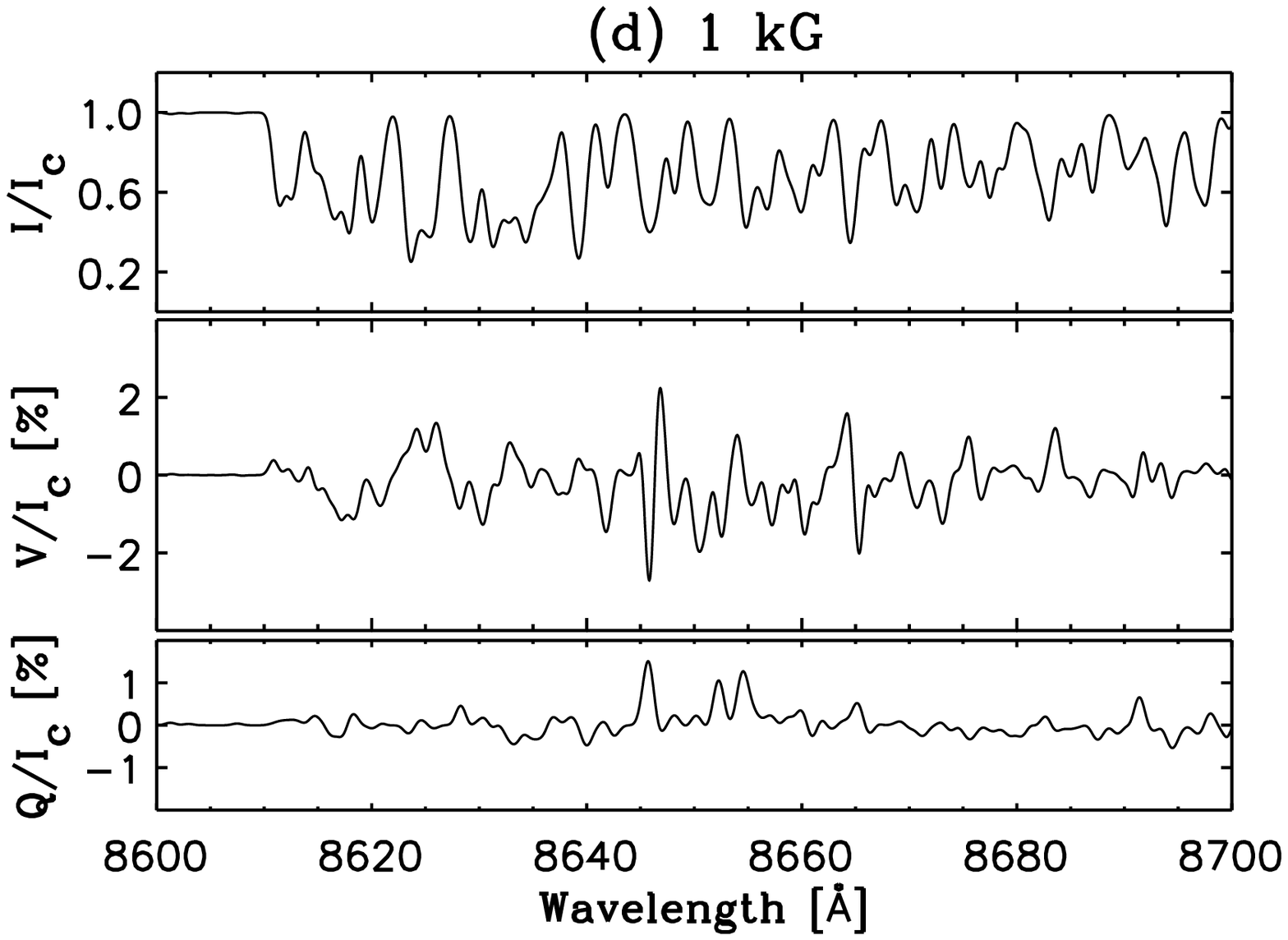}}\\ %[$0.01$ kG\label{fig: sp7a}]
 \vspace{-0.25cm}
 \subfloat{\includegraphics[scale=0.37]{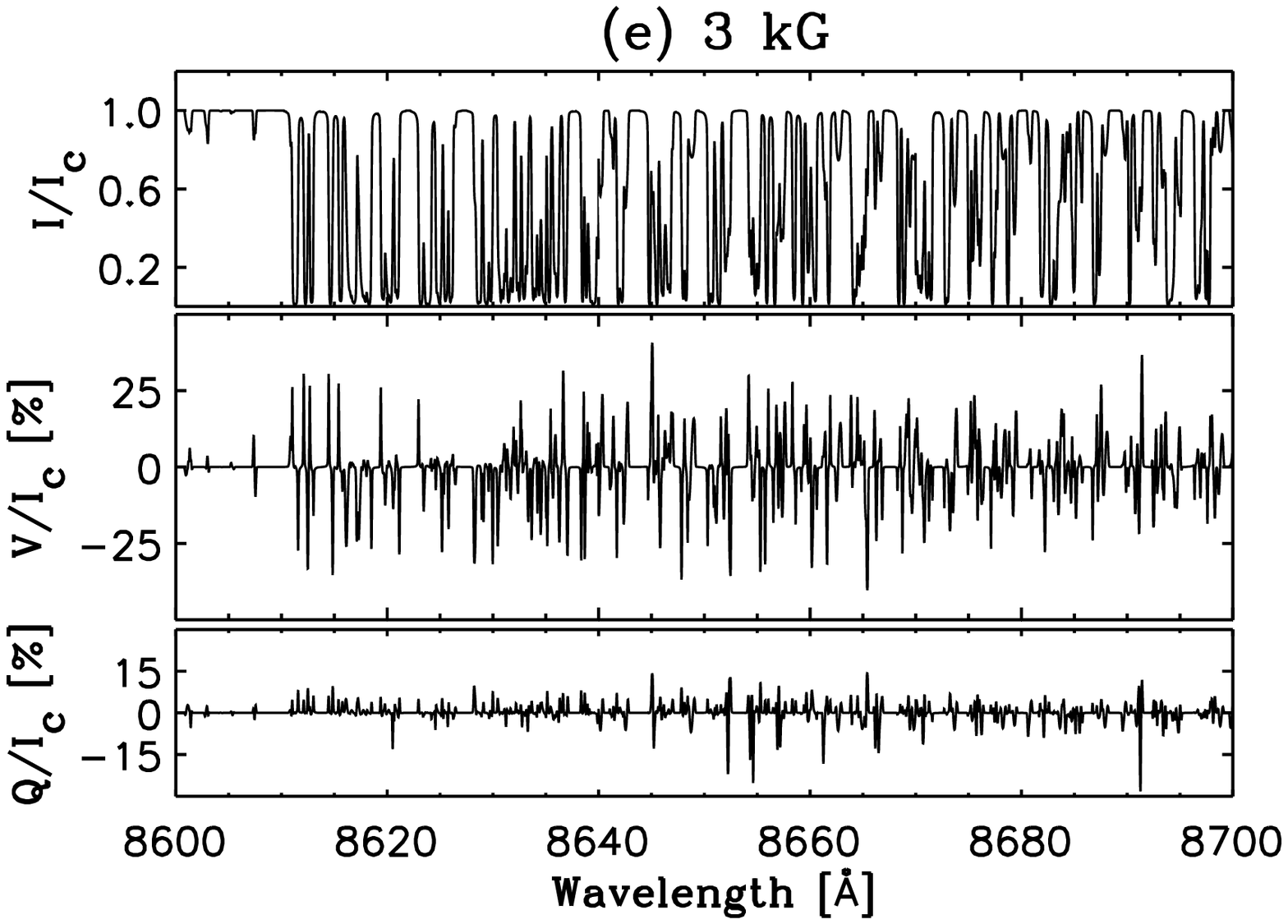}} %[$0.05$ kG\label{fig: zp8}]
 \hspace{2cm}
 \subfloat{\includegraphics[scale=0.37]{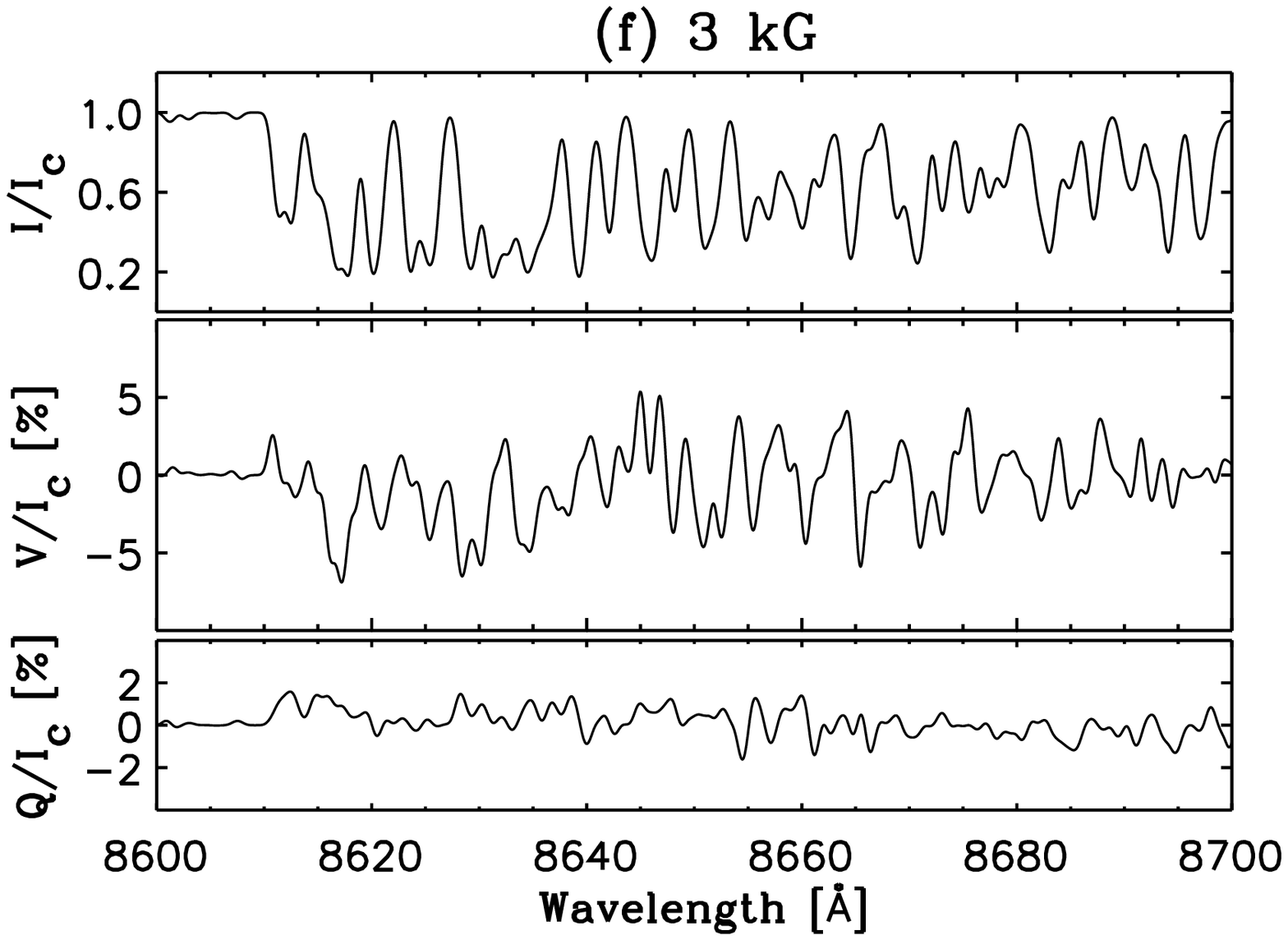}}\\ %[$0.05$ kG \label{fig: sp8}]
  \vspace{-0.25cm}
 \subfloat{\includegraphics[scale=0.37]{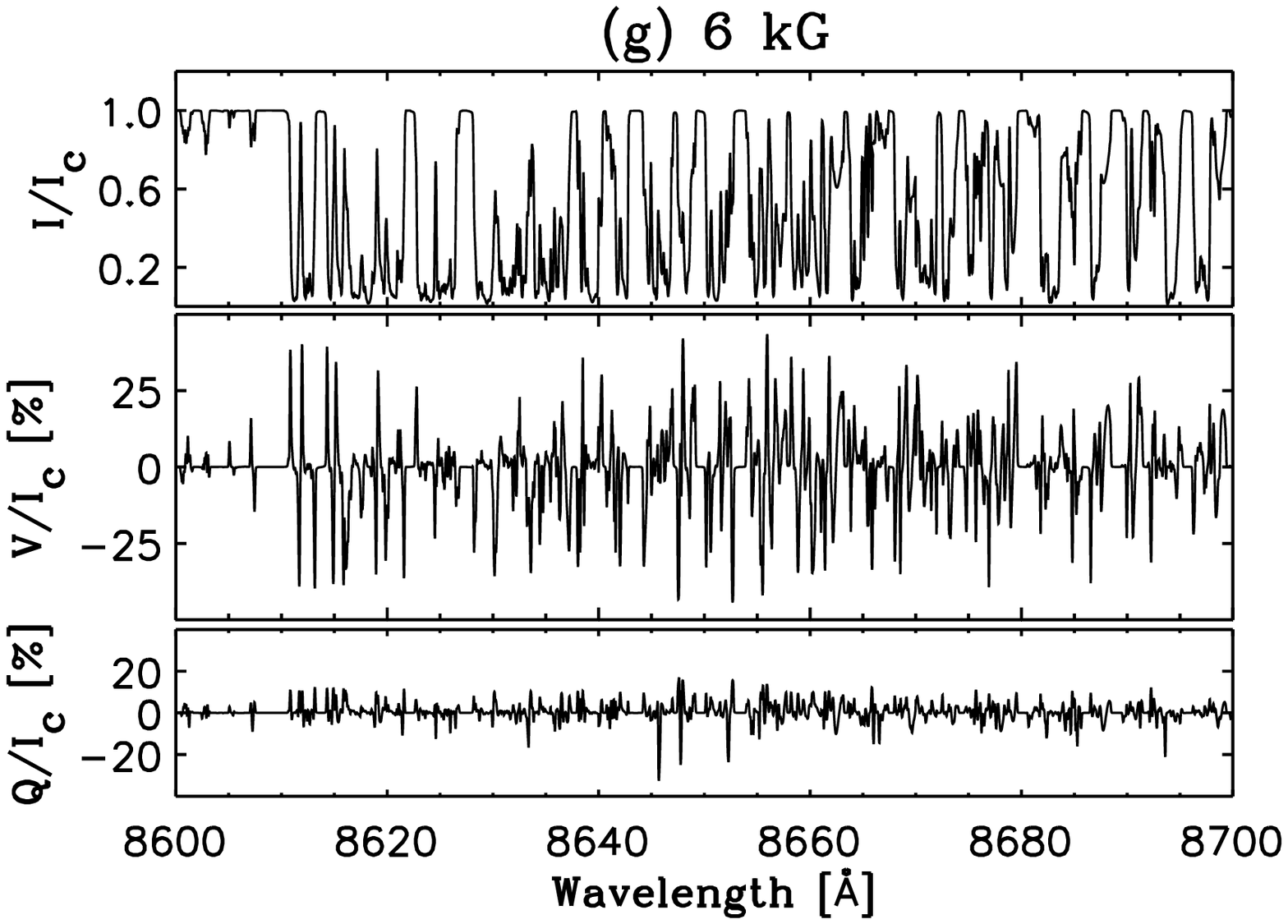}} %[$0.1$ kG\label{fig: zp8a}]
 \hspace{2cm}
 \subfloat{\includegraphics[scale=0.37]{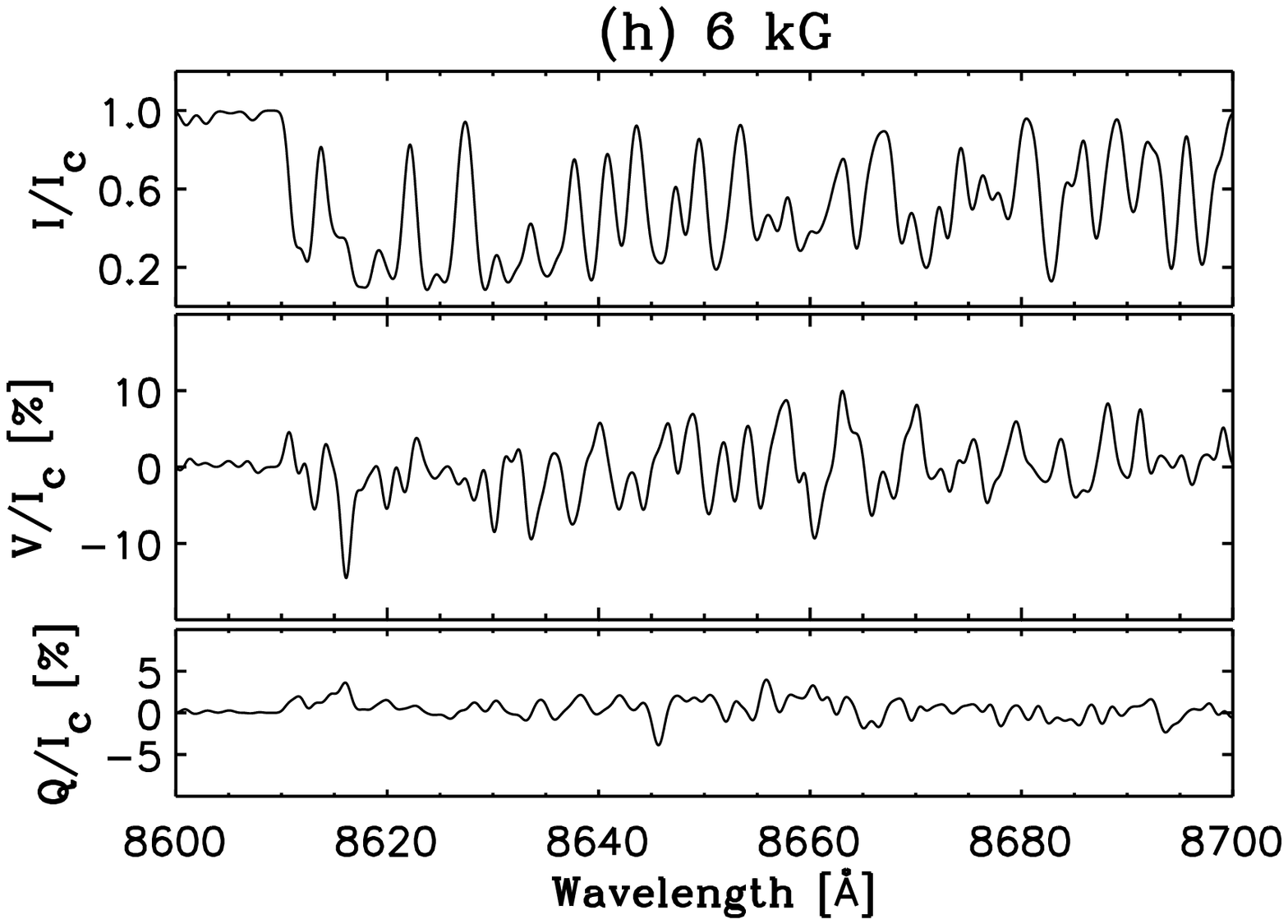}}\\ %[$0.1$ kG\label{fig: sp8a}]
   \vspace{-0.25cm}
 \subfloat{\includegraphics[scale=0.37]{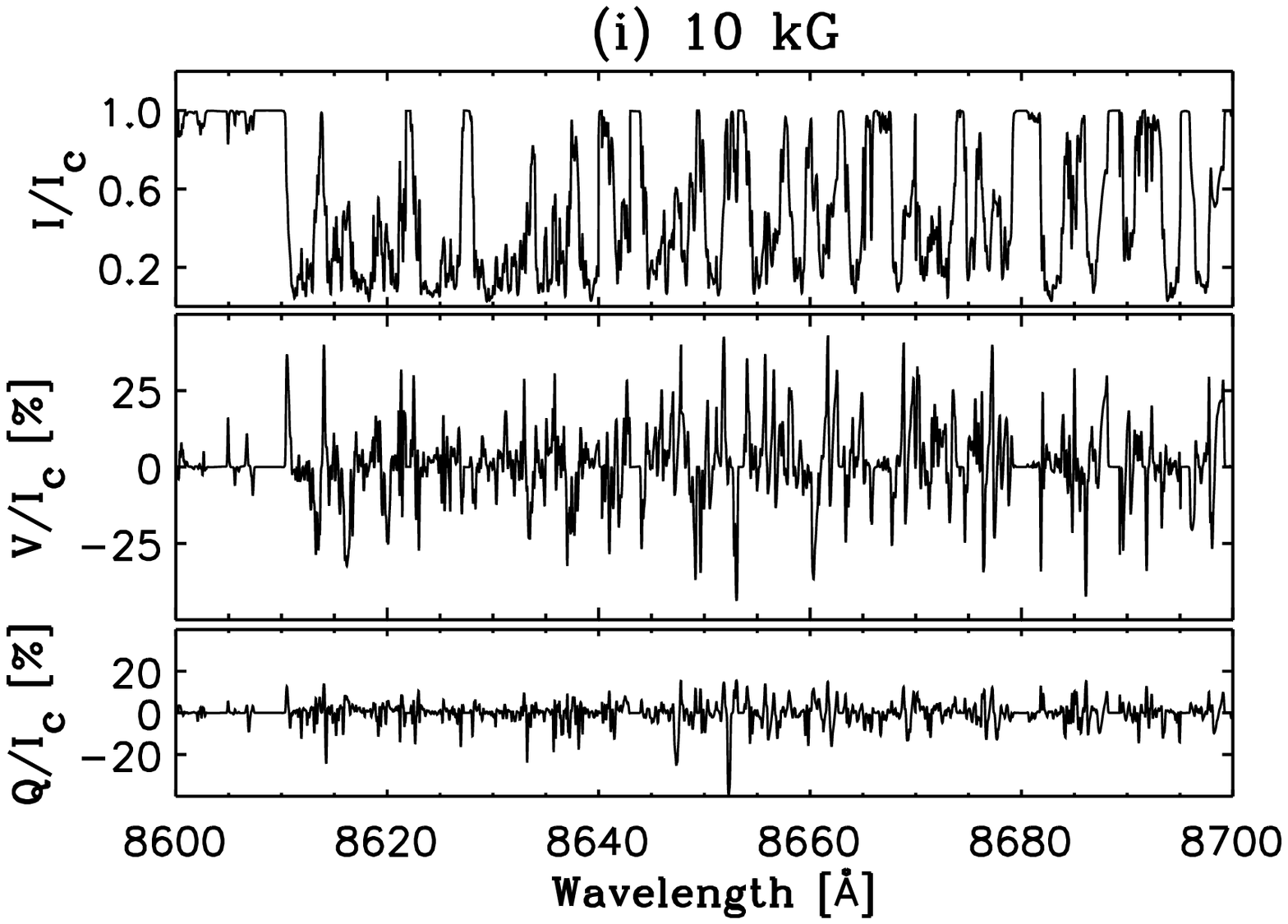}} %[$1$ kG\label{fig: zp9a}]
 \hspace{2cm}
 \subfloat{\includegraphics[scale=0.37]{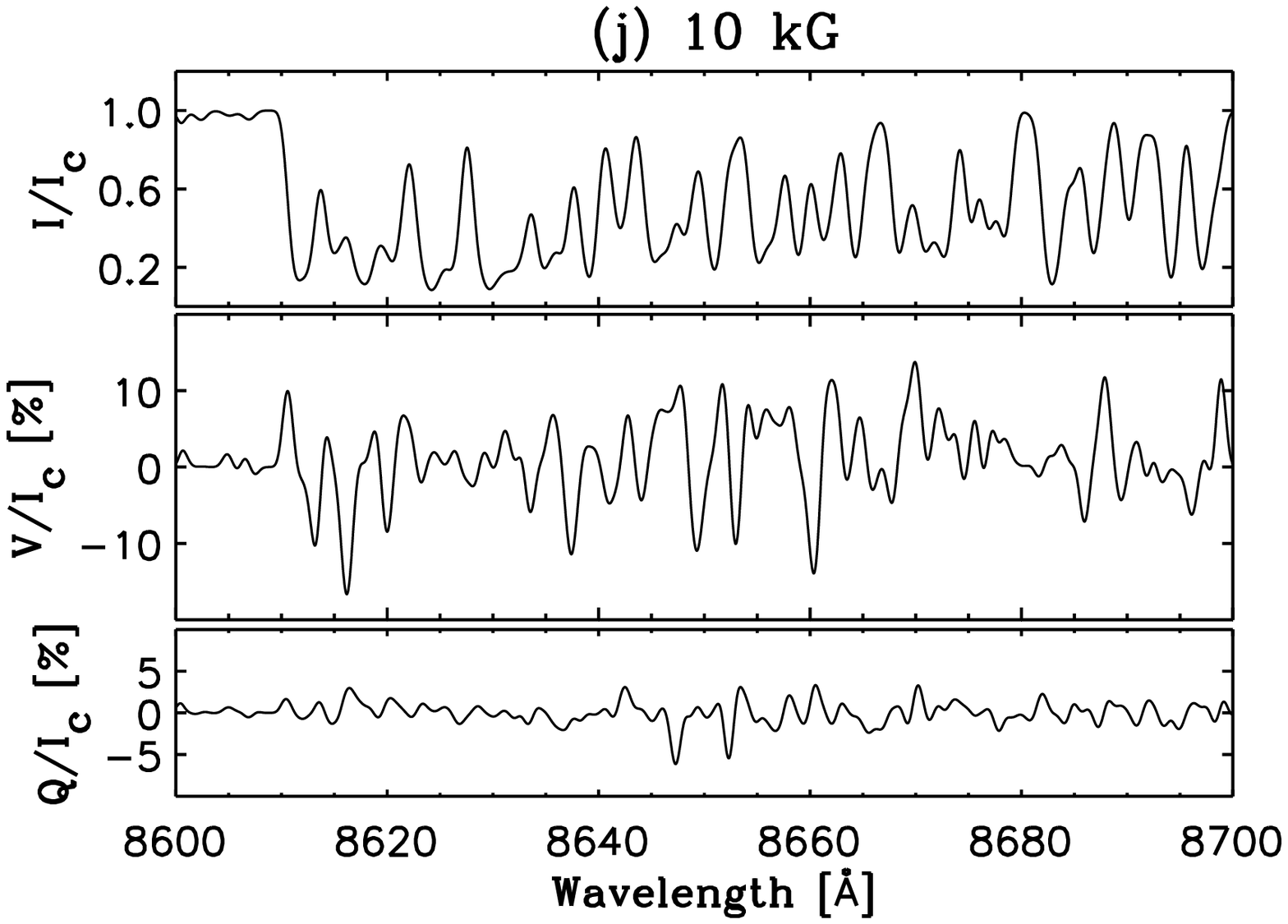}}%[$1$ kG\label{fig: sp9a}]

 \caption{Same as Fig. \ref{fig: crh00_2500K}, but for $T_\mathrm{eff}=2000$~K.}
 \label{fig: crh00_2000K}
\end{figure*}

\begin{figure*}%[]
 \centering
 %use [scale=0.3] for the referee format, and [scale=0.37] for the normal one.
 \subfloat{\includegraphics[scale=0.37]{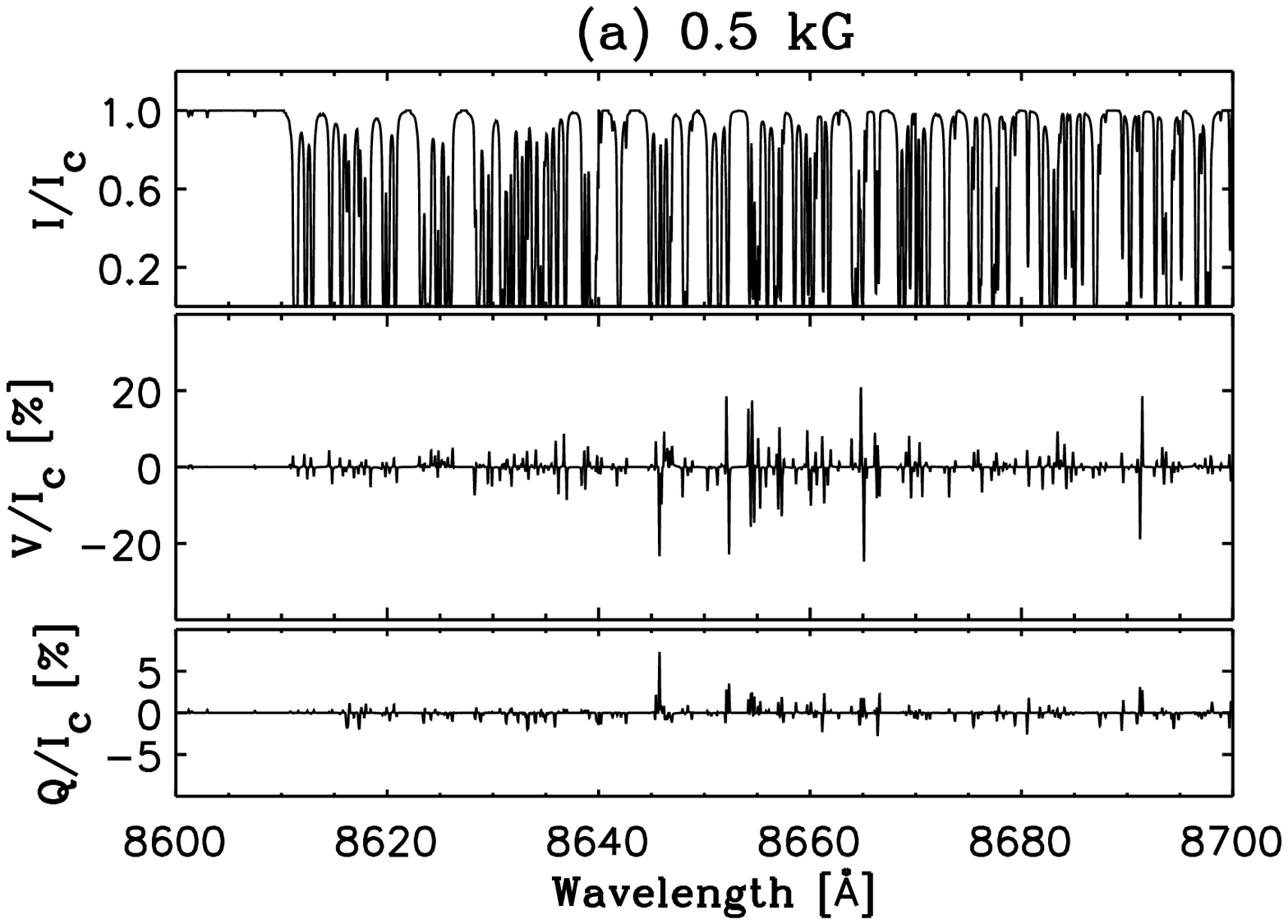}}%crh00_1kG_8600-8700_form.eps
 \hspace{2cm}
 \subfloat{\includegraphics[scale=0.37]{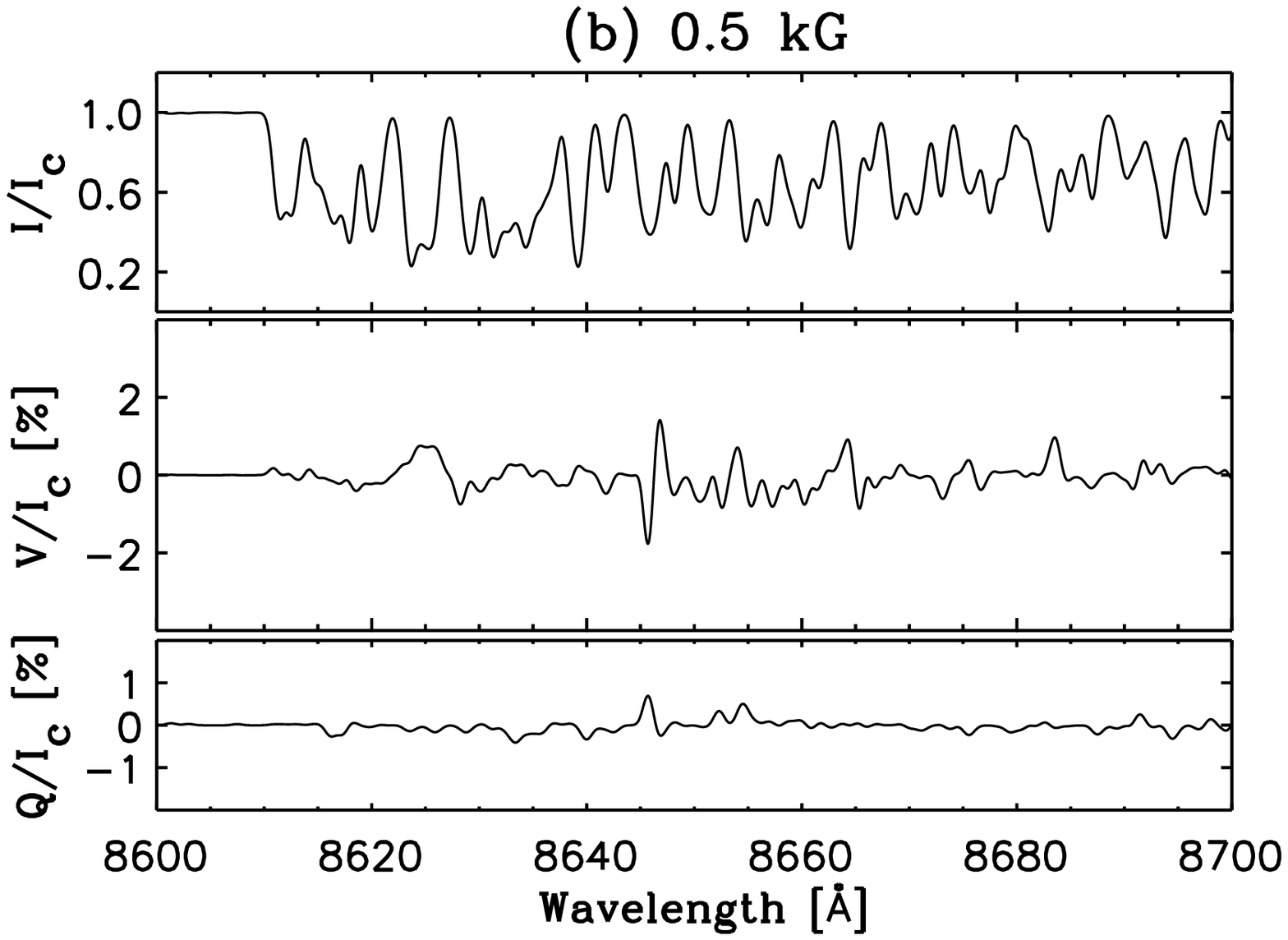}}\\ %[$0.001$ kG \label{fig: sp7}]
 \vspace{-0.25cm}
 \subfloat{\includegraphics[scale=0.37]{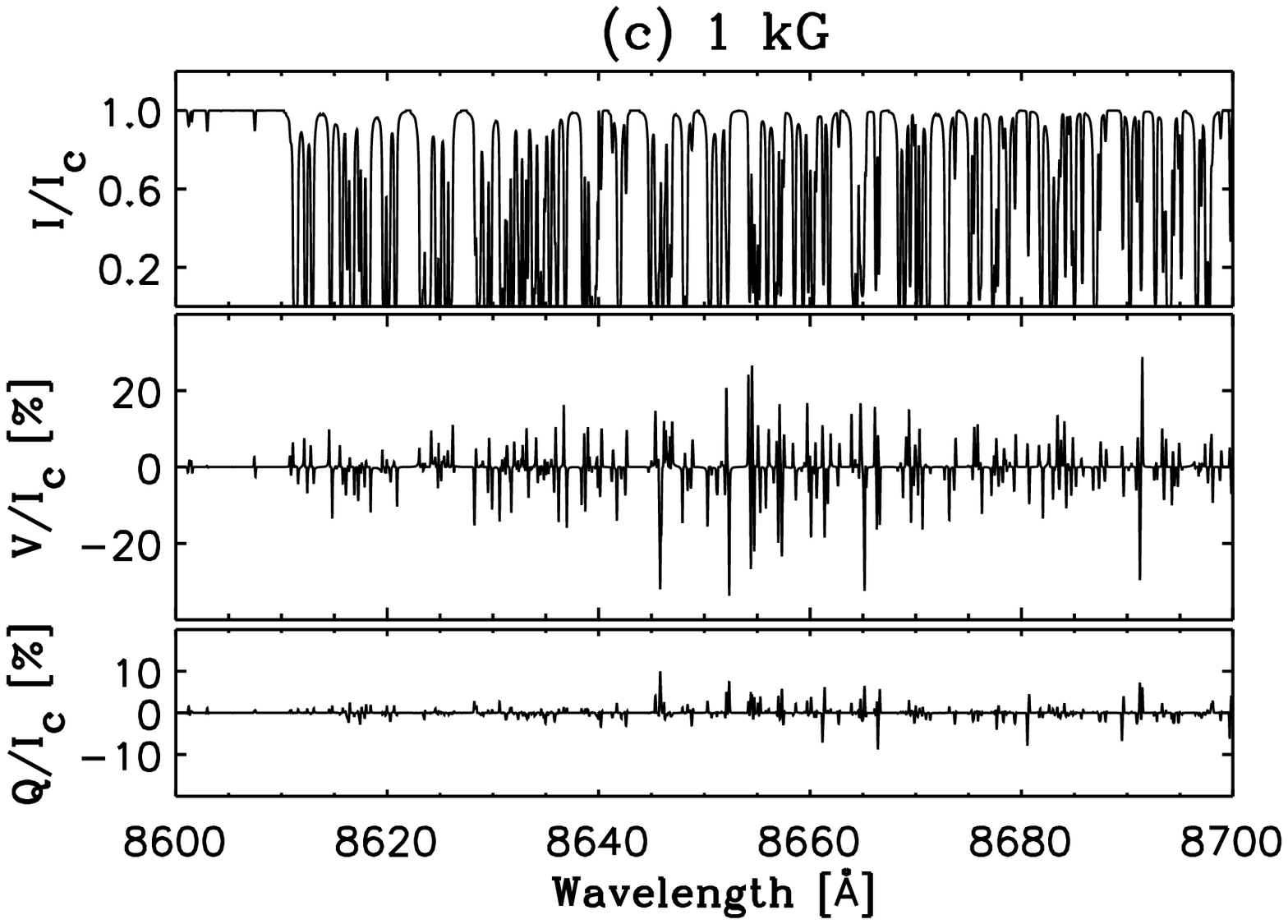}} %[$0.01$ kG\label{fig: zp7a}]
 \hspace{2cm}
 \subfloat{\includegraphics[scale=0.37]{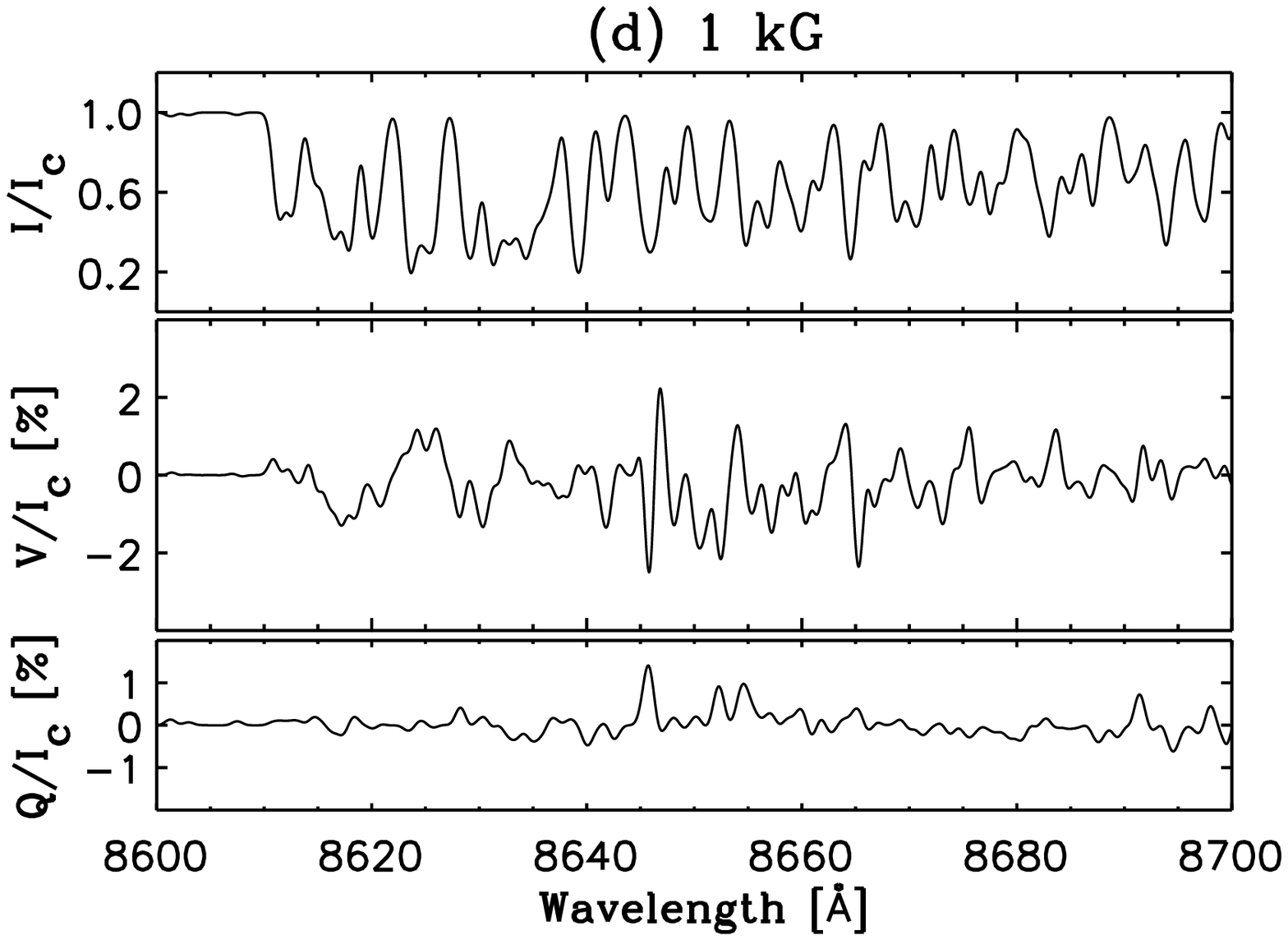}}\\ %[$0.01$ kG\label{fig: sp7a}]
 \vspace{-0.25cm}
 \subfloat{\includegraphics[scale=0.37]{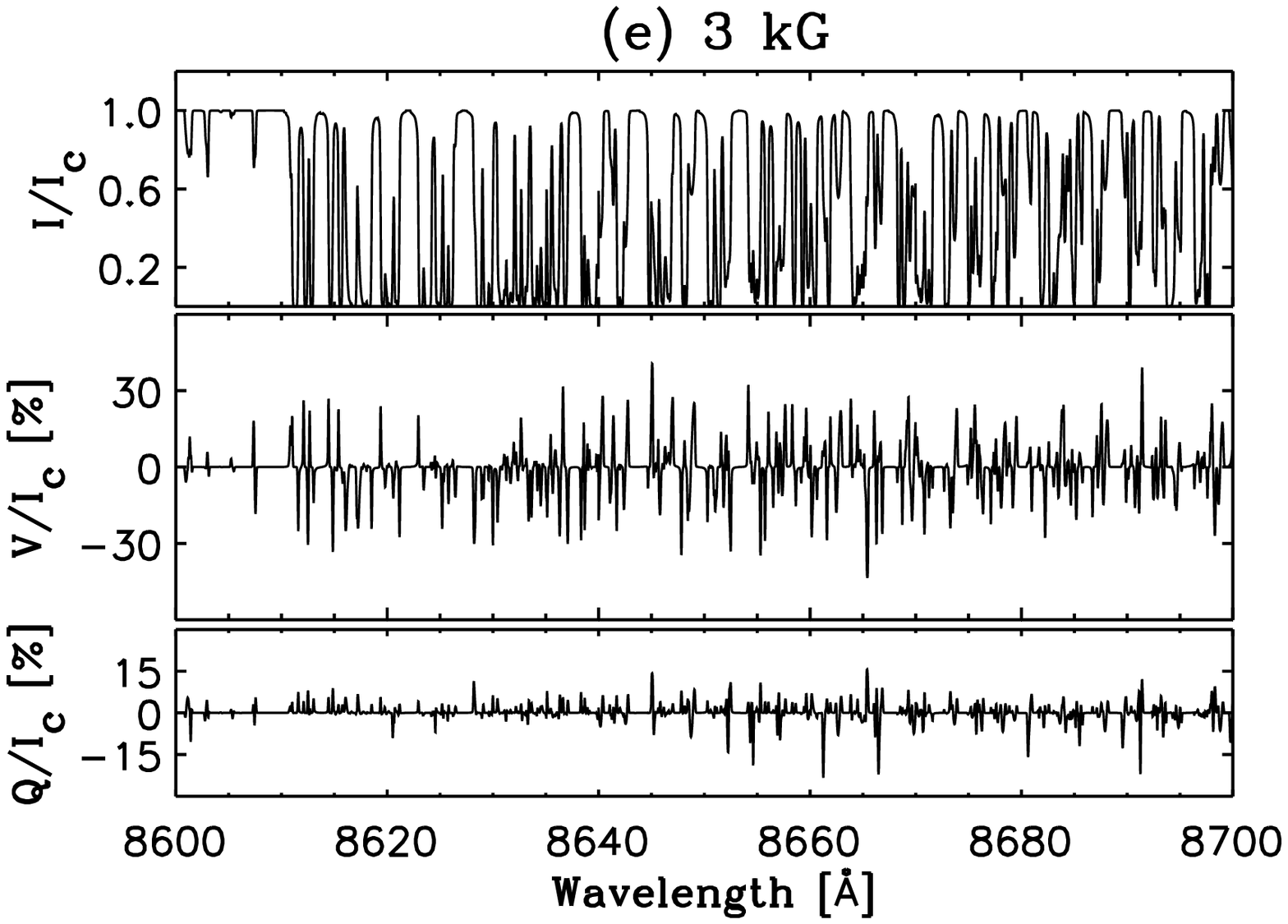}} %[$0.05$ kG\label{fig: zp8}]
 \hspace{2cm}
 \subfloat{\includegraphics[scale=0.37]{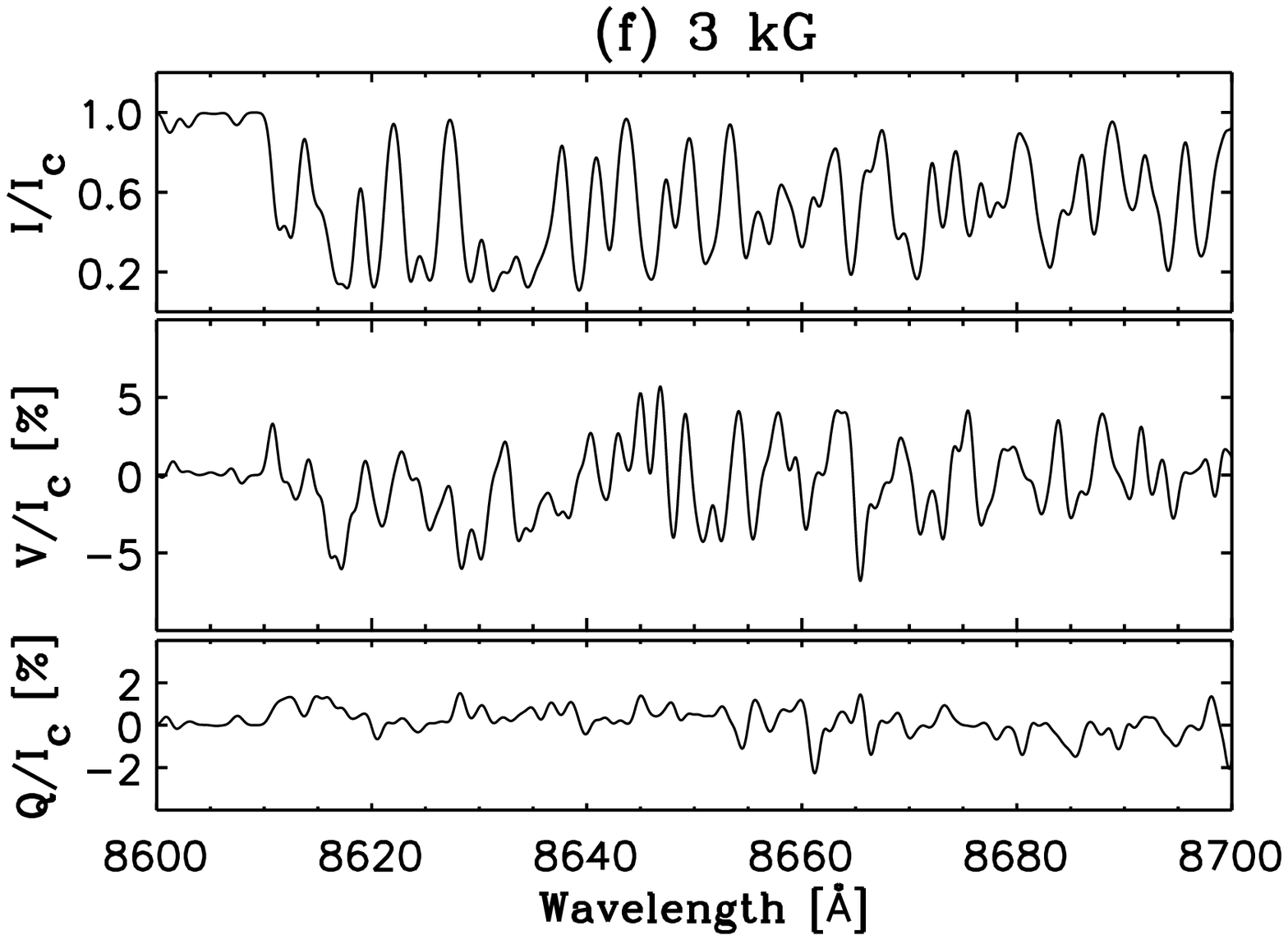}}\\ %[$0.05$ kG \label{fig: sp8}]
  \vspace{-0.25cm}
 \subfloat{\includegraphics[scale=0.37]{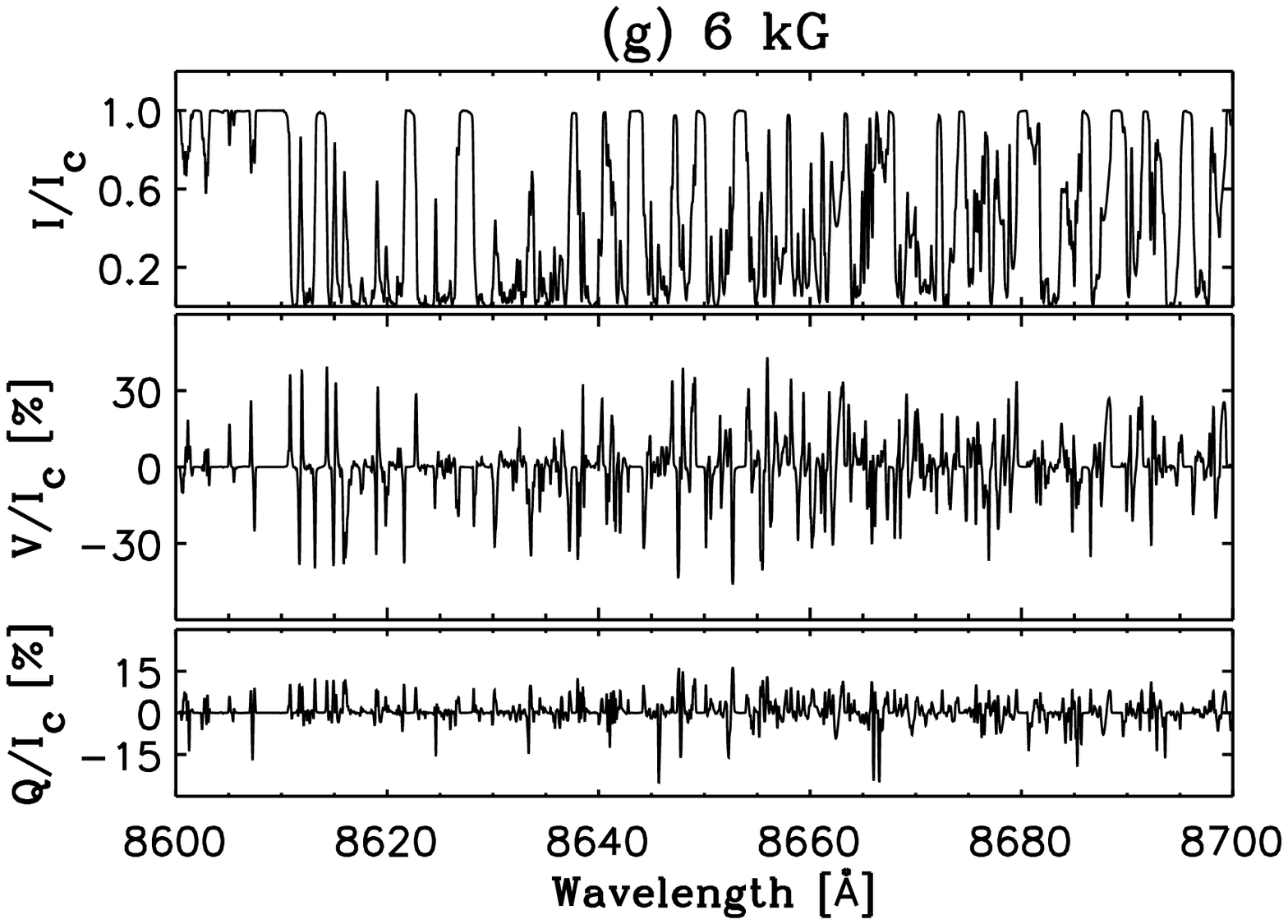}} %[$0.1$ kG\label{fig: zp8a}]
 \hspace{2cm}
 \subfloat{\includegraphics[scale=0.37]{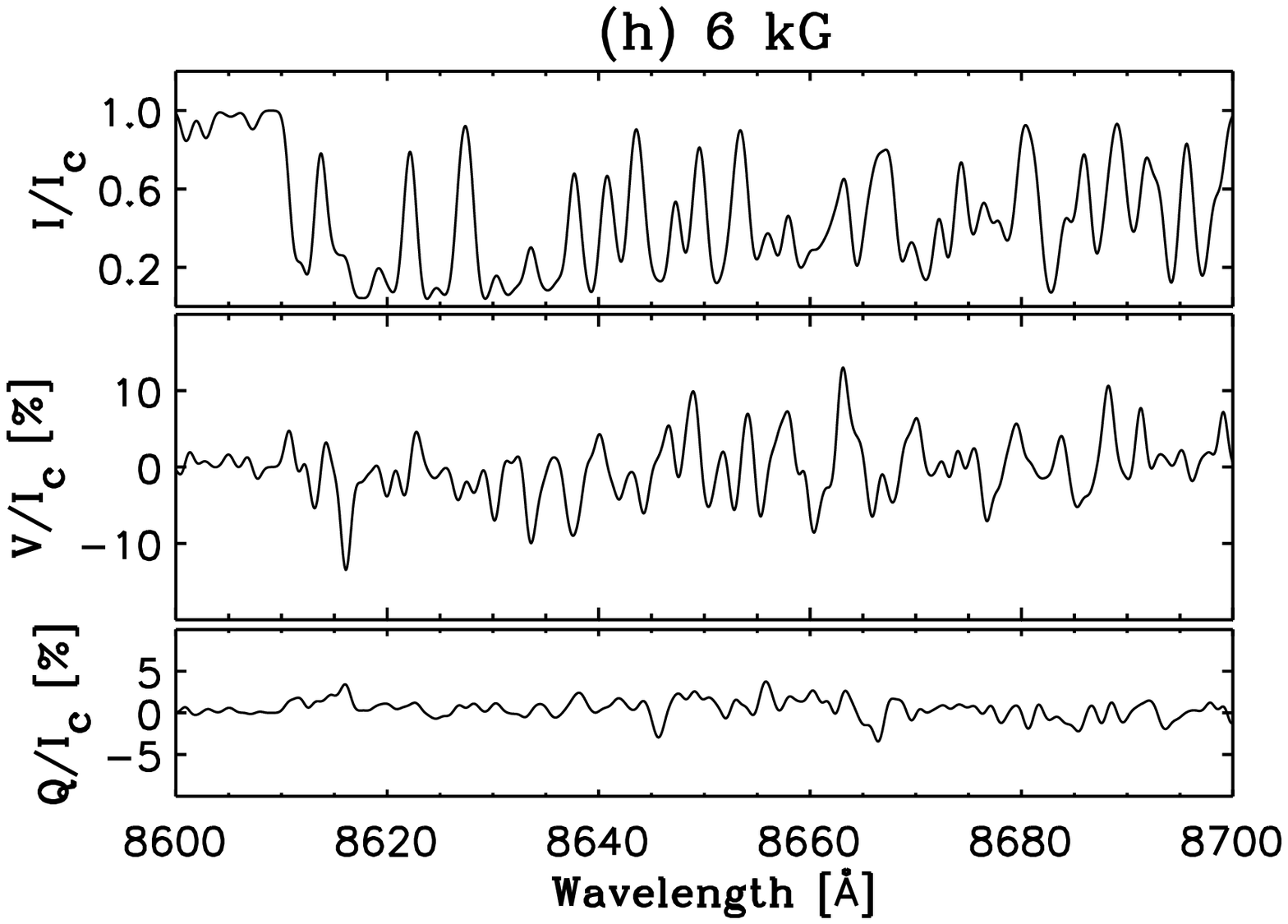}}\\ %[$0.1$ kG\label{fig: sp8a}]
   \vspace{-0.25cm}
 \subfloat{\includegraphics[scale=0.37]{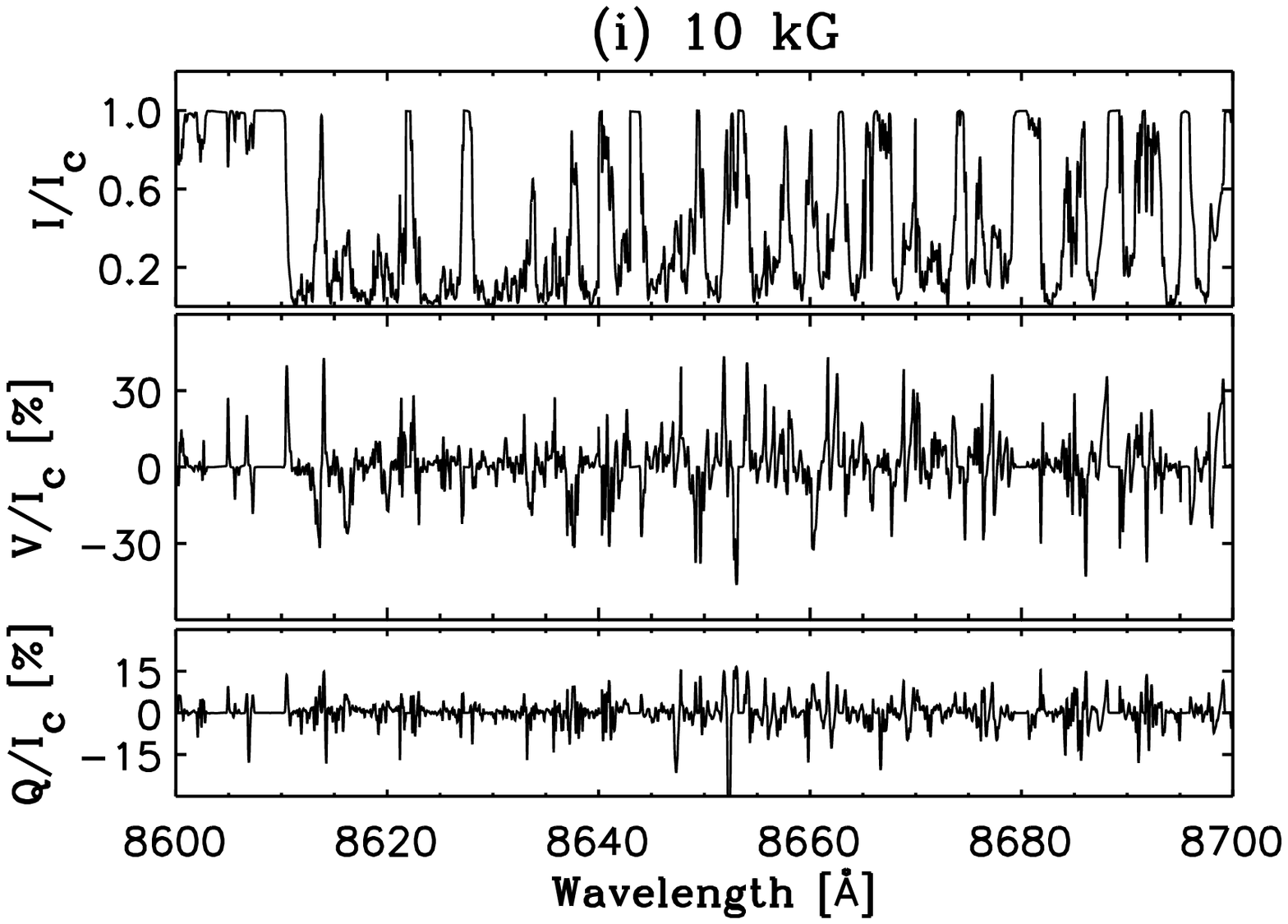}} %[$1$ kG\label{fig: zp9a}]
 \hspace{2cm}
 \subfloat{\includegraphics[scale=0.37]{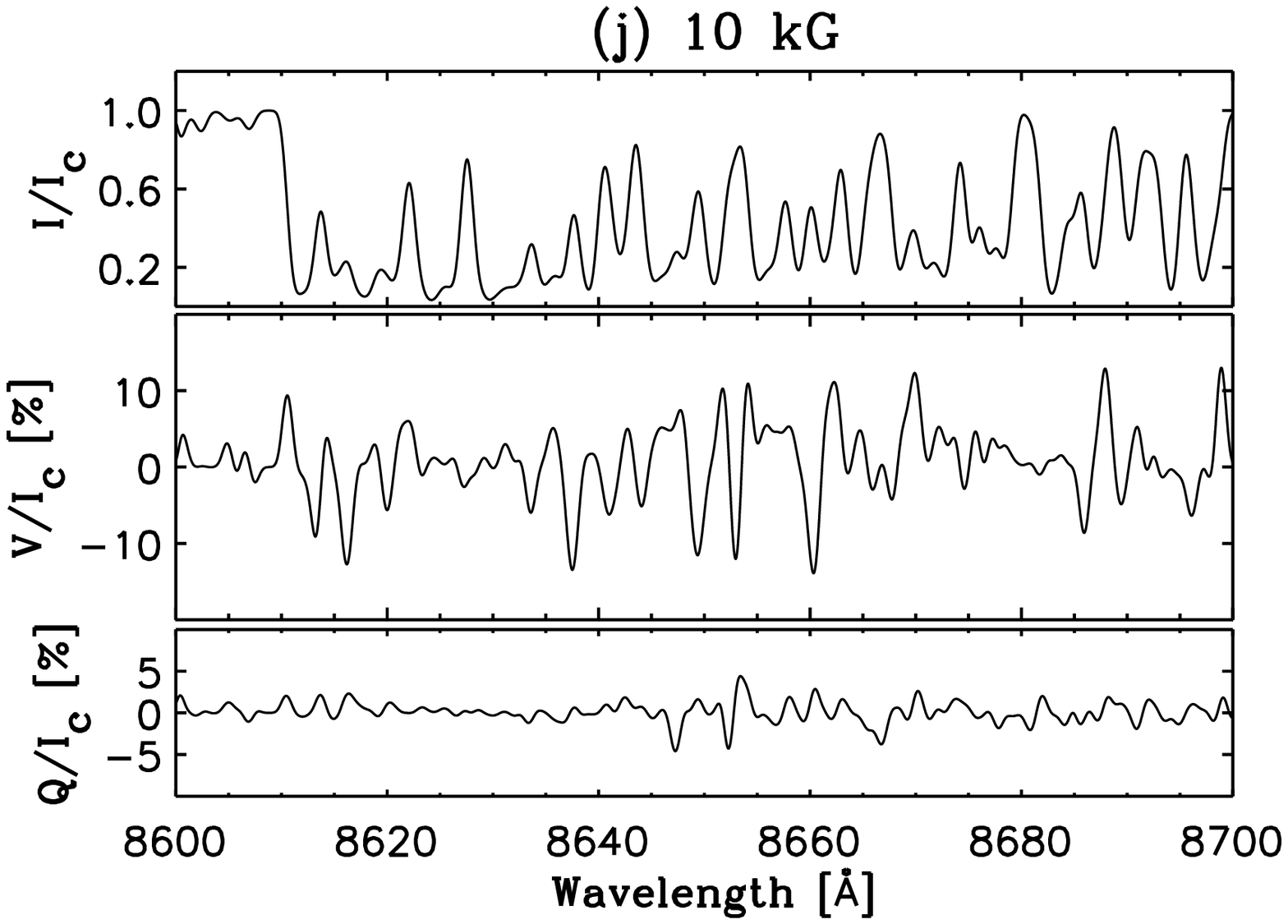}}%[$1$ kG\label{fig: sp9a}]

 \caption{Same as Fig. \ref{fig: crh00_2500K}, but for $T_\mathrm{eff}=1500$~K.}
 \label{fig: crh00_1500K}
\end{figure*}

%-------------------------------------------------
\subsection{Stokes profiles}\label{subsec: stokes_profiles}
%-------------------------------------------------

First we calculate the Stokes profiles for the individual lines at the magnetic field strengths $0.001$, $0.01$, $0.1$, $1$, and $10$~kG. As an example, the inclination of the magnetic field $\gamma$ is set to $45\,^\circ$ and the azimuth $\chi$ to $0\,^\circ$ in all calculations presented in this paper. Hence, both Stokes $V$ and $Q$ are distinct from zero, while the Stokes $U$ vanishes. We normalize the Stokes parameters to the continuum intensity $I_{\rm c}$.

 Figures~\ref{fig: zeeman} and \ref{fig: pb} show the Stokes profiles for the transition in Eq.~(\ref{eq: transition}) assuming the ZR and PBR, respectively. A remarkable feature in Figs. \ref{fig: pb}b, \ref{fig: pb}d, \ref{fig: pb}f, \ref{fig: pb}h, and \ref{fig: pb}j is that neither of the Stokes $Q/I_{\rm c}$ shows a symmetric shape as it does in Fig. \ref{fig: Bild2}. This originates because the states contributing to the transition (at least one of them) are in the PBR. Even at a weak magnetic field (about $0.001$~kG) the ZR is inappropriate for many lines from the \mbox{(0,0)} band, as revealed by the non-symmetric Stokes~$Q/I_c$ (Fig. \ref{fig: pb}b). Indeed, the deviation of the magnetic levels from their degenerate position depends linearly on the magnetic field strength as long as the sub-levels with the same magnetic quantum number do not come close to each other. Otherwise, the sub-levels with the same number $M$ begin to interact (repel), so that the linear dependence from the magnetic field strength, demonstrated by Eq.~(\ref{eq: zeeman_regime}), is no longer guaranteed. Because the separation of the fine structure levels that split in the presence of a magnetic field is not equidistant, and some levels can lie very close together, as shown in Fig.~\ref{fig: magfield}, the ZR fails for such levels even at very small magnetic field strengths. Our careful investigation shows that the Stokes~$Q/I_c$ is especially sensitive to the shifts of the magnetic transitions with respect to their frequency in the absence of an external magnetic field. In other words, even tiny deviations from Eq.~(\ref{eq: zeeman_regime}) cause a non-symmetric Stokes~$Q/I_c$.

Figures~\ref{fig: crh00_2500K}, \ref{fig: crh00_2000K}, and \ref{fig: crh00_1500K} show the Stokes profiles calculated for the entire \mbox{(0,0)} band for $T_\mathrm{eff}=2500$~K, $2000$~K, and $1500$~K, respectively, in the wavelength range $8600 - 8700$~\AA\, at the magnetic field strengths $0.5$, $1$, $3$, $6$, and $10$~kG. The line list used includes about $500$ CrH lines calculated using the theory given in Sect. \ref{sec: energy_levels}. Very weak forbidden and satellite lines were sorted out from the list. Natural and pressure induced line broadening were included in the calculations. In addition, effects of low spectral resolution and possible fast stellar rotation were taken into account.

We have found that the Stokes $V/I_c$ spectra show significant signals at the field strengths considered. In general, polarization signals increase for stronger magnetic fields from a few percent at $0.5$~kG to more than 20\%\ at fields of several kG when observed at high spectral resolution. In addition, the polarization signal arising from some lines or groups of lines (e.g., near $8625$ or $8685$ \AA) shows a considerable asymmetry due to the Paschen-Back effect in these lines (cf. panels a and b in Figs.~\ref{fig: crh00_2500K}, \ref{fig: crh00_2000K}, and \ref{fig: crh00_1500K}). Furthermore, the satellite lines ($\Delta J \ne \Delta N$) gain in strength, while the main lines ($\Delta J = \Delta N$) become weaker (cf. Stokes~$I/I_c$ in Figs.~\ref{fig: crh00_2500K}a, \ref{fig: crh00_2500K}c, \ref{fig: crh00_2500K}e, \ref{fig: crh00_2500K}g, \ref{fig: crh00_2500K}i). Thus, the polarization signal coming from a great number of satellite lines becomes comparable with that coming from the main lines. It reaches $25$~\% in Stokes~$V/I_c$ and $15$~\% in Stokes~$Q/I_c$ at $10$~kG (cf. Figs.~~\ref{fig: crh00_2500K}a, \ref{fig: crh00_2500K}c, \ref{fig: crh00_2500K}e, \ref{fig: crh00_2500K}g, \ref{fig: crh00_2500K}i). Moreover, while the magnetic field strength increases, the absorption intensity (Stokes~$I/I_c$) decreases noticeably. This is caused by redistribution of the radiative energy between main and satellite lines. 

In the strong field regime, the Stokes $V/I_c$ undergoes a crucial alteration: it becomes asymmetric over a wide range of wavelengths (cf. panels f, h, and j in Figs.~\ref{fig: crh00_2500K}, \ref{fig: crh00_2000K}, and \ref{fig: crh00_1500K}) leading to a large scale broad-band polarization, which can be detected even at a much lower spectral resolution. For instance, magnetic fields of 6 kG and stronger could be detected on very faint brown dwarfs using this feature.

%To demonstrate these effects at low spectral resolution, we take into account the instrumental broadening of $0.8$~\AA\ as well as the broadening due to stellar rotation of $v\sin i=20$~kms$^{-1}$ (see Figs.~\ref{fig: crh00}b, \ref{fig: crh00}d, \ref{fig: crh00}f, \ref{fig: crh00}h and \ref{fig: crh00}j).

%%%%%%%%%%%%%%%%%%%%%%%%%%%%%%%%%%%%%%%%%%%%%%%%%%%%%%%%%
\section{Conclusions}\label{sec: conclusions}
%%%%%%%%%%%%%%%%%%%%%%%%%%%%%%%%%%%%%%%%%%%%%%%%%%%%%%%%%

We have explored the magnetic sensitivity of the \mbox{(0,0)} vibrational band from the A$^6\Sigma^+$-- X$^6\Sigma^+$ system of the CrH molecule and developed a new diagnostic for magnetic field measurements on very cool stars, brown dwarfs, and, potentially, on hot Jupiters. 

Our quantum mechanical calculations of the energy level structure of the CrH (0,0) band in the absence of an external magnetic field reproduce well the results of \citet{ram1993} and are in a good agreement with the results obtained by \citet{kleman1959} and \citet{burrows2002}.

Employing the approach by \citet{berdyugina2005} for predicting the magnetic level structure of electronic states of any multiplicity in diatomic molecules, we have calculated for the first time the CrH level structure in the presence of an external magnetic field. Transition frequencies and their strengths are obtained in both the ZR and PBR.

We confirm the following general behavior of transition strengths in the PBR as the magnetic field strength increases \citep{berdyugina2005}. First, satellite lines become stronger, while the main branch lines weaken. In addition, $\sigma^+$ components ($\Delta M=1$) become stronger than $\sigma^-$ components ($\Delta M=-1$) in the $R$ branch, while the opposite is true for the $P$ branch. 

Our detailed study of a number of Stokes profiles of the individual CrH lines shows that even at weak magnetic fields ($\sim 1$~G) Eq.~(\ref{eq: zeeman_regime}) for the energy shifts in the ZR becomes inadequate for many of these lines. In general, the energy shifts due to an external magnetic field must be considered non-linear at any magnetic field strength, even though the deviations from the ZR are approaching zero at a very weak magnetic field.

A spectral synthesis of the CrH \mbox{(0,0)} band at different magnetic field strengths ($0.5-10$~kG) has revealed the following general behavior of the spectra as the field strength increases:
\begin{itemize}
\item the band profile varies with the magnetic field strength;
\item absorption in Stokes $I/I_c$ decreases;
\item polarization, particularly Stokes $V/I_c$, increases; 
\item Stokes profiles become asymmetric;
\item integral polarization over the band is distinct from zero (broad-band polarization).
\end{itemize}

The asymmetric shape of the Stokes~$V/I_c$ is the main effect that provides the sensitivity to the magnetic fields on substellar objects.
%(in the PBR the width of this asymmetry can be even smaller than the instrumental broadening). 
Based on our results, fields of $\sim 100$~G and stronger can be detected with existing instruments. This lower limit has two origins: i) at weaker magnetic fields the polarization degree is too small to be detected, and ii) most of lines in the band are not yet in the complete PBR to produce a considerable asymmetry in the Stokes $V/I_c$ signal.

The calculations presented do not include a filling factor for magnetic regions on the stellar surface because it is not well known for brown dwarfs; however, its effect is linear. Taking into account a filling factor, which is a positive rational number between $0$ and $1$, will proportionally reduce the intensity of a polarization signal. However, this does not affect the asymmetry and the shape of the signal. Therefore, an analysis of the line (or band) profiles in the PBR can provide unambiguous estimates of both the magnetic field strength and its filling factor. This is in contrast to the ZR, when only a product of these two quantities can be inferred.

The wavelength range considered ($8600 - 8700$\AA) includes blends other than CrH lines (e.g., FeH and TiO lines). These have to be taken into account when evaluating and interpreting the observational data. 

Overall, we conclude that the CrH  A$^6\Sigma^+$-- X$^6\Sigma^+$ system is a very sensitive magnetic diagnostic for cool substellar objects, and we will employ it for our studies of brown dwarfs in the near future.

\begin{acknowledgements}
We are grateful to P.~Bernath for his helpful advice on the calculation of the rotational structure of the~CrH. We thank Ch.~Helling and S.~Witte for providing us with the Drift-Phoenix model atmospheres and the corresponding dust properties. This work was supported by the grant of the Leibniz Association SAW-2011-KIS-7, ERC Advanced Grant HotMol, and the NASA Astrobiology Institute.
\end{acknowledgements}

\bibliographystyle{aa} %style aa.bst
\bibliography{crhpaper} %references in crhpaper.bib

\begin{thebibliography}{26}
\expandafter\ifx\csname natexlab\endcsname\relax\def\natexlab#1{#1}\fi

\bibitem[{Afram {et~al.}(2008)Afram, Berdyugina, Fluri, {et~al.}}]{afram2008}
Afram, N., Berdyugina, S., Fluri, D., {et~al.} 2008, \aap, 482, 387

\bibitem[{Allard \& Hauschildt(1995)}]{allard1995}
Allard, F. \& Hauschildt, P. 1995, \apj, 445, 433

\bibitem[{Berdyugina(2011)}]{berdyugina2011}
Berdyugina, S.~V. 2011, ASPC, 437, 219

\bibitem[{Berdyugina {et~al.}(2007)Berdyugina, Berdyugin, \&
  Piriiola}]{berd2007}
Berdyugina, S.~V., Berdyugin, A.~V., \& Piriiola, V. 2007, \prl, 99, 1

\bibitem[{Berdyugina {et~al.}(2005)Berdyugina, Braun, \&
  Fluri}]{berdyugina2005}
Berdyugina, S.~V., Braun, P., \& Fluri, D. 2005, \aap, 444, 947

\bibitem[{Berdyugina \& Solanki(2002)}]{berdyugina2002}
Berdyugina, S.~V. \& Solanki, S. 2002, \aap, 385, 701

\bibitem[{Berdyugina {et~al.}(2003)Berdyugina, Solanki, \&
  Frutiger}]{berdyugina2003}
Berdyugina, S.~V., Solanki, S., \& Frutiger, C. 2003, \aap, 412, 513

\bibitem[{Bernath(2009)}]{bernath2009}
Bernath, P. 2009, International {R}eviews in {P}hysical {C}hemistry, 28, 681

\bibitem[{Bohren \& Huffman(1998)}]{bohren1998}
Bohren, C. \& Huffman, D. 1998, Absorption and Scattering of Light by Small
  Particles (John Wiley \& Sons)

\bibitem[{Brown \& Carrington(2003)}]{brown2003}
Brown, J. \& Carrington, A. 2003, Rotational Spectroscopy of Diatomic Molecules
  (Cambridge University Press)

\bibitem[{Burrows {et~al.}(2002)Burrows, Ram, Bernath, {et~al.}}]{burrows2002}
Burrows, A., Ram, R.~S., Bernath, P., {et~al.} 2002, \apj, 577, 986

\bibitem[{Edl\'{e}n(1966)}]{edlen1966}
Edl\'{e}n, B. 1966, Metrologia, 2, 71

\bibitem[{Engvold {et~al.}(1980)Engvold, W\"ohl, \& Brault}]{engvold1980}
Engvold, O., W\"ohl, H., \& Brault, J.~W. 1980, \aaps, 42, 209

\bibitem[{Frutiger {et~al.}(1999)Frutiger, Solanki, Fligge,
  {et~al.}}]{frutiger1999}
Frutiger, C., Solanki, S., Fligge, M., {et~al.} 1999, Solar polarization, 281

\bibitem[{Kirkpatrick {et~al.}(1999)Kirkpatrick, Reid, Liebert,
  {et~al.}}]{kirkpatrick1999}
Kirkpatrick, J., Reid, I., Liebert, J., {et~al.} 1999, \apj, 519, 802

\bibitem[{Kleman \& Uhler(1959)}]{kleman1959}
Kleman, B. \& Uhler, U. 1959, Canadian Journal of Physics, 37, 537

\bibitem[{Landau \& Lifshitz(1960)}]{landau8}
Landau, L. \& Lifshitz, E. 1960, Course of Theoretical Physics, Vol. 8.
  Electrodynamics Of Continuous Media (Pergamon Press)

\bibitem[{Mie(1908)}]{mie1908}
Mie, G. 1908, Annalen der {P}hysik, 330, 377

\bibitem[{Nakajima {et~al.}(1995)Nakajima, Oppenheimer, Kulkarni,
  {et~al.}}]{nakajima1995}
Nakajima, T., Oppenheimer, B., Kulkarni, S., {et~al.} 1995, \nat, 378, 463

\bibitem[{Pavlenko(1999)}]{pavlenko1999}
Pavlenko, Y. 1999, Astronomy Reports, 43, 748

\bibitem[{Ram {et~al.}(1993)Ram, Jarman, \& Bernath}]{ram1993}
Ram, R., Jarman, C., \& Bernath, P. 1993, Journal of Molecular Spectroscopy,
  161, 445

\bibitem[{Schadee(1978)}]{schadee1978}
Schadee, A. 1978, \jqsrt, 19, 517

\bibitem[{Shulyak {et~al.}(2010)Shulyak, Reiners, Wende,
  {et~al.}}]{shulyak2010}
Shulyak, D., Reiners, A., Wende, S., {et~al.} 2010, \aap, 523

\bibitem[{Solanki(1987)}]{solanki1987}
Solanki, S. 1987, PhD thesis, ETH Zuerich

\bibitem[{Sriramachandran \& Shanmugavel(2011)}]{sriramachandran2011}
Sriramachandran, P. \& Shanmugavel, R. 2011, \apss, 336, 379

\bibitem[{Witte {et~al.}(2009)Witte, Helling, \& Hauschildt}]{witte2009}
Witte, S., Helling, C., \& Hauschildt, P. 2009, \aap, 506, 1367

\end{thebibliography}

\end{document}